\documentclass[preprint,12pt]{elsarticle}
\usepackage{epsfig}
\usepackage{graphicx}
\def\figfig#1#2#3{\centerline{\epsfig{figure=figures/#2.eps,height=#1}}%
\caption{\small #3}\label{fig:#2}}

\journal{Icarus}
\begin{document}

\title{Asteroid families classification: \\ exploiting very large data sets}

\author[pi]{Andrea Milani}
\ead{milani@dm.unipi.it}
\author[to]{Alberto Cellino}
\author[bg]{Zoran Kne\v zevi\'c}
\author[bu]{Bojan Novakovi\'c}
\author[pi]{Federica Spoto} 
\author[pf]{Paolo Paolicchi} 

\address[pi]{Dipartimento di Matematica, Universit\`a di Pisa,
        Largo Pontecorvo 5,
        56127 Pisa, Italy}
\address[to]{INAF--Osservatorio Astrofisico di Torino,
10025 Pino Torinese, Italy}
\address[bg]{Astronomical Observatory, Volgina 7,
         11060 Belgrade 38, Serbia}
\address[bu]{Department of Astronomy, Faculty of Mathematics,
  University of Belgrade, Studenski trg 16, 11000 Belgrade, Serbia}
\address[pf]{Dipartimento di Fisica, Universit\`a di Pisa,
        Largo Pontecorvo 3,
        56127 Pisa, Italy}

\date{submitted, ???, 2013}
\begin{abstract}

The number of asteroids with accurately determined orbits
increases fast, and this increase is also accelerating.  The catalogs
of asteroid physical observations have also increased, although the
number of objects is still smaller than in the orbital catalogs.  Thus
it becomes more and more challenging to perform, maintain and update a
classification of asteroids into families.
To cope with these challenges we developed a new approach to the
asteroid family classification by combining the Hierarchical
Clustering Method (HCM) with a method to add new members to existing
families. This procedure makes use of the much larger amount of
information contained in the proper elements catalogs, with respect to
classifications using also physical observations for a smaller number
of asteroids.

Our work is based on the large catalog of the high accuracy synthetic
proper elements (available from AstDyS), containing data for $>
330\,000$ numbered asteroids. By selecting from the catalog a much
smaller number of large asteroids, we first identify a number of core
families; to these we attribute the next layer of smaller
objects. Then, we remove all the family members from the catalog, and
reapply the HCM to the rest.  This gives both halo families which
extend the core families and new independent families, consisting
mainly of small asteroids. These two cases are discriminated by
another step of attribution of new members and by merging intersecting
families. This leads to a classification with 128 families and
currently 87095 members; the number of members can be increased
automatically with each update of the proper elements catalog.

By using information from absolute magnitudes, we take advantage of
the larger size range in some families to analyze their shape in the
proper semimajor axis vs. inverse diameter plane.  This leads to a new
method to estimate the family age, or ages in cases where we identify
internal structures. The analysis of the plot above evidences some
open problems but also the possibility of obtaining further
information of the geometrical properties of the impact
process. The results from the previous steps are then analyzed, using
also auxiliary information on physical properties including WISE
albedos and SDSS color indexes.  This allows to solve some difficult
cases of families overlapping in the proper elements space but
generated by different collisional events.

The families formed by one or more cratering events are found to be
more numerous than previously believed because the fragments are
smaller.  We analyze some examples of cratering families (Massalia,
Vesta, Eunomia) which show internal structures, interpreted as
multiple collisions. We also discuss why Ceres has no family.

\end{abstract}

\maketitle


\section{Introduction}
\label{sec:intro}

Asteroid families are a powerful tool to investigate the collisional
and dynamical evolution of the asteroid belt. If they are correctly
identified, they allow to describe the properties of the parent
body, the collisional event(s) generating the family and the subsequent
evolution due to chaotic dynamics, non-gravitational perturbations,
and secondary collisions.

However, asteroid families are statistical entities. They can be
detected by density contrast with respect to a background in the phase
space of appropriate parameters, but their reliability strongly
depends upon the number of members, the accuracy of the parameters and
the method used for the classification. The exact membership cannot be
determined, because the portion of phase space filled by the family
can have already been occupied by some background asteroids, thus
every family can and usually does have
interlopers \cite{Migliorinietal95}. Moreover, the exact boundary of
the region occupied by the family corresponds to marginal density
contrast and thus cannot be sharply defined.

The problem is, the purpose of a classification is not just to get as
many families as possible, but to obtain families with enough members,
with a large enough spread in size, and accurate enough data to derive
quantities such as number of family generating collisional events, age
for each of them, size of parent bodies, size distribution of
fragments (taking into account observational bias), composition, and
flow from the family to Near Earth Asteroid (NEA) and/or cometary type
orbits.

This has three important implications. First, the quality of a family
classification heavily depends on the number of asteroids used, and
upon the accuracy of the parameters used in the process. The number of
asteroids is especially critical, because small number statistics
masks all the information content in small families. Moreover, there
is no way of excluding that families currently with few members are
statistical flukes, even if they are found likely to be significant by
statistical tests.

Second, different kinds of parameters have very different value in a
family classification.  This can be measured by the information
content, which is, for each parameter, the base 2 logarithm of the
inverse of relative accuracy, summed for all asteroids for which such
data are available; see Section~\ref{sec:dataset}. By using this
measure, it is easy to show that the dynamical parameters, such as the
proper elements\footnote{As an alternative the corresponding
  frequencies $n,g,s$ can be used, the results should be the same if
  the appropriate metric is used \cite{propfreq}.}  $a,e,I$ have a
much larger information content than the physical observations,
because the latter are either available only for much smaller catalogs
of asteroids, or with worse relative accuracy, or both. As an example,
absolute magnitudes are available for all asteroids with good orbits
and accurate proper elements, but they are computed by averaging over
inhomogeneous data with poorly controlled quality. There are catalogs
with albedos such as WISE, described in \cite{WISE, NEOWISE}, and
color indexes such as the Sloan Digital Sky Survey (SDSS), described in
\cite{sdss}, but the number of asteroids included is smaller by
factors 3 to 5 with respect to proper elements catalogs, and this
already includes many objects observed with marginal S/N, thus poor
relative accuracy.

Third, catalogs with asteroid information grow with time at a rapid
and accelerating pace: e.g., we have used for this paper a catalog,
distributed from
AstDyS\footnote{http://hamilton.dm.unipi.it/astdys2/index.php?pc=5}
with $336\, 219$ numbered asteroids with synthetic proper elements computed
up to November 2012. By April 2013 the same catalog had grown to
$354\, 467$ objects, i.e., $5.4\%$ in 5 months. If the rate of asteroid
numbering was to continue at this rate, a pessimistic assumption,
in less than 8 years the catalog would be doubled. Catalogs of
physical observations are also likely to grow, although in a less
regular and predictable way. Thus the question is whether the increase
of the catalogs will either confirm or contradict the conclusions we
draw at present; or better, the goal should be to obtain robust
conclusions which will pass the test of time.

As a consequence our purpose in this paper is not to make "yet
another asteroid family classification", but to setup a
classification which can be automatically updated every time the
dataset is increased.  We are going to use the proper elements
first, that is defining "dynamical families", then use information
from absolute magnitudes, albedos and generalized color indexes,
when available and useful, as either confirmation or rejection, e.g,
identification of interlopers and possibly of intersections of
different collisional families.  We will continue to update
the classification by running at short intervals (few months) a
fully automated "classification" computer program which attaches
newly numbered objects to existing families.

This will not remove the need to work on the identification of new
families and/or on the analysis of the properties of the already known
families. Every scientist will be practically enabled to do
this at will, since for our dataset and our classification we follow a
strict open data policy, which is already in place when this paper is
submitted: all the data on our family classification is already
available on AstDyS, and in fact it has already been updated with
respect to the version described in this paper. 

We have also recently made operational a new graphic interface on
AstDyS providing figures very much like the examples shown in this
paper\footnote{http://hamilton.dm.unipi.it/astdys2/Plot/index.html}.
The senior authors of this paper have learned much on the books
Asteroids I and Asteroids II which contained Tabulation sections with
large printed tables of data. This has been abolished in Asteroids
III, because by the the year 2000 the provision of data in electronic
form, of which AstDyS is just one example, was the only service
practically in use. It is now time to reduce the number of figures,
especially in color, and replacing them with ``graphic servers''
providing figures based upon the electronically available data and
composed upon request from the users: this is what the AstDyS graphic
tool is proposing as an experiment which may be followed by others.
Readers are warmly invited to take profit of this tool, although the
most important statements we do in the discussion of our results are
already supported by a (forcibly limited) number of explanatory
figures.

The main purpose of this paper is to describe and make available some
large data sets, only some of the interpretations are given, mostly as
examples to illustrate how the data could be used. We would like to
anticipate one major conceptual tool, which will be presented in the
paper. This concerns the meaning of the word ``family'', which has
become ambiguous because of usage by many authors with very different
background and intent.

We shall use two different terms: since we believe the proper elements
contain most of the information, we perform a family classification
based only upon them, thus defining \textit{dynamical families}. A
different term \textit{collisional families} shall be used to indicate
a set of asteroids in the catalog which have been formed at once by a
single collision, be it either a catastrophic fragmentation or a
cratering event. Note that there is no reason to think that the
correspondence between dynamical families and collisional families
should be one to one. A dynamical family may correspond to multiple
collisional events, both on the same and on different parent bodies. A
collisional family may appear as two dynamical families because of a
gap due to dynamical instability. A collisional family might have
dissipated long ago, leaving no dynamical family. A small dynamical
family might be a statistical fluke, with no common origin for its
members. In this paper we shall give several examples of these
non-correspondences.

\section{Dataset}
\label{sec:dataset}

In this section we list all datasets used in our classification and in
its analysis contained in this paper. 

\subsection{Proper elements}

Proper elements $a, e, \sin{i}$ are three very accurate parameters,
and we have computed, over many years up to November 2012, synthetic
proper elements \cite{synthpro1,synthpro2} for $336\,319$ asteroids.
We made a special effort to recompute synthetic proper elements for
asteroids missing in our database because of different reasons: being
close to strong resonances, suffering close encounters with major
planets, or having high eccentricities and inclinations. In particular
we aimed at increasing the number of objects in the high $e,I$
regions, in order to improve upon the sparse statistics and the
reliability of small families therein. We thus integrated orbits of a
total of $2\,222$ asteroids, out of which the proper elements
computation failed for only $62$ of them. The rest are now included in
the AstDyS synthetic proper elements files. This file is updated a few
times per year.

For each individual parameter in this dataset there are available both
standard deviation and maximum excursion obtained from the analysis of
the values computed in sub-intervals \cite{synthpro1}. If an asteroid
has large instabilities in the proper elements, as it happens for high
proper $e, \sin{I}$, then  the family classification can be dubious.

The same catalog contains also absolute magnitudes and estimates of
Lyapounov Characteristic Exponents, discussed in the following subsections.

%bad=sa>3e-4|LCE>50;

\subsection{Absolute magnitudes}

Another piece of information is the set of absolute magnitudes
available for all numbered asteroids computed by fitting the apparent
magnitudes obtained incidentally with the astrometry, thus stored in
the Minor Planet Center (MPC) astrometric database. The range of
values for all numbered asteroids is $15.7$, the accuracy is difficult
to be estimated because the incidental photometry is very
inhomogeneous in quality, and the information on both S/N and
reduction method is not available.

The sources of error in the absolute magnitude data are not just the
measurement errors, which are dominant only for dim asteroids, but
also star catalog local magnitude biases, errors in conversion from
star-calibrated magnitudes to the assumed V magnitude used in absolute
magnitudes, and the lightcurve effect.  Moreover, the brightness of an
asteroid changes at different apparitions as an effect of shape and
spin axis orientation, so strictly speaking the absolute magnitude is
not a real constant.  Last but not least, the absolute magnitude is
defined at zero phase angle (ideal solar opposition) thus some
extrapolation of observations obtained in different illumination
conditions is always needed, and this introduces other significant
errors, especially for high phase angles. The standard deviation of
the apparent magnitude residuals has a distribution peaked at $0.5$
mag: since numbered asteroids have in general many tens of magnitude
observations, the formal error in the absolute magnitude, which is
just a corrected average, is generally very small, but biases can be
very significant. Thus we do not have a reliable estimate of the
accuracy for the large dataset of $336\,219$ absolute magnitudes
computed by AstDyS, we can just guess that the standard deviation
should be in the range $0.3\div 0.5$ magnitudes.

The more optimistic estimate gives a large information content (see
Table~\ref{tab:infocount}), which would be very significant, but there are
many quality control problems. Other databases of photometric
information with better and more consistent information are available,
but the number of objects included is much smaller, e.g., $583$
asteroids with accuracy better than $0.21$ magnitudes
\cite{pravec2012}: these authors also document the existence of
serious systematic size-dependent biases in the H values.

\subsection{WISE and SDSS catalogs of physical observations}

The WISE catalog of albedos\footnote{Available at URL
  http://wise2.ipac.caltech.edu/staff/bauer/NEOWISE\_pass1} has
information on $94\,632$ numbered asteroids with synthetic proper
elements, but the relative accuracy is moderate: the average standard
deviation (STD) is $0.045$, but the average ratio between STD and
value is $0.28$; e.g., the asteroids in the WISE catalog for which the
albedo measured is $<3$ times the STD are $26\%$ (we are going to use
measure $>3$ times STD as criterion for using WISE data in
Section~\ref{sec:use_physical}).

This is due to the fact that WISE albedos, in particular for small
asteroids, have been derived from a \textit{measured} infrared flux
and \textit{an estimated} visible light flux derived from an adopted
nominal value of absolute magnitude. Both terms are affected by random
noise increasing for small objects, and by systematics in particular
in the visible, as outlined in the previous subsection. In principle
one should use a value of absolute magnitude not only accurate, but
also corresponding to the same observation circumstances of the
thermal IR observations, which is seldom the case.  For comparatively
large asteroids, albedo can be constrained also from the ratios
between different infrared channels of WISE, thus the result may be
less affected by the uncertainty of the absolute magnitude.

The 4th release of the SDSS MOC\footnote{Available at URL
  http://www.astro.washington.edu/users/ivezic/sdssmoc/} contains data
for 471 569 moving objects, derived from the images taken in five
filters, $u$, $g$, $r$, $i$, and $z$, ranging from 0.3 to 1.0
$\mu$m. Of those, 123 590 records refer to asteroids present in our
catalog of proper elements. As many of these objects have more than
one record in the SDSS catalog, a total number of asteroids present in
both catalogs is 61 041. The latter number was then further decreased
by removing objects having error in magnitude larger than 0.2 mag in
at least one band (excluding the $u$-band which was not used in our
analysis). This, non-conservative limit, is used to remove only
obviously degenerate cases. Thus, we used the SDSS MOC 4 data for 59
975 numbered asteroids.

It is well known that spectrophotometric data is of limited accuracy,
and should not be used to determine colors of single objects. Still, 
if properly used, the SDSS data could be useful in some situations,
e.g. to detect presence of more than one collisional family inside a dynamical family,
or to trace objects escaping from the families.

Following \cite{parker2008} we have used $a^{*}$, the first principal
component\footnote{According to \cite{sdss} the first principal
  component in the $r-i$ vs $g-r$ plane is defined as $a^{*} =
  0.89(g-r) + 0.45(r-i) - 0.57$.} in the $r-i$ versus $g-r$
color-color plane, to discriminate between $C$-type asteroids
($a^{*}<0$) and $S$-type asteroids ($a^{*}\geq0$). Then, among the
objects having ($a^{*}\geq0$) the $i-z$ values can be used to further
discriminate between $S$- and $V$-type, with the latter one
characterized by the $i-z < -0.15$.  The average standard deviation of
data we used is $0.04$ for $a^*$, $0.08$ for $i-z$.

\subsection{Resonance identification}

Another source of information available as an output of the numerical
integrations used to compute synthetic proper elements is an estimate
of the maximum Lyapounov Characteristic Exponent (LCE). 
The main use of this is to identify asteroids affected by chaos over
comparatively short time scales (much shorter than the ages of the
families)\footnote{Every asteroid is affected by chaotic effects over
  timescales comparable to the age of the solar system, but this does
  not matter for family classification.}. These are mostly due to mean
motion resonances with major planets (only resonances with Jupiter,
Saturn and Mars are affecting the Main Belt at levels which are
significant for family classification). Thus we use as criterion to
detect these ``resonant/chaotic'' asteroids the occurrence of at least
one of the following: either a LCE $> 50$ per Million years (that is a
Lyapounov time $<20\,000$ years) or $STD(a)>3\times 10^{-4}$ au.

Note that the numerical integrations done to compute proper elements
use a dynamical model not including any asteroid as perturber. This is
done for numerical stability reasons, because all asteroids undergo
mutual close approaches and these would need to be handled accurately,
which is difficult while integrating hundreds of asteroids
simultaneously. Another reason for this choice is that we wish to
identify specifically the chaos which is due to mean motion resonances
with the major planets. As shown by \cite{laskar1}, if even a few
largest asteroids are considered with their mass in the dynamical
model, then all asteroids are chaotic with Lyapounov times of a few
$10\,000$ years. However, the long term effect of such chaos
endogenous to the asteroid belt is less important than the chaos
generated by the planetary perturbations \cite{laskar2}. 

The asteroid perturbers introduce many new frequencies, resulting in an
enormous increase of the Arnold web of resonances, to the point of
leaving almost no space for conditionally periodic orbits, and the
Lyapounov times are short because the chaotic effects are driven by
mean motion resonances. However, these resonances are extremely weak,
and they do not result in large scale instability, not even over time
spans of many thousands of Lyapounov times, the so called ``stable
chaos'' phenomenon \cite{helga}. In particular, locking in a stable
resonance with another asteroid is almost impossible, the only known
exception being the $1/1$ resonance with Ceres, see the discussion
about the Hoffmeister family in Section~\ref{sec:haloproblem} and
Figure~\ref{fig:Hoffmeister_aI}. This implies that the
(size-dependent) Yarkovsky effect, which accumulates secularly in time
in semimajor axis, cannot have secular effects in eccentricity and
inclination, as it happens when there is capture in resonance.

We have developed a sensitive detector of mean motion resonances with
the major planets, but we would like to know which resonance,
which is the integer combination of mean motions forming the ``small
divisor''. For this we use the catalog of asteroids in mean motion
resonances by \cite{smirnov}, which has also been provided to us by
the authors in an updated and computer readable form. This catalog
will continue to be updated, and the information will be presented
through the AstDyS site.

Asteroid families are also affected by secular resonances, with
``divisor'' formed with an integer combination of frequencies
appearing in the secular evolution of perihelia and nodes, namely $g,
g_5, g_6$ for the perihelia of the asteroid, Jupiter and Saturn, and
$s, s_6$ for the ones in the nodes of the asteroid and
Jupiter\footnote{In the Hungaria region even some resonances involving
  the frequencies $g_3,g_4, s_4$ for Earth and Mars can be significant
  \cite{hungaria}.}. The data on the asteroids affected by secular
resonances can be found with the analytic proper elements, computed by
us with methods developed in the 90s \cite{propel1, propel2,
  propel3}. In these algorithms, the small divisors associated with
secular resonances appear as obstruction to the convergence of the
iterative algorithm used to compute proper elements, thus error codes
corresponding to the secular resonances are reported with the proper
elements\footnote{We must admit these codes are not user friendly,
  although a Table of conversion from the error codes to the small
  divisors is given in \cite[table 5.1]{propel1}. We shall try to
  improve the AstDyS user interface on this.}.

Note that we have not used analytic proper elements as a primary set
of parameters for family classification, since they are significantly
less accurate (that is, less stable in time over millions of years)
than the synthetic ones, by a factor about 3 in the low $e$ and $I$
portion of the main belt. The accuracy becomes even worse for high
$e,I$, to the point that for $\sqrt{e^2+\sin^2{I}}>0.3$ the analytical
elements are not even computed \cite{compareproper}; they are also
especially degraded in the outer portion of the main belt, for $a>3.2$
au, due to too many mean motion resonances.  On the other hand,
analytic proper elements are available for multiopposition asteroids,
e.g., for $98\, 926$ of them in November 2012, but these would be more
useful in the regions where the number density of numbered asteroids
is low, which coincide with the regions of degradation: high $e,I$ and
$a>3.2$ au.  It is also possible to use proper elements for
multiopposition asteroids to confirm and extend the results of family
classification, but for this purpose it is, for the moment,
recommended to use ad hoc catalogs of synthetic proper elements
extended to multiopposition, as we have done in
\cite{hungaria,bojan_highi}.

\subsection{Amount of information}
 
For the purpose of computing the information content of each entry of
the catalogs mentioned in this section, we use as definition of
inverse relative accuracy the ratio of two quantities:

1. for each parameter, the useful range, within which most ($>
99\%$) of the values are contained;

2. the standard deviation, as declared in each catalog,
for each individual computed parameter.

Then the content in bit of each individual parameter is the base 2
logarithm of this ratio. These values are added for each asteroid in
the catalog, thus forming a total information content, reported in the
last column of Table~\ref{tab:infocount} in Megabits. For statistical
purposes we give also the average number of bits per asteroid in the
catalog.

\begin{table}[h]
\footnotesize
 \centering
  \caption{An estimate of the information content of catalogs. The
    columns give: parameters contained in the catalogs, minimum and
    maximum values and range of the parameters, average information
    content in bits for a single entry in the catalog, number of
    entries in the catalog and total information content.}
  \label{tab:infocount}
\medskip
  \begin{tabular}{lrrrrrr}
  \hline
parameter &  min  &  max   & range  &  av.bits  & number &  tot Mb \\
 \hline
\\
a (au)  & 1.80  & 4.00  & 2.20 & 16.7 &336219 & 5.63\\

e       & 0.00 & 0.40  & 0.40 & 10.7 &336219 & 3.59\\

sin I   & 0.00 & 0.55  & 0.55 & 15.1 &336219 &5.08\\

total& & & & & & 14.39 \\
\\
\hline
\\
H       & 3.30 & 19.10  & 15.8 & 5.7 &336219& 1.92 \\
\\
\hline
\\
albedo  & 0.00   & 0.60   & 0.60 & 4.5 &94632& 0.43\\
\\
\hline
\\
a*     & -0.30 & 0.40   & 0.70 & 4.4 &59975 & 0.26\\

i-z    & -0.60 & 0.50   & 1.10 & 4.0 &59975 & 0.24\\ 

total  & &  & & & & 0.50\\

\\
\hline
\end{tabular}
\end{table}

Note that for the absolute magnitude we have assumed a standard
deviation of $0.3$ magnitudes for all, although this might be
optimistic.

With the numbers in Table~\ref{tab:infocount} we can estimate the
information content of our synthetic proper element catalog to be
about $14$ Megabits, the absolute magnitudes provide almost $2$
megabits with somewhat optimistic assumptions, the physical data
catalogs WISE and SDSS are $1$ Megabit together.

\section{Method for large dataset classification} \label{sec:method}

Our adopted procedure for family identification is largely based on
applications of the classical Hierarchical Clustering Method (HCM)
already adopted in previous families searches performed since the
pioneering work by \cite{zapetal90}, and later improved in a number
of papers \cite{zapetal94, zapetal95, hungaria,
  bojan_highi}. Since the method has been already extensively
explained in the above papers, here we will limit ourselves to a very
quick and basic description.

We have divided the asteroid belt in zones, corresponding to different
intervals of heliocentric distance, delimited by the most important
mean-motion resonances with Jupiter. These resonances correspond to
Kirkwood gaps wide enough to exclude family classification across the
boundaries.

\begin{table}[h]
\footnotesize
 \centering
  \caption{Summary of the relevant parameters for application of the HCM.}
  \label{tab:zones}
\medskip
  \begin{tabular}{lccrrrrr}
  \hline
\\
Zone &  $\sin{I}$ &     range $a$  &  $N$ (total) & $H_{completeness}$ & $N(H_{completeness})$ &
$N_{min}$  & \multicolumn{1}{c}{QRL} \\
     &            &                &              & \multicolumn{1}{c}{(when used)} &
     \multicolumn{1}{c}{(when used)} &  & \multicolumn{1}{c}{(m/s)} \\
\\
 \hline
\\
  1  &     $>0.3$  &   $1.600\div 2.000$  &   4249 &     &      & 15  &  70 \\
  2  &     $<0.3$  &   $2.065\div 2.501$  & 115004 & 15.0& \multicolumn{1}{c}{15888} & 17 & 70 \\
  2  &     $>0.3$  &   $2.065\div 2.501$  &   2627 &     &      &  11 & 130 \\
  3  &     $<0.3$  &   $2.501\div 2.825$  & 114510 & 14.5& \multicolumn{1}{c}{16158} & 19 & 90 \\
  3  &     $>0.3$  &   $2.501\div 2.825$  &   3994 &     &      &   9 & 140 \\
  4  &     $<0.3$  &   $2.825\div 3.278$  &  85221 & 14.0& \multicolumn{1}{c}{14234} & 17 & 100 \\
  4  &     $>0.3$  &   $2.825\div 3.278$  &   7954 &     &      & 12  & 80 \\
  5  &     all     &   $3.278\div 3.700$  &    991 &     &      & 10  & 120 \\
  6  &     all     &   $3.700\div 4.000$  &   1420 &     &      & 15  & 60 \\
\\
 \hline
\end{tabular}
\end{table}

As shown in Table~\ref{tab:zones}, our ``zone 1'' includes objects
having proper semi-major axes between 1.6 and 2 au. In this region,
only the so-called Hungaria objects at high inclination ($\sin I \ge
0.3$) are dynamically stable \cite{hungaria}. Our ``zone 2''
includes objects with proper orbital elements between 2.067 and
2.501 au. The ``zone 3'' is located between 2.501 and 2.825 au, and
``zone 4'' between 2.825 and 3.278 au. Zones 2, 3 and 4 were already
used in several previous analyzes by our team. In addition, we use
also a ``zone 5'', corresponding to the interval between 3.278 and
3.7 au., and a ``zone 6'', extending between 3.7 and 4.0 au. (Hilda
zone).

Moreover, some semi-major axis zones have been also split by the value
of proper sin I, between a moderate inclination region $\sin I < 0.3$
and a high inclination region $\sin I > 0.3$. In some zones the
boundary value corresponds to a gap due to secular resonances and/or
stability boundaries. E.g., in zone 1 the moderate inclination region
is almost empty and contains very chaotic objects (interacting
strongly with Mars). In zone 2 the $g - g_6$ secular resonance clears a
gap below the Phocaea region. In zones 3 and 4 there is no natural
dynamical boundary which could be defined by inclination only, and
indeed we have problems with families found in both regions.
In zones 5 and 6 there are few high inclination asteroids, and
a much smaller population.

A metric function has been defined to compute the mutual distances
of objects in the proper element space. Here, we have adopted the
so-called ``standard metric'' $d$ proposed by \cite{zapetal90}, and
since then adopted in all HCM-based family investigations. We remind
that using this metric the distances between objects in the proper
element space correspond to differences of velocity and are given in
m/s.

Having adopted a metric function, the HCM algorithm allows the users
to identify all existing groups of objects which, at any given value
of mutual distance $d$, are linked, in the sense that for each
member of a group there is at least one other member of the same
group which is closer than $d$. The basic idea is therefore to
identify groups which are sufficiently populous, dense and compact
(i.e., include large numbers of members down to small values of
mutual distance) to be reasonably confident, in a statistical sense,
that their existence cannot be a consequence of random fluctuations
of the distribution of the objects in the proper element space.

In this kind of analysis, which is eminently statistical, a few
parameters play a key role. The most important ones are the minimum
number of objects $N_{min}$ required for groups to be considered as
candidate families, and the critical level of distance adopted for
family identification. As for $N_{min}$, it is evident that its
choice depends on the total number of objects present in a given
region of the phase space. At the epoch of the pioneering study by
\cite{zapetal90}, when the total inventory of asteroids with
computed proper elements included only about 4,000 objects in the
whole main belt, its value was chosen to be 5. Since then, in
subsequent analyzes considering increasingly bigger datasets, the
adopted values of $N_{min}$ were scaled as the square root of the
ratio between the numbers of objects in the present dataset and the
one in some older sample in the same volume of the proper element
space. We follow this procedure also in the present paper, so we
chose the new $N_{min}$ values by scaling with respect to the
$N_{min}$ adopted by \cite{zapetal95} for the low-I portions of
zones 2, 3, and 4, and \cite{bojan_highi} for the high-I portions of
the same zones. Zone 5 and 6 were analyzed for the first time, they
contain relatively low numbers of objects, and for them we adopted
$N_{min}$ values close to $1\%$ of the sample, after checking that
this choice did not affect severely the results.

As for the critical distance level, it is derived by generating
synthetic populations (``Quasi-Random Populations'') of fictitious
objects occupying the same volume of proper element space, and
generated in such a way as to have {\em separately} identical
distributions of $a$, $e$ and $\sin I$ as the real asteroids present
in the same volume. An HCM analysis of these fictitious populations is
then performed, and this makes it possible to identify some critical
values of mutual distance below which it is not possible to find
groupings of at least $N_{min}$ quasi-random objects. All groups of
real objects which include $N_{min}$ members at distance levels below
the critical threshold, are then considered as dynamical
families. Note also that we always looked at groupings found at
discrete distance levels separated by steps of 10 m/s, starting from a
minimum value of 10 m/s.

As for the practical application of the method described above, one
might paradoxically say that this is a rare case, in the domain of
astrophysical disciplines, in which the abundance of data, and not
their scarcity, starts to produce technical problems. The reason is
that the inventory of asteroids for which the orbital proper
elements are available is today so big, that difficult problems of
overlapping between different groupings must be faced. In other
words, a simple application of the usual HCM procedures developed in
the past to deal with much smaller asteroid samples would be highly
problematic now in many respects, especially the phenomenon of
\emph{chaining} by which obviously independent families get attached
together by a thin chain of asteroids.

For these reasons, when necessary (i.e., in the most populous zones
of the asteroid belt) we have adopted in this paper a new,
multistep procedure, allowing us to deal at each step with
manageable numbers of objects, and developing appropriate
procedures to link the results obtained in different steps.

\subsection{Step 1: Core families}
\label{sec:hcm}

In order to cope with the challenge posed by the need of analyzing
very big samples of objects in the most populous regions of the
belt, namely the low-inclination portions of zones 2, 3 and 4, as a
first step we look for the cores of the most important families
present in these regions.

In doing this, we take into account that small asteroids below one or
few km in size are subject to comparatively fast drifts in semi-major
axis over time as the consequence of the Yarkovsky effect. Due to this
and other effects (including low-energy collisions, see
\cite{Delloroetal12}) the cloud of smallest members of a newly born
family tends to expand in the proper element space and the family
borders tend to ``blur'' as a function of time. For this reason, we
first removed from our HCM analysis of the most populous regions the
small asteroids. In particular, we removed all objects having absolute
magnitudes $H$ fainter than a threshold roughly corresponding to the
completeness limit in each of the low-I portions of zones 2, 3 and
4. These completeness limits, listed in Table~\ref{tab:zones}, were
derived from the cumulative distributions of $H$; for the purposes of
our analysis, the choice of this threshold value is not
critical. Having removed the objects having $H$ fainter than the
threshold value, we were left with much more manageable numbers of
asteroids, see Table~\ref{tab:zones}.

To these samples, we applied the classical HCM analysis procedures.
As a preliminary step we considered in each zone samples of $N$
completely random synthetic objects ($N$ being the number of real
objects present in the zone), in order to determine a distance value
$RL$ at which these completely random populations could still
produce groups of $N_{min}$ members. Following \cite{zapetal95}, in
order to smooth a little the quasi-Random populations to be created
in each region, groups of the real population having more than
$10\%$ of the total population at $RL$ were removed and substituted
by an equal number of fully-random objects. The reason of this
preliminary operation is to avoid that in the real
population, a few very big and dense groups could be exceedingly
dominant and could affect too strongly the distributions of proper
elements in the zone, obliging some bins of the $a, e, \sin{I}$
distribution, from which the QRL population is built (see below), to
be over-represented. This could affect the generation of
Quasi-Random objects, producing an exceedingly deep (low)
Quasi-Random level ($QRL$) of distance, leading to a too severe
criterion for family acceptance. In Zones 3 and 4 a few groups
(only one group in Zone 3 and two groups in Zone 4) were
first removed and substituted by equal numbers of
randomly generated clones. In Zone 2, the $RL$ turned out to be
exceedingly high: 160 m/s, corresponding to a distance level
at which practically the whole real population merges in a unique
group. Removing real objects at that level would have meant to
substitute nearly the whole real population by fully-random objects.
So this substitution was \textit{not} done in Zone 2.

After that, we ran the classical generations of quasi-random
populations. In each zone, the distributions of proper a, e, and sin
I to be mimicked by the quasi-random populations were subdivided in
number of bins, as already done in previous papers. In each zone, we
determined the minimum level of distance for which groupings of
$N_{min}$ objects could be found. We considered as the critical
Quasi Random level $QRL$ in each zone the minimum value obtained in
ten generations. The $QRL$ values adopted in each zone are also
listed in Table~\ref{tab:zones}.

Then we run the HCM algorithm on the real proper elements.  Families
were defined as the groups having at least $N_{min}$ members at
distance levels 10 m/s lower than $QRL$, or just reaching $QRL$, but
with a number of members $N \ge N_{min} + 2 \sqrt{N_{min}}$. The
families obtained in this first step of our procedure include only a
small subset, corresponding to the largest objects, of the asteroids
present in the low-I portion of zones 2, 3 and 4. For this reason, we
call them ``core families'': they represent the inner ``skeletons'' of
larger families present in these zones, whose additional members are
then identified by the following steps of the procedure (see below
Figure~\ref{fig:20_vshapea}).

In the case of the high-I portions of zones 2, 3 and 4, and the entire
zones 1, 5 and 6, the number of asteroids is not extremely large, and
we identified families within them by applying the classical HCM
procedure, without multistep procedure. For each family, the members
were simply all the objects found to belong to it at the resulting QRL
value of the zone. In other words, we did not adopt a case-by-case
approach based on looking at the details of the varying numbers of
objects found within each group at different distance levels, as was
done by \cite{bojan_highi} to derive memberships among
high-inclination families.

\subsection{Step 2: Attaching smaller asteroids to core families}

The second step of the procedure in the low-I portions of zones 2, 3
and 4 was to \textit{classify} individual asteroids, which had not
been used in the core classification, by attaching some of them to the
established family cores. For this we used the same QRL distance
levels used in the identification of the family cores, but we allowed
only single links for attachment, because otherwise we would get
chaining, with the result of merging most families together. In other
words, in step 2 we attribute to the core families the asteroids
having a distance from at least one member (of the same core family)
not larger than the QRL critical distance. The result is that the
families are extended in the absolute magnitude/size dimension, not
much in proper elements space, especially not in proper $a$ (see
Figure~\ref{fig:20_vshapea}).

Since this procedure has to be used many times (see
Section~\ref{automatic}), it is important to use an efficient
algorithm. In principle, if the distance has to be $d<QRL$, we need to
compute all distances, e.g., with $M$ proper elements we should
compute $M\cdot (M-1)/2$ distances, then select the ones $<QRL$. The
computational load can be reduced by the partition into regions, but
with zones containing $>100\,000$ asteroids with proper elements it is
anyway a time consuming procedure.

This problem has a solution which is well known, although it may not
have been used previously in asteroid family classification. Indeed,
the problem is similar to the one of comparing the computed
ephemerides of a catalog of asteroids to the available observations
from a large survey \cite[Section 11.3]{orbdet}. We select one
particular dimension in the proper elements space, e.g, $\sin{I}$; we
sort the catalog by the values of this proper element. Then, given the
value of $\sin{I_0}$ for a given asteroid, we find the position in the
sorted list, then scan the list starting from this position up and
down, until the values $\sin{I_0}+QRL$ and $\sin{I_0}-QRL$,
respectively, are exceeded. In this way the computational complexity
for the computation of the distances $<QRL$ for $M$ asteroids is of
the order of $M\,log_2(M)$, instead of $M^2$. The large distances are
not computed, even less stored in the computer memory.

%In practice, about 1 hour of CPU on a single processor is
%enough for the step 2.

\subsection{Step 3: Hierarchical clustering of intermediate background}

As an input to the third step in the low-I portion of zones 2, 3, and
4, we use the ``intermediate background'' asteroids, defined as the
set of all the objects not attributed to any family in steps 1 and
2. The HCM procedure was then applied to these objects, separately in
each of the three zones.

The numbers of objects left for step 3 of our analysis were 99399 in
zone 2, 94787 in zone 3 and 57422 in zone 4. The corresponding values
of $N_{min}$ were 42, 46 and 34, respectively, adopting the same
criterion \cite{zapetal95} already used for core families. The same
HCM procedures were applied, with only a few differences.  In
computing the critical $QRL$ distance threshold, we did not apply any
preliminary removal of large groupings of real objects, because {\em a
  priori} we were not afraid to derive in this step of our procedure
rather low values of the QRL distance level threshold. The reason is
that, dealing with very large numbers of small asteroids, we
adopted quite strict criteria for family acceptance, in order to
minimize the possible number of false grouping, and to
reduce the chance of spurious family merging.

By generating in each of the three zones 10 synthetic quasi-random
populations, and looking for the deepest groups of $N_{min}$ objects
to determine the $QRL$, we obtained the following $QRL$ values: 50, 60
and 60 m/s for zones 2, 3 and 4, respectively. Following the same
criteria used for core families, step 3 families need to be found as
groupings having at least $N_{min}$ members at 40 m/s in zone 2, and
50 m/s in the zones 3 and 4.

On the other hand, as mentioned above, in identifying step 3 families
we are forced to be quite conservative. This is due to the known
problems of overlapping between different families as a consequence of
the intrinsic mobility of their smallest members in the proper element
space, as a consequence of (primarily) Yarkovsky as well as of
low-energy collisions. For this reason, we adopted a value of 40 m/s
for step 3 family identification in all three zones. We also
checked that adopting a distance level of 50 m/s in zones 3 and 4
would tend to produce an excessive effect of chaining, which would
suggest merging of independent families.

If collisional processes produce overlapping of members of different
families in the proper element space, we can reduce this effect
on our family classification only at the cost of being somewhat
conservative in the identification of family memberships.

Families identified at this step are formed by the population of
asteroids left after removing from the proper elements dataset family
members identified in steps 1 and 2 of our procedure. There are
therefore essentially two possible cases: ``step 3'' families can
either be fully independent, new families having no relation with the
families identified previously, or they may be found to overlap ``step
1+2'' families, and to form ``haloes'' of smaller objects surrounding
some family cores. The procedure adopted to distinguish these two
cases is described in the following.

\subsection{Step 4: Attaching background asteroids to all families}

After adding the step 3 families to the list of core families of step
1, we repeat the procedure of attribution with the same algorithm of
step 2.  The control value of distance $d$ for attribution to a given
family is the same used in the HCM method as QRL for the same family;
thus values are actually different for step 1 and step 3 families, even
in the same zone.

If a particular asteroid is found to be attributed to more than one
family, it can be considered as part of an intersection. A small
number of these asteroids with double classification is unavoidable,
as a consequence of the statistical nature of the procedure. However,
the concentration of multiple intersections between particular families
requires some explanation.

One possible explanation is due to the presence of families at the
boundaries between high and low inclination regions in zones 3 and 4,
where there is no gap between them. Indeed, the classification has
been done for proper $\sin{I}>0.29$ for the high inclination regions,
for $\sin{I}<0.3$ for the low inclination. This implies that in the
overlap strip $0.29< \sin{I}<0.30$ some families are found twice,
e.g,, family 729. In other cases two families are found with
intersections in the overlap strip. This is obviously an artifact of
our decomposition in zones and needs to be corrected by merging the
intersecting families.

\subsection{Step 5: Merging halo families with core families}

Another case of family intersections, created by step 4, is the ``halo
families''. This is the case when a new family appears as an extension of an
already existing family identified at steps 1 and 2, with intersections 
near the boundary.

In general for the merging of two families we require as a criterion
multiple intersections. Visual inspection of the three planar
projections of the intersecting families proper elements is used as
a check, and sometimes allows to assess ambiguous cases.

Of the 77 families generated by HCM in step 3, we have considered 34
to be halo. Even 2 core families have been found to be halo of other
core families and thus merged. There are of course dubious cases, with
too few intersections, as discussed in
Section~\ref{sec:haloproblem}. Still the number of asteroids belonging
to intersections decreases sharply, e.g., in the two runs of single-step
linkage before and after the step 5 mergers, the number of asteroids
with double classification decreased from $1\,042$ to $29$.

Note that the notion of ``halo'' we are using in this paper does not
correspond to other definitions found in the literature, e.g.,
\cite{brozmorby,carruba}. The main difference is that we have on
purpose set up a procedure to attach ``halo families'' formed on
average by much smaller asteroids, thus we have used the absolute
magnitude information to decompose the most densely populated regions.
This is consistent with the idea that the smaller asteroids are more
affected by the Yarkovsky effect, which is inversely proportional to
the diameter: it spreads the proper elements $a$ by the direct secular
effect, $e, \sin{I}$ by the interaction with resonances encountered
during the drift in $a$. This has very important consequences on the
family classification precisely because we are using a very large
proper elements catalog, which contains a large proportion of newly
discovered smaller asteroids.

On the other hand, the other 43 families resulting from step 3 have
been left in our classification as independent families, consisting
mostly of smaller asteroids. As discussed in
Sections~\ref{sec:mediumfam} and \ref{sec:smallfam}, some of them are
quite numerous and statistically very significant, some are not large
and may require confirmation, but overall the step 3 families give an
important contribution.

\subsection{Automatic update}
\label{automatic}

The rapid growth of the proper elements database due to the fast rate
of asteroid discoveries and follow-up using modern observing
facilities, and to the efficiency of the computation of proper
elements, results in any family classification becoming quickly
outdated. Thus we devised a procedure for an automatic update of this
family classification, to be performed periodically, e.g., two-three
times per year.

The procedure consists in repeating the attribution of asteroids to
the existing families every time the catalog of synthetic proper
elements is updated. What we repeat is actually step 4, thus the lists
of core families members (found in step 1) and of members of smaller
families (from step 3) are the same, and also the list of mergers
(from step 5) is unchanged. Thus newly discovered asteroids, after they
have been numbered and have proper elements, automatically are added
to the already established families when this is appropriate.

There is a step which we do not think can be automated, and that is
step 5: in principle, as the list of asteroids attached to established
families grows, the intersection can increase. As an example, with the
last update of the proper elements catalog with $18\,149$ additional
records, we have added $3\,586$ new ``step 4'' family members.  Then
the number of intersections, that is members of two families, grows
from 29 to 36.  In some cases the new intersections support some merge
previously not implemented because considered dubious, some open new
problems, in any case to add a new merger is a delicate decision which
we think should not be automated. 

As time goes by, there will be other changes, resulting from the
confirmation/contradiction of some of our interpretations as a
result of new data: as an example, some small families will be
confirmed by the attribution of new members and some will not. At
some point we may be able to conclude that some small families are
flukes and decide to remove them from the classification, but this
is a complicated decision based on assessment of the statistical
significance of the lack of increase.

In conclusion we can only promise we will keep an eye on the
situation as the classification is updated and try to perform
non-automated changes whenever we believe there is enough evidence
to justify them. The purpose of both automated and non-automated
upgrades of the classification is to maintain the validity of the
information made public for many years, without the need for
repeating the entire procedure from scratch. This is not only to
save our effort, but also to avoid confusing the users with the need
of continuously resetting their perception of the state of the art.

\subsection{Some methodological remarks}

As it should be clear after the description of our adopted techniques,
our approach to asteroid family identification is based on procedures
which are in some relevant details different with respect to other
possible approaches previously adopted by other authors.

In particular, we do not use any systematic family classifications in
$>3$ dimensional spaces, such as the ones based either upon proper
elements and albedo, or proper elements and color indexes, or all
three datasets. We make use in our procedure, when dealing with very
populous zones, of the absolute magnitude, but only as a way to
decompose into steps the HCM procedure, as discussed in
Subsection~\ref{sec:hcm}. Any other available data are used only
\emph{a posteriori}, as verification and/or improvement, after a
purely dynamical classification has been built. The reasons for this
are explained in Table~\ref{tab:infocount}: less objects, each with a
set of $4\div 6$ parameters, actually contain less information.

We acknowledge that other approaches can be meaningful and give
complementary results. The specific advantage of our datasets and of
our methods is in the capability of handling large numbers of small
asteroids. This allows to discover, or at least to measure better,
different important phenomena. The best example of this is the radical
change in perception about the cratering families, which have been more
recently discovered and are difficult to appreciate in any approach
biased against the use of information provided by small
asteroids, as it is the case for classifications which require the
availability of physical data (see Section~\ref{sec:craters}).

Moreover, we do not make use of naked eye taxonomy, that is of purely
visual perception as the main criterion to compose families.  This is
not because we disagree on the fact that the human eye is an extremely
powerful instrument, but because we want to provide the users with data
as little as possible affected by subjective judgments. We
have no objection on the use of visual appreciation of our proposed
families by the users of our data, as shown by the provision of a
dedicated public graphic tool. But this stage of the work needs to be
kept separate, after the classification has been computed by objective
methods.

\section{Results from dynamical classification}
\label{sec:result_dyn}

\subsection{Large families}
\label{sec:bigfam}

By ``large families'' we mean those having, in the classification
presented in this paper, $> 1000$ members. There are $19$ such
families, and some of their properties are listed in
Table~\ref{tab:bigfam}.

\begin{table}[ht]
\footnotesize
 \centering
  \caption{Large families with $> 1000$ members sorted by \# tot. The
    columns give: family, zone, QRL distance (m/s), number of family
    members classified in steps 1, 3, 2+4 and the total number of
    members, family boundaries in terms of proper $a$, $e$ and
    $\sin{I}$. }
  \label{tab:bigfam}
\medskip
  \begin{tabular}{lcrrrrrcccccc}
  \hline
family&zone& QRL & 1 & 3 & 2+4 & tot& $a_{min}$ &  $a_{max}$ & $e_{min}$& $e_{max}$& $sI_{min}$& $sI_{max}$\\
\hline
\\
135 &   2&    70&   1141&   5001&   5286\phantom{0}&  11428&  2.288&  2.478&  0.134&  0.206&  0.032&  0.059\\
221 &   4&   100&   3060&    310&   6966\phantom{0}&  10336&  2.950&  3.146&  0.022&  0.133&  0.148&  0.212\\
4   &   2&    70&   1599&    925&   5341\phantom{0}&   7865&  2.256&  2.482&  0.080&  0.127&  0.100&  0.132\\
15  &   3&    90&   2713&      0&   4132\phantom{0}&   6845&  2.521&  2.731&  0.117&  0.181&  0.203&  0.256\\
158 &   4&   100&    930&      0&   4671\phantom{0}&   5601&  2.816&  2.985&  0.016&  0.101&  0.029&  0.047\\
20  &   2&    70&     86&   3546&   1126\phantom{0}&   4758&  2.335&  2.474&  0.145&  0.175&  0.019&  0.033\\
24  &   4&   100&   1208&      0&   2742\phantom{0}&   3950&  3.062&  3.240&  0.114&  0.192&  0.009&  0.048\\
10  &   4&   100&    511&     50&   1841\phantom{0}&   2402&  3.067&  3.241&  0.100&  0.166&  0.073&  0.105\\
5   &   3&    90&     27&   1743&    350\phantom{0}&   2120&  2.552&  2.610&  0.146&  0.236&  0.054&  0.095\\
847 &   3&    90&    176&    175&   1682\phantom{0}&   2033&  2.713&  2.819&  0.063&  0.083&  0.056&  0.076\\
170 &   3&    90&    785&      0&   1245\phantom{0}&   2030&  2.523&  2.673&  0.067&  0.128&  0.231&  0.269\\
93  &   3&    90&    641&      0&   1192\phantom{0}&   1833&  2.720&  2.816&  0.115&  0.155&  0.147&  0.169\\
145 &   3&    90&    327&      0&   1072\phantom{0}&   1399&  2.573&  2.714&  0.153&  0.181&  0.193&  0.213\\
1726&   3&    90&     84&    159&   1072\phantom{0}&   1315&  2.754&  2.818&  0.041&  0.053&  0.066&  0.088\\
2076&   2&    70&    140&    528&    477\phantom{0}&   1145&  2.254&  2.323&  0.130&  0.153&  0.088&  0.106\\
490 &   4&   100&    187&     46&    903\phantom{0}&   1136&  3.143&  3.196&  0.049&  0.079&  0.151&  0.172\\
434 &   1&    70&    662&      0&    455\phantom{0}&   1117&  1.883&  1.988&  0.051&  0.097&  0.344&  0.378\\
668 &   3&    90&    259&      0&    842\phantom{0}&   1101&  2.744&  2.811&  0.188&  0.204&  0.129&  0.143\\
1040&   4&   100&    226&      0&    870\phantom{0}&   1096&  3.083&  3.174&  0.176&  0.217&  0.279&  0.298\\
\\
\hline
\end{tabular}
\end{table}

The possibility of finding families with such large number of members
results from our methods, explained in the previous section, to attach
to the \textit{core families} either families formed with smaller
asteroids or individual asteroids which are suitably close to the
core.  The main effect of attaching individual asteroids is to extend
the family to asteroids with higher $H$, that is smaller. The main
effect of attaching families formed with smaller objects is to extend
the families in proper $a$, which can be understood in terms of the
Yarkovsky effect, which generates a drift $da/dt$ inversely
proportional to the diameter $D$.

As the most spectacular increase in the family size, the
core family of (5) Astraea is very small, with only 27 members,
growing to $2\,120$ members with steps 2--5: almost all the family
members are small, i.e., $H>14$.

For example: among the largest families, the ones with namesakes (135)
Hertha and (4) Vesta are increased significantly in both ways, by
attaching families with smaller asteroids on the low $a$ side (the
high $a$ side being eaten up, in both cases, by the $3/1$ resonance
with Jupiter) and by attaching smaller asteroids to the core.  In both
cases the shape of the family, especially when projected on the proper
$a, e$ plane, clearly indicates a complex structure, which would be
difficult to model with a single collisional event. There are two
different reasons why these families contain so many asteroids: family
135 actually contains the outcome of the disruption of two different
parents (see Section~\ref{sec:use_physical}); family 4 is the product
of two or more cratering events, but on the same parent body (see
Section~\ref{sec:absol_mag}).

\begin{figure}[h!]
  \figfig{12cm}{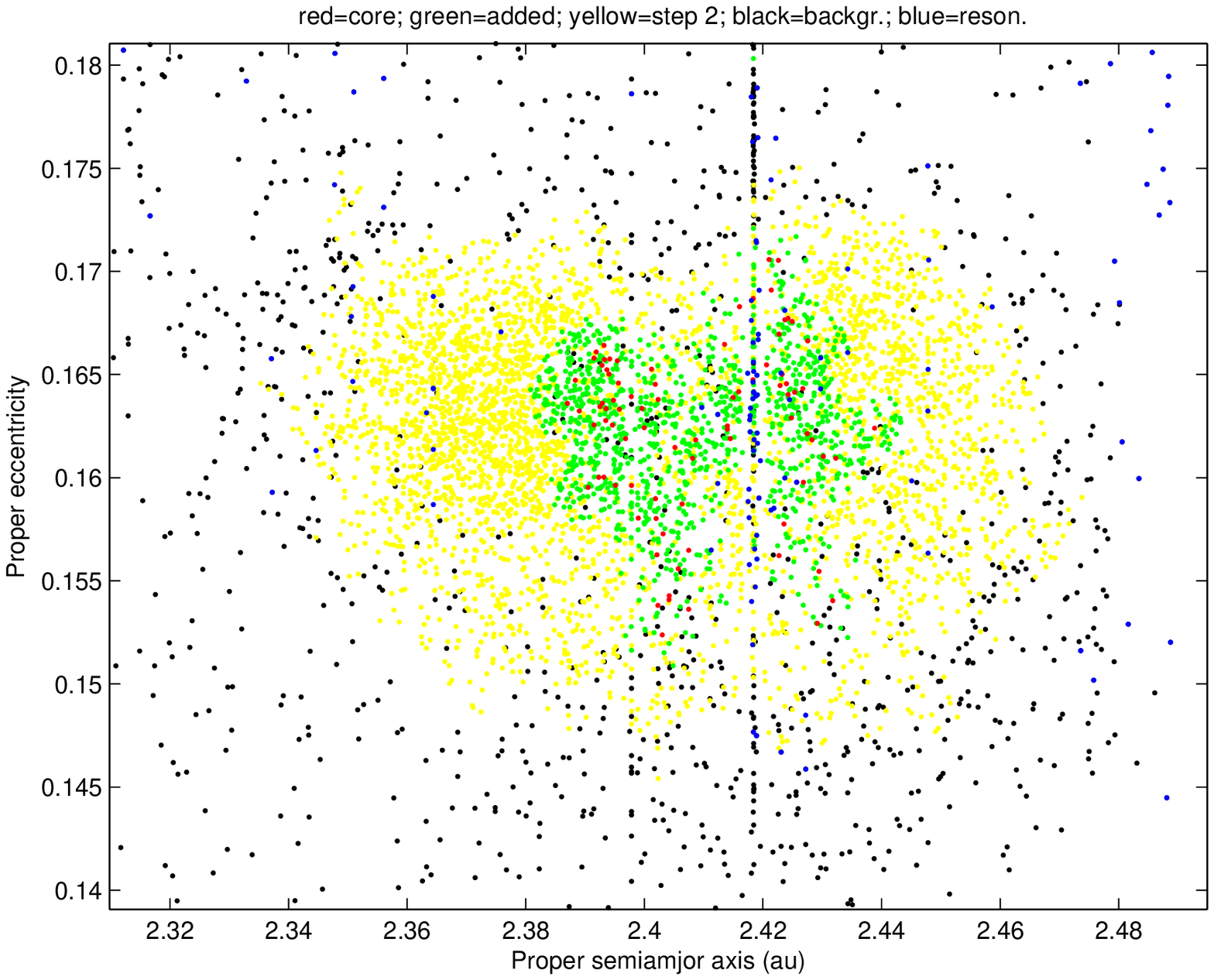}{The family of Massalia as it appears in
    the $a - e$ plane. Red dots indicate objects belonging to the
    family core. Green dots refer to objects added in step 2 and 4 of
    the classification procedure, while yellow points refer to objects
    linked at step 3 (see text). Black dots are not recognized as
    nominal family members, although some of them might be, while
    others are background objects. Blue dots are chaotic objects,
    affected by the $2/1$ resonance with Mars.}
\end{figure}

Another spectacular growth with respect to the core family is the
one shown by the family of (20) Massalia, which is also due to a
cratering event. The region of proper elements space occupied by the
family has been significantly expanded by the attribution of the
``halo families'', in yellow in Figure~\ref{fig:massalia_ae}, on
both the low $a$ and the high $a$ side. The shape is somewhat
bilobate, and this, as already reported by \cite{vokyorp}, is due to
the $1/2$ resonance with Mars which is clearly indicated by chaotic
orbits (marked in blue) and also by the obvious line of diffusion
along the resonance. The border of the family on the high proper $a$
side is very close to the $3/1$ resonance with Jupiter; the
instability effects due to this extremely strong resonance may be
responsible for the difficulties of attributing to the family a
number of objects currently classified as background (marked in
black). We can anyway suggest that Massalia can be a significant
source of NEA and meteorites through a chaotic path passing though
the $3/1$. On the contrary, there is no resonance affecting the
border on the low $a$.

The families of (221) Eos and (15) Eunomia have also been increased
significantly by our procedure, although the core families were
already big. In both cases there is a complex structure, which
makes it difficult to properly define the family boundaries as well
as to model the collisional event(s).

On the contrary, the family of (158) Koronis was produced by an impact
leading to complete disruption of the original parent body, since this
family does not show any dominant largest member.  Family 158 has no
halo: this is due to being sandwiched between the $5/2$ resonance and
the $7/3$ resonance with Jupiter. The same lack of halo occurs for the
family of (24) Themis: the $2/1$ resonance with Jupiter explains the
lack of halo on the high $a$, the $11/5$ resonance has some influence
on the low $a$ boundary.

There has been some discussion in the past on the family 490, but as
pointed out already in \cite{chaosclock} the asteroid (490) Veritas is
in a very chaotic orbit resulting in transport along a resonance
(later identified as $5J-2S-2A$), thus it currently appears to be far
away in proper $e$ from the center of the family, but still can be
interpreted as the parent body. A significant fraction of family
members are locked in the same resonance, thus giving the strange
shape which can be seen in the right low portion of
Figure~\ref{fig:eos_aI_otherfam}.

We note that in our analysis we do not identify a family associated
with (8) Flora. A Flora family was found in some previous family
searches, but always exhibited a complicated splitting behavior which
made the real membership to appear quite uncertain
\cite{zapetal95}. We find (8) Flora to belong to a step 1 grouping
which is present at a distance level of 110 m/s, much higher than the
adopted QRL for this zone (70 m/s). This grouping merges with both (4)
and (20) at 120 m/s, obviously meaningless. In a rigorous analysis,
the QRL cannot be increased arbitrarily just to accept as a family
groupings like this one, which do not comply with our criteria.

\subsubsection{Halo problems}
\label{sec:haloproblem}

We are not claiming that our method of attaching ``halo'' families to
larger ones can be applied automatically and/or provide an absolute
truth.  There are necessarily dubious cases, most of which can be
handled only by suspending our judgment. Here we are discussing all
the cases in which we have found problems, resulting in a total of 29
asteroids belonging to family intersections.

For the family of (15) Eunomia, the $3/1$ resonance with Jupiter opens
a wide gap on the low $a$ side. The $8/3$ resonance appears to control
the high $a$ margin, but there is a possible appendix beyond the
resonance, which is classified as the family of (173) Ino: we have
found four intersections 15--173.  A problem would occur if we were to
merge the two families, because the proper $a=2.743$ au of the large
asteroid (173) appears incompatible with the dynamical evolution of a
fragment from (15). The only solution could be to join to family 15
only the smaller members of 173, but we do not think that such a merge
could be considered reliable at the current level of information.

\begin{figure}[h!]
  \figfig{12cm}{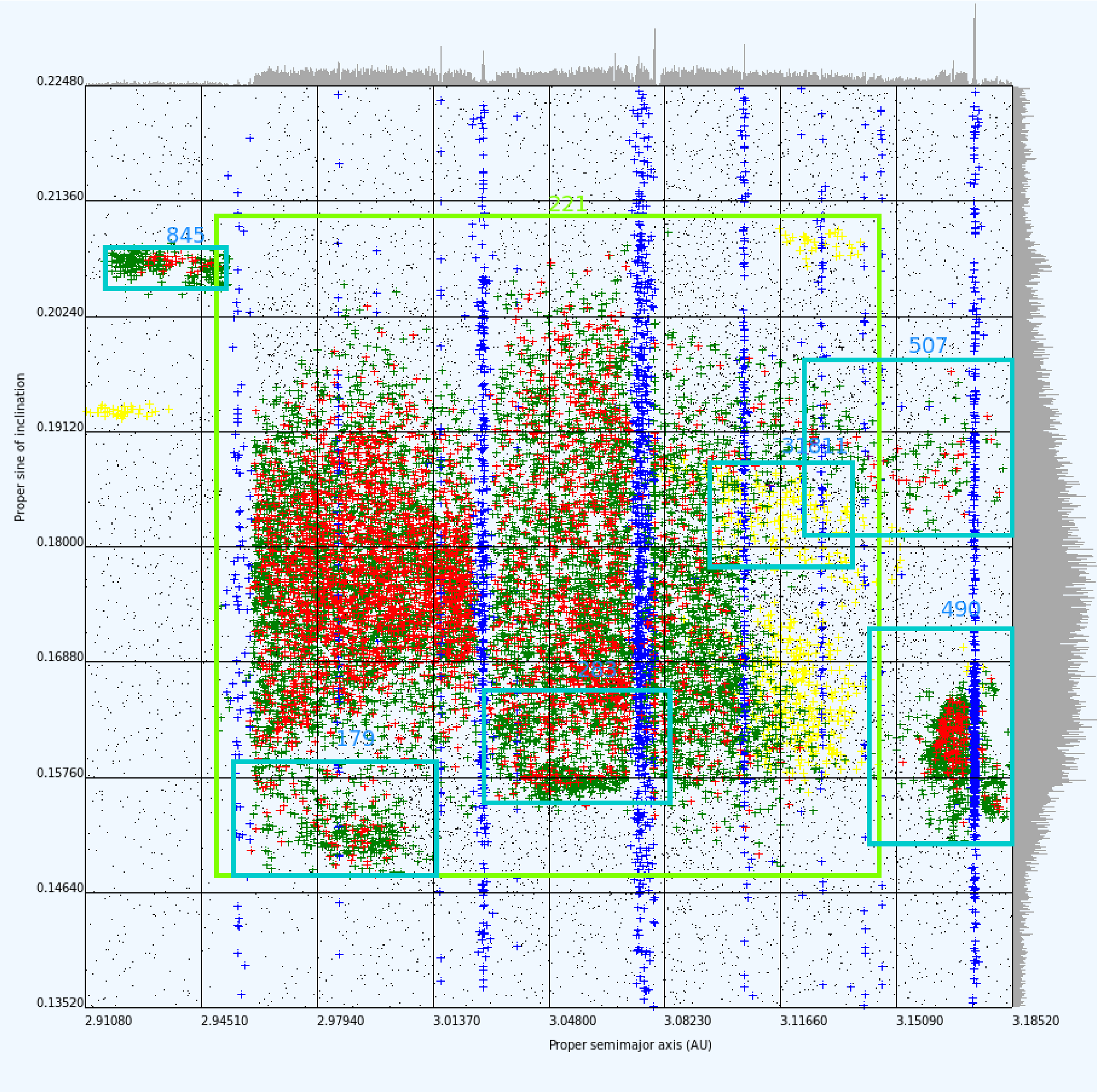}{The region surrounding the family of
    (221) Eos in the proper $a$, proper $\sin{I}$ plane. Boxes are
    used to mark the location of families, some of which overlap
    family 221 in this projection but not in others, such a $a,
    e$. (This figure has been generated with the new AstDyS graphics
    server.)}
\end{figure}

The family of (221) Eos appears to end on the lower $a$ side at the
$7/3$ resonance with Jupiter, but the high $a$ boundary is less
clear. There are two families 507 and 31811 having a small number
(six) of intersections with 221: they could be interpreted as a
continuation of the family, which would have a more complex
shape. However, for the moment we do not think there is enough
information to draw this conclusion. Other families in the same region
have no intersections and appear separate: the $a, \sin{I}$ projection
of Figure~\ref{fig:eos_aI_otherfam} shows well the separation of core
families 179, 490, 845, and small families 1189 and 8737, while 283 is
seen to be well separate by using an $e, \sin{I}$ projection.

The family of (135) Hertha has few intersections (a total of four)
with the small families 6138 (48 members) 6769 (45 members) 7220 (49
members).
%Family 7220 has a peculiar shape, due to the effect of the $2/1$
%resonance with Mars, which affects also 6769. Family 6138 has higher
%proper $e$ than the members of 135; we know that (6138) is affected by
%the resonance $4J-3S-1A$.  
All three are unlikely to be separate collisional families, but we
have not yet merged them with 135 because of too little evidence.

The family of (2076) Levin appears to be limited in the low $a$ side
by the $7/2$ resonance with Jupiter. It has few (three) intersections with
families 298 and 883. 883 is at lower $a$ than the $7/2$ resonance,
and could be interpreted as a halo, with lower density due to the
ejection of family members by the resonance. Although this is an
interesting interpretation which we could agree with, we do not feel
this can be considered proven, thus we have not merged 2076--883. As
for the family of (298) Baptistina, again the merge with 2076 could be
correct, but the family shape would become quite complex, thus we have
not implemented it for now. Note that a halo family, with lowest
numbered (4375), has been merged with 2076 because of 38 intersection,
resulting in a much larger family.

The family of (1040) Klumpkea has an upper bound of the proper
$\sin{I}$ very close to $0.3$, that is to the boundary between the
moderate inclination and the high inclination zones, to which the HCM
has been applied separately. This boundary also corresponds to a sharp
drop in the number density of main belt asteroids (only $5.3\%$ have
proper $\sin{I}>0.3$), which is one reason to separate the HCM
procedure because of a very different level of expected background
density. The small family of (3667) Anne-Marie has been found in a
separate HCM run for high proper $\sin{I}$, but there are ten
intersections with family 1040. The two families could have a common
origin, but if we were to merge them the shape of the family would be
bizarre, with a sharp drop in number density inside the family. This
could have explanation either because of a stability boundary or
because of an observational bias boundary. However, this would need to
be proven by a very detailed analysis, thus we have not implemented
this merge.

The family of (10) Hygiea has two intersections with family
1298, but the two are well separated in the $e,\sin{I}$
plane, thus they have not been merged.

\begin{figure}[h!]
  \figfig{12cm}{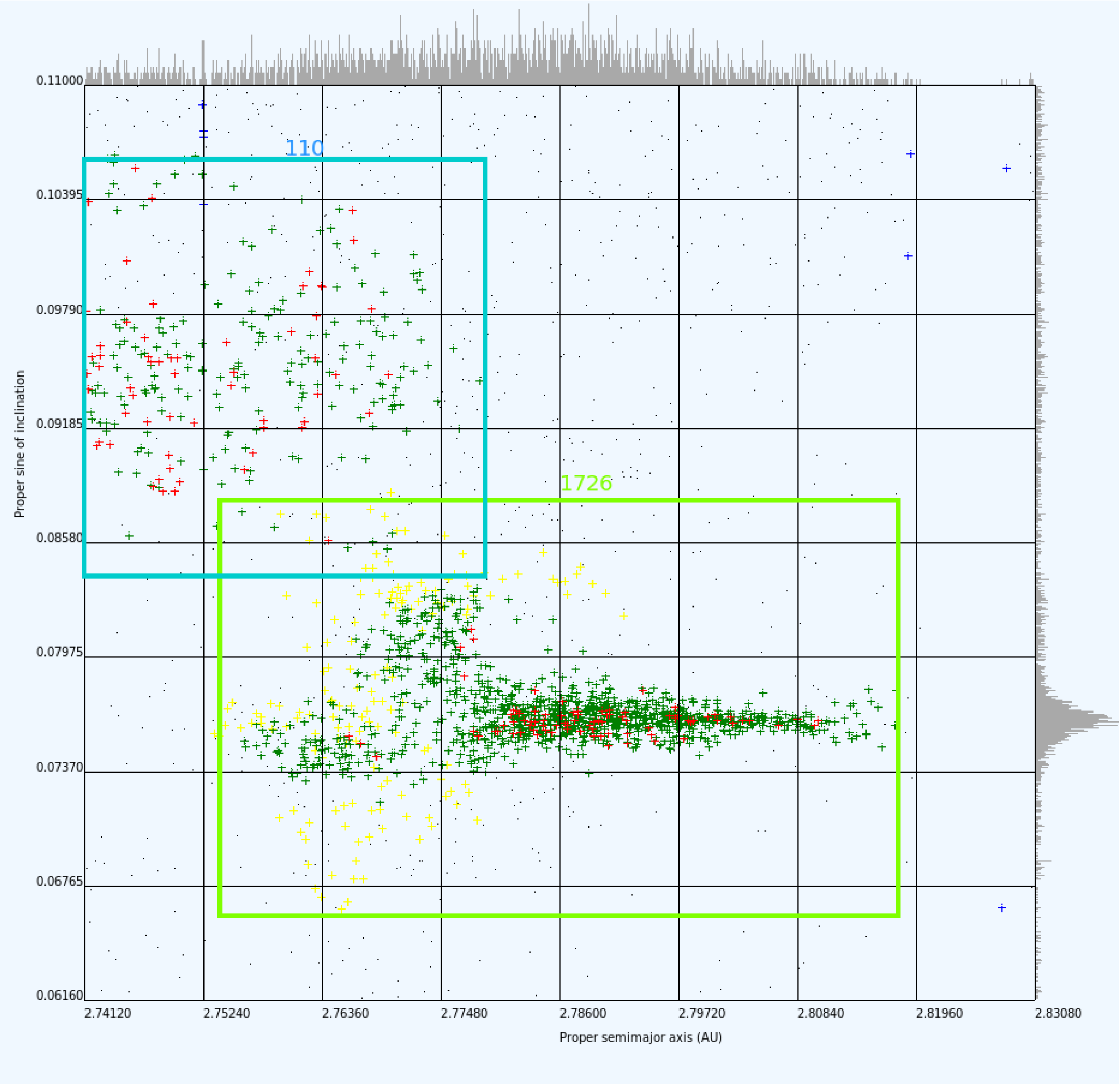}{The strange shape of the family of
    (126) Hoffmeister is shown in the proper $a,\sin{I}$
    projection. Some gravitational perturbations affect the low $a$
    portion of the family, including the halo family 14970. The family
    of (110) Lydia is nearby, but there is no intersection.}
\end{figure}

The family of (1726) Hoffmeister has twenty intersection with the
family 14970, formed with much smaller asteroids. Given such an
overlap, merging the two families appears fully consistent with our
procedure as defined in Section 3. However, the merged family has a
strange shape, see Figure~\ref{fig:Hoffmeister_aI}, in particular with
a protuberance in the $\sin{I}$ direction which would not be easy to
reconcile with a standard model of collisional disruption followed by
Yarkovsky effect. Moreover, the strange shape already occurs in the
core family, that is for the few largest asteroids, and thus should be
explained by using perturbations not depending upon the size,
that is gravitational ones.

Indeed, by consulting the database of analytic proper elements, it is
possible to find that (14970) has a ``secular resonance flag'' $10$,
which can be traced to the effect of the secular resonance
$g+s-g_6-s_6$, see also \cite[Figure 7]{propel3}; the same flag is $0$
for (1726). Indeed, the value of the ``divisor'' $g+s-g_6-s_6$
computed from the synthetic proper elements is $0.1$ arcsec/y for
(14970), $0.65$ for (1726). On top of that, the proper semimajor axis
of (1) Ceres is $2.7671$ au, which is right in the range of $a$ of the
protuberance in proper $\sin{I}$, thus it is clear that close
approaches and the $1/1$ resonance with Ceres can play a significant
role in destabilizing the proper elements \cite{laskar2}. We do not
have a rigorous answer fully explaining the shape of the family, but
we have decided to apply this merger because the number of
intersection is significant and the strange shape did not appear as a
result of the procedure to enlarge the core family.

From these examples, we can appreciate that it is not possible to
define some algorithmic criterion, like a fixed minimum number of
intersections, to automatize the process.  All of the above cases can
be considered as still open problems, to be more reliably solved by
acquiring more information.

\subsection{Medium families} 
\label{sec:mediumfam}

By ``medium families'' we mean families we have found to have more
than $100$ and no more than $1\,000$ members; the properties of the 41
families in this group are given in Table~\ref{tab:mediumfam}.

\begin{table}[p]
\footnotesize
 \centering
  \caption{The same as in Table~\ref{tab:bigfam} but for medium
    families with $100< \# \leq 1000$ members.}
  \label{tab:mediumfam}
\medskip
  \begin{tabular}{lcrrrrrcccccc}
  \hline
family&zone& QRL & 1 & 3 & 2+4 & tot& $a_{min}$ &  $a_{max}$ & $e_{min}$& $e_{max}$& $sI_{min}$& $sI_{max}$\\
\hline
\\
31    &      4&    80&    968&      0&      0\phantom{00}&    968&  3.082&  3.225&  0.150&  0.231&  0.431&  0.459\\
25    &      2&   130&    944&      0&      0\phantom{00}&    944&  2.261&  2.415&  0.160&  0.265&  0.366&  0.425\\
480   &      3&   140&    839&      0&      0\phantom{00}&    839&  2.538&  2.721&  0.008&  0.101&  0.364&  0.385\\
808   &      3&    90&     72&    166&    567\phantom{00}&    805&  2.705&  2.805&  0.125&  0.143&  0.080&  0.093\\
3     &      3&    90&     45&    257&    462\phantom{00}&    764&  2.623&  2.700&  0.228&  0.244&  0.225&  0.239\\
110   &      3&    90&    168&      0&    561\phantom{00}&    729&  2.696&  2.779&  0.026&  0.061&  0.084&  0.106\\
3827  &      3&    90&     29&    310&    332\phantom{00}&    671&  2.705&  2.768&  0.082&  0.096&  0.080&  0.094\\
3330  &      4&   100&     63&      0&    537\phantom{00}&    600&  3.123&  3.174&  0.184&  0.212&  0.171&  0.184\\
1658  &      3&    90&     98&    172&    288\phantom{00}&    558&  2.546&  2.626&  0.165&  0.185&  0.123&  0.142\\
375   &      4&   100&    229&      0&    273\phantom{00}&    502&  3.096&  3.241&  0.059&  0.130&  0.264&  0.299\\
293   &      4&   100&     38&      0&    405\phantom{00}&    443&  2.832&  2.872&  0.119&  0.133&  0.256&  0.264\\
10955 &      3&    40&      0&    428&      0\phantom{00}&    428&  2.671&  2.739&  0.005&  0.026&  0.100&  0.113\\
163   &      2&    40&      0&    392&      0\phantom{00}&    392&  2.332&  2.374&  0.200&  0.218&  0.081&  0.098\\
569   &      3&    40&      0&    389&      0\phantom{00}&    389&  2.623&  2.693&  0.169&  0.183&  0.035&  0.045\\
1128  &      3&    40&      0&    389&      0\phantom{00}&    389&  2.754&  2.817&  0.045&  0.053&  0.008&  0.018\\
283   &      4&   100&     49&      0&    320\phantom{00}&    369&  3.029&  3.084&  0.107&  0.124&  0.155&  0.166\\
179   &      4&   100&     60&      0&    306\phantom{00}&    366&  2.955&  3.015&  0.053&  0.080&  0.148&  0.159\\
5026  &      2&    40&      0&    346&      0\phantom{00}&    346&  2.368&  2.415&  0.200&  0.217&  0.082&  0.096\\
3815  &      3&    40&      0&    283&      0\phantom{00}&    283&  2.563&  2.583&  0.138&  0.143&  0.145&  0.164\\
1911  &      6&    60&    280&      0&      0\phantom{00}&    280&  3.964&  3.967&  0.159&  0.222&  0.041&  0.055\\
845   &      4&   100&     29&      0&    224\phantom{00}&    253&  2.917&  2.953&  0.029&  0.041&  0.205&  0.209\\
194   &      3&   140&    235&      0&     17\phantom{00}&    252&  2.522&  2.691&  0.154&  0.196&  0.293&  0.315\\
396   &      3&    40&      0&    242&      0\phantom{00}&    242&  2.731&  2.750&  0.164&  0.170&  0.057&  0.062\\
12739 &      3&    40&      0&    240&      0\phantom{00}&    240&  2.682&  2.746&  0.047&  0.060&  0.031&  0.041\\
778   &      4&   100&     29&      0&    200\phantom{00}&    229&  3.158&  3.191&  0.240&  0.261&  0.243&  0.253\\
945   &      3&   140&    219&      0&      0\phantom{00}&    219&  2.599&  2.659&  0.190&  0.289&  0.506&  0.521\\
1303  &      4&    80&    179&      0&      0\phantom{00}&    179&  3.193&  3.236&  0.106&  0.144&  0.310&  0.337\\
752   &      2&    70&     27&     90&     41\phantom{00}&    158&  2.421&  2.484&  0.084&  0.095&  0.085&  0.092\\
18466 &      3&    40&      0&    155&      0\phantom{00}&    155&  2.763&  2.804&  0.171&  0.182&  0.229&  0.236\\
173   &      3&    90&     29&      0&    125\phantom{00}&    154&  2.708&  2.770&  0.159&  0.180&  0.229&  0.239\\
606   &      3&    40&      0&    153&      0\phantom{00}&    153&  2.573&  2.594&  0.179&  0.183&  0.166&  0.168\\
507   &      4&   100&     38&      0&    111\phantom{00}&    149&  3.124&  3.207&  0.049&  0.075&  0.181&  0.198\\
13314 &      3&    40&      0&    146&      0\phantom{00}&    146&  2.756&  2.801&  0.170&  0.183&  0.069&  0.078\\
302   &      2&    40&      0&    143&      0\phantom{00}&    143&  2.385&  2.418&  0.104&  0.111&  0.056&  0.060\\
1298  &      4&   100&     69&      0&     74\phantom{00}&    143&  3.088&  3.220&  0.105&  0.123&  0.104&  0.123\\
87    &      5&   120&    119&      0&     20\phantom{00}&    139&  3.459&  3.564&  0.046&  0.073&  0.162&  0.179\\
883   &      2&    70&     46&      0&     86\phantom{00}&    132&  2.213&  2.259&  0.140&  0.151&  0.092&  0.102\\
298   &      2&    70&     43&      0&     88\phantom{00}&    131&  2.261&  2.288&  0.146&  0.161&  0.100&  0.114\\
19466 &      3&    40&      0&    125&      0\phantom{00}&    125&  2.724&  2.761&  0.007&  0.020&  0.103&  0.111\\
1547  &      3&    40&      0&    108&      0\phantom{00}&    108&  2.641&  2.650&  0.267&  0.270&  0.211&  0.212\\
1338  &      2&    70&     38&      0&     66\phantom{00}&    104&  2.259&  2.302&  0.119&  0.130&  0.075&  0.091\\
\\
\hline
\end{tabular}
\end{table}

Of course the exact boundaries $100$ and $1\,000$ are chosen just for
convenience: still the distinction between families based on these
boundaries has some meaning. The medium families are such that the
amount of data may not be enough for detailed studies of the family
structure, including family age determination, size distribution,
detection of internal structures and outliers. However, they are
unlikely to be statistical flukes, they represent some real
phenomenon, but some caution needs to be used in describing it.

In this range of sizes it is necessary to analyze each individual family
to find out what can be actually done with the information they
provide. As for the ones near the lower boundary for the number of
members, they are expected to grow as the family classification
procedure is applied automatically by the AstDyS information system to
larger and larger proper elements datasets.  As a result, they should
over a time span of several years grow to the point that more
information on the collisional and dynamical evolution process is
available. The only other expected outcome is that some of them can
become much stronger candidates for merging with big families (e.g.,
family 507 cited above as a possible appendix to 221).  If some others
were not to grow at all, even over a timespan in which there has been
a significant increase in number density (e.g, $50\%$) in the region,
this would indicate a serious problem in the classification and would
need to be investigated.

Note that 14 of the medium families have been generated in step 3,
that is they are formed with the intermediate background after removal
of step 1 and 2 family members, roughly speaking with ``smaller''
asteroid.

\subsubsection{Some remarkable medium families}

The families of (434) Hungaria, (25) Phocaea, (31) Euphrosyne, and
(480) Hansa, clustered around the $1\,000$ members boundary, are the
largest high inclination families, one for each semimajor axis
zone\footnote{Hungaria is included in our list of large
families. Zones 5 and 6 have essentially no stable high
  inclination asteroids}. Given the lower number density for proper
$\sin{I}>0.3$ the numbers of members are remarkably high, and suggest
that it may be possible to obtain information on the collisional
processes which can occur at higher relative velocities. These four
have been known for some time
\cite{hungaria,bojan_highi,synthpro2,phocea_carruba}, but now it has
become possible to investigate their structure.

For family 480 the proper $e$ can be very small: this results in a
difficulty in computing proper elements (especially $e$ and the proper
frequency $g$) due to "paradoxical libration".  We will try to fix our
algorithm to avoid this, but we have checked that it has no influence
on the family membership, because $e=0$ is not a boundary.

The largest family in zone 5 is the one of (87) Sylvia, which is well
defined, but with a big central gap corresponding to the resonance
$9/5$ with Jupiter. This family has in (87) such a dominant largest
member that it can be understood as a cratering event, even if we do
not have a good idea of how many fragments have been removed by the
$9/5$ and other resonances\footnote{This family is interesting also
  because (87) Sylvia has been the first recognized triple asteroid
  system, formed by a large primary and two small satellites
  \cite{Marchis2005}}. 

The largest family in zone 6, that is among the Hilda asteroid locked
in the $3/2$ resonance with Jupiter, is the one of (1911) Schubart. By
the way, proper elements for Hildas, taking into account the
resonance, have been computed by \cite{schubart}, but we have used as
input to the HCM (step 1) procedure the synthetic proper elements
computed without taking into account the resonance, thus averaging
over the libration in the critical argument. This is due to the need
to use the largest dataset of proper elements, and is a legitimate
approximation because the contribution of even the maximum libration
amplitude to the metrics similar to $d$, to be used for resonant
asteroids, is more than an order of magnitude smaller than the one due
to eccentricity.

\subsection{Small families}
\label{sec:smallfam}

The families we rate as ``small'' are those in the range between $30$
and $100$ members; data for 43 such families are in Table~\ref{tab:smallfam}.

\begin{table}[p]
\footnotesize
 \centering
  \caption{The same as in Table~\ref{tab:bigfam} but for small families with  $30< \#\leq 100$ members.}
  \label{tab:smallfam}
\medskip
  \begin{tabular}{lcrrrrrcccccc}
  \hline
family&zone& QRL & 1 & 3 & 2+4 & tot& $a_{min}$ &  $a_{max}$ & $e_{min}$& $e_{max}$& $sI_{min}$& $sI_{max}$\\
\hline
\\
96   &       4&   100&     38&      0&     62\phantom{00}&    100&  3.036&  3.070&  0.176&  0.189&  0.280&  0.289\\
148  &       3&   140&     95&      0&      0\phantom{00}&     95&  2.712&  2.812&  0.116&  0.150&  0.420&  0.430\\
410  &       3&    90&     55&      0&     38\phantom{00}&     93&  2.713&  2.761&  0.238&  0.265&  0.146&  0.160\\
2782 &       3&    90&     21&      0&     71\phantom{00}&     92&  2.657&  2.701&  0.185&  0.197&  0.061&  0.072\\
31811&       4&    40&      0&     89&      1\phantom{00}&     90&  3.096&  3.138&  0.060&  0.075&  0.178&  0.188\\
3110 &       3&    40&      0&     86&      0\phantom{00}&     86&  2.554&  2.592&  0.134&  0.145&  0.049&  0.065\\
18405&       4&    40&      0&     85&      0\phantom{00}&     85&  2.832&  2.858&  0.103&  0.110&  0.158&  0.162\\
7744 &       3&    40&      0&     78&      0\phantom{00}&     78&  2.635&  2.670&  0.069&  0.075&  0.042&  0.049\\
1118 &       4&   100&     47&      0&     30\phantom{00}&     77&  3.145&  3.246&  0.035&  0.059&  0.252&  0.266\\
729  &       3&    90&     73&      0&      2\phantom{00}&     75&  2.720&  2.814&  0.110&  0.144&  0.294&  0.305\\
17392&       3&    40&      0&     75&      0\phantom{00}&     75&  2.645&  2.679&  0.059&  0.070&  0.036&  0.042\\
4945 &       3&    40&      0&     71&      0\phantom{00}&     71&  2.570&  2.596&  0.235&  0.244&  0.087&  0.096\\
63   &       2&    40&      0&     70&      0\phantom{00}&     70&  2.383&  2.401&  0.118&  0.127&  0.107&  0.118\\
16286&       4&    40&      0&     68&      0\phantom{00}&     68&  2.846&  2.879&  0.038&  0.047&  0.102&  0.111\\
1222 &       3&   140&     68&      0&      0\phantom{00}&     68&  2.769&  2.803&  0.068&  0.113&  0.350&  0.359\\
11882&       3&    40&      0&     66&      0\phantom{00}&     66&  2.683&  2.708&  0.059&  0.066&  0.031&  0.040\\
21344&       3&    40&      0&     62&      0\phantom{00}&     62&  2.709&  2.741&  0.150&  0.159&  0.046&  0.050\\
3489 &       2&    40&      0&     57&      0\phantom{00}&     57&  2.390&  2.413&  0.090&  0.096&  0.103&  0.109\\
6124 &       6&    60&     57&      0&      0\phantom{00}&     57&  3.966&  3.967&  0.186&  0.212&  0.146&  0.159\\
29841&       3&    40&      0&     53&      0\phantom{00}&     53&  2.639&  2.668&  0.052&  0.059&  0.033&  0.040\\
25315&       3&    40&      0&     53&      0\phantom{00}&     53&  2.575&  2.596&  0.243&  0.251&  0.090&  0.096\\
3460 &       4&   100&     28&      0&     24\phantom{00}&     52&  3.159&  3.218&  0.187&  0.209&  0.016&  0.028\\
2967 &       4&    80&     52&      0&      0\phantom{00}&     52&  3.150&  3.224&  0.092&  0.124&  0.295&  0.303\\
8905 &       3&    40&      0&     49&      0\phantom{00}&     49&  2.599&  2.620&  0.181&  0.190&  0.084&  0.091\\
7220 &       2&    40&      0&     48&      1\phantom{00}&     49&  2.418&  2.424&  0.183&  0.195&  0.026&  0.036\\
3811 &       3&    40&      0&     49&      0\phantom{00}&     49&  2.547&  2.579&  0.101&  0.110&  0.185&  0.190\\
6138 &       2&    40&      0&     46&      2\phantom{00}&     48&  2.343&  2.357&  0.204&  0.215&  0.039&  0.045\\
32418&       3&    40&      0&     48&      0\phantom{00}&     48&  2.763&  2.795&  0.255&  0.261&  0.152&  0.156\\
53546&       3&    40&      0&     47&      0\phantom{00}&     47&  2.709&  2.735&  0.170&  0.174&  0.247&  0.251\\
43176&       4&    40&      0&     47&      0\phantom{00}&     47&  3.109&  3.152&  0.065&  0.074&  0.174&  0.183\\
618  &       4&    40&      0&     46&      0\phantom{00}&     46&  3.177&  3.200&  0.056&  0.059&  0.270&  0.277\\
28804&       3&    40&      0&     46&      0\phantom{00}&     46&  2.589&  2.601&  0.146&  0.156&  0.063&  0.070\\
7468 &       4&    40&      0&     45&      0\phantom{00}&     45&  3.031&  3.075&  0.087&  0.091&  0.060&  0.061\\
6769 &       2&    40&      0&     44&      1\phantom{00}&     45&  2.398&  2.431&  0.148&  0.155&  0.051&  0.056\\
159  &       4&    40&      0&     45&      0\phantom{00}&     45&  3.091&  3.131&  0.111&  0.117&  0.084&  0.090\\
5651 &       4&   100&     20&      0&     22\phantom{00}&     42&  3.097&  3.166&  0.112&  0.128&  0.231&  0.241\\
21885&       4&    40&      0&     42&      0\phantom{00}&     42&  3.079&  3.112&  0.026&  0.035&  0.184&  0.188\\
780  &       4&    80&     41&      0&      0\phantom{00}&     41&  3.085&  3.129&  0.060&  0.074&  0.310&  0.314\\
22241&       4&    40&      0&     40&      0\phantom{00}&     40&  3.082&  3.096&  0.126&  0.133&  0.087&  0.096\\
2    &       3&   140&     38&      0&      0\phantom{00}&     38&  2.756&  2.791&  0.254&  0.283&  0.531&  0.550\\
1189 &       4&    40&      0&     38&      0\phantom{00}&     38&  2.904&  2.936&  0.071&  0.075&  0.192&  0.194\\
8737 &       4&    40&      0&     37&      0\phantom{00}&     37&  3.116&  3.141&  0.112&  0.121&  0.207&  0.211\\
3438 &       4&   100&     20&      0&     14\phantom{00}&     34&  3.036&  3.067&  0.176&  0.186&  0.249&  0.255\\
\hline
\end{tabular}
\end{table}

Note that 29 out 41 of these ``small families'' have been added in
step 3, and have not been absorbed as halo families. 

The families in this category have been selected on the basis of
statistical tests, which indicates they are unlikely to be statistical
flukes. Nevertheless, most of them need some confirmation, which may
not be available due to small number statistics. Thus most of these
proposed families are there waiting for confirmation, which may
require waiting for more observational data. 

The possible outcomes of this process, which requires a few years, are
as follows: (i) the family is confirmed by growing in membership, as a
result of automatic attachment of new numbered asteroids; (ii) the
family is confirmed by the use of physical observations and/or
modeling; (iii) the family grows and become attached as halo to a
larger family; (iv) the family is found not to exist as collisional
family because it does not increase with new smaller members; (v) the
family is found not to exist as collisional family because of enough
physical observations showing incompatible composition.

In other words, the tables published in this paper are to be used as
reference and compared, at each enlargement of the proper elements
database, with the automatically generated new table based on more
asteroids, to see which family is growing\footnote{The need for fixed
  lists to be cited over many years as a comparison with respect to
  the ``current'' number of members explains why these tables need to
  be published in printed form. The current family table can be
  downloaded from AstDyS at
  http://hamilton.dm.unipi.it/\~{ }astdys2/propsynth/numb.famtab}.

However, there are cases in which some of these outcomes appear more
likely, and we shall comment on a few of them.

\subsubsection{Small but convincing families}

The family of (729) Watsonia has been obtained by joining a high $I$
and a low $I$ families. 
%[MIXED QRL 90-140: what should we say? NOTHING. Let explain 
% these details only if requested by a referee].
We are convinced that it is robust, but it may grow unevenly in the
high $I$ and in the low $I$ portions because of the drop in number
density, whatever its cause. Other results on this family are given in
Section~\ref{sec:use_physical}.

The family of (2) Pallas has only 38 members, but it is separated, in
proper $a$, by a gap from the tiny family 14916. The gap can be
explained as the effect of resonances, the main one being the $3J -1S
-1A$ 3-body resonance. Given the large escape velocity from Pallas,
families 2 and 14916 together would be within the range of proper
elements obtained from the ejection of the fragments following the
cratering event.  That is, the distribution of proper elements could
be due to the initial spread of velocities rather than to Yarkovsky,
implying that they would contain no evidence for the age of the
family. However, we have not merged these two families because this
argument, although we believe it is correct, arises from a model, not
from the data as such.

\subsubsection{Small families which could belong to the halo of large families}

On top of the small families already listed in
Subsection~\ref{sec:haloproblem}, which are considered as possible
halos because of intersections, there are other cases in which small
families are very close to large ones, and thus could become
candidates for merging as the size of the proper elements catalog
increases.

To identify these cases we have used for each family the ``box''
having sides corresponding to the ranges in proper $a, e, \sin{I}$
listed in Tables~\ref{tab:bigfam}--\ref{tab:tinyfam}, and we
analyzed all the overlaps between them. The parameter we use as an
alert of proximity between two families is the ratio between the
volume of the intersection to the volume of the box for the smaller
family. If this ratio is $100\%$ then the smaller family is fully
included within the box of the larger one; we have found 12 such
cases. We found another 17 cases with ratio $>20\%$.  By removing
the cases with intersections, or anyway already discussed in
Sections~\ref{sec:bigfam} and \ref{sec:mediumfam}, we are left with
$17$ cases to be analyzed.

One case of these overlapping-box families is about two medium
families, namely family 10955 (with 428 members the largest of the
step 3 families) and family 19466, which has $40\%$ of its box
contained in the box of 10955. The possibility of future merger cannot
be excluded.

Among small/tiny families with boxes overlapping larger ones, we have
found 10 cases we think could evolve into mergers with more data:
4-3489, 5-8905, 5-28804, 10-159, 10-22241, 221-31811, 221-41386,
375-2967, 480-34052, 1040-29185.

In two of the above cases there is already, from the first automatic
update, supporting evidence: of the 7 new intersections, one is
10-22241 and another is 375-2967. We are not claiming this is enough
evidence for a merge, but it shows how the automatic upgrade of the
classification works.

In three cases we do not think there could be future mergers:
15-145, 15-53546, 221-21885.  In another three cases the situation
is too complicated to allow us to make any guess; 24-3460, 31-895,
4-63.

The conclusion from this discussion is clear: a significant fraction
of the small families of Table~\ref{tab:smallfam}, and few from
Table~\ref{tab:tinyfam}, could be in the future included in the halo
of larger families. Others could be confirmed as independent
families, and some could have to be dismissed.

\subsection{Tiny families}
\label{sec:tinyfam}

The ``tiny families'' are the ones with $<30$ members; of course
their number is critically dependent upon the caution with which the
small clusters have been accepted as proposed families. In
Table~\ref{tab:tinyfam} we are presenting a set of $25$ such
families.
%[please comment on the statistical arguments here]

\begin{table}[h]
\footnotesize
 \centering
  \caption{The same as in Table~\ref{tab:bigfam} but for tiny families with  $< 30$ members.}
  \label{tab:tinyfam}
\medskip
  \begin{tabular}{lcrrrrrcccccc}
  \hline
family&zone& QRL & 1 & 3 & 2+4 & tot& $a_{min}$ &  $a_{max}$ & $e_{min}$& $e_{max}$& $sI_{min}$& $sI_{max}$\\
\hline
\\
3667  &      4&    80&     25&      0&      3\phantom{00,}&     28&  3.087&  3.125&  0.184&  0.197&  0.294&  0.301\\
895   &      4&    80&     25&      0&      0\phantom{00,}&     25&  3.202&  3.225&  0.169&  0.183&  0.438&  0.445\\
909   &      5&   120&     23&      0&      1\phantom{00,}&     24&  3.524&  3.568&  0.043&  0.058&  0.306&  0.309\\
29185 &      4&    80&     23&      0&      0\phantom{00,}&     23&  3.087&  3.116&  0.196&  0.209&  0.295&  0.304\\
4203  &      3&   140&     22&      0&      0\phantom{00,}&     22&  2.590&  2.648&  0.124&  0.135&  0.473&  0.486\\
34052 &      3&   140&     21&      0&      0\phantom{00,}&     21&  2.641&  2.687&  0.073&  0.087&  0.368&  0.377\\
5931  &      4&    80&     19&      0&      0\phantom{00,}&     19&  3.174&  3.215&  0.160&  0.172&  0.302&  0.313\\
22805 &      4&    80&     17&      0&      0\phantom{00,}&     17&  3.136&  3.159&  0.165&  0.175&  0.301&  0.308\\
1101  &      4&    80&     17&      0&      0\phantom{00,}&     17&  3.229&  3.251&  0.030&  0.037&  0.363&  0.375\\
10369 &      3&   140&     17&      0&      0\phantom{00,}&     17&  2.551&  2.609&  0.105&  0.118&  0.470&  0.482\\
3025  &      4&    80&     16&      0&      0\phantom{00,}&     16&  3.192&  3.221&  0.059&  0.066&  0.368&  0.378\\
14916 &      3&   140&     16&      0&      0\phantom{00,}&     16&  2.710&  2.761&  0.270&  0.282&  0.537&  0.542\\
3561  &      6&    60&     15&      0&      0\phantom{00,}&     15&  3.962&  3.962&  0.127&  0.133&  0.149&  0.156\\
45637 &      5&   120&     14&      0&      1\phantom{00,}&     15&  3.344&  3.369&  0.103&  0.123&  0.142&  0.151\\
260   &      5&   120&     11&      0&      4\phantom{00,}&     15&  3.410&  3.464&  0.081&  0.088&  0.100&  0.108\\
58892 &      4&    80&     14&      0&      0\phantom{00,}&     14&  3.121&  3.154&  0.153&  0.162&  0.300&  0.308\\
6355  &      4&    80&     13&      0&      0\phantom{00,}&     13&  3.188&  3.217&  0.088&  0.097&  0.374&  0.378\\
40134 &      3&   140&     13&      0&      0\phantom{00,}&     13&  2.715&  2.744&  0.223&  0.235&  0.429&  0.438\\
116763&      3&   140&     13&      0&      0\phantom{00,}&     13&  2.621&  2.652&  0.236&  0.246&  0.463&  0.468\\
10654 &      4&    80&     13&      0&      0\phantom{00,}&     13&  3.207&  3.244&  0.051&  0.056&  0.368&  0.374\\
10000 &      3&   140&     13&      0&      0\phantom{00,}&     13&  2.562&  2.623&  0.260&  0.273&  0.316&  0.325\\
7605  &      4&    80&     12&      0&      0\phantom{00,}&     12&  3.144&  3.153&  0.065&  0.073&  0.447&  0.453\\
69559 &      4&    80&     12&      0&      0\phantom{00,}&     12&  3.202&  3.219&  0.196&  0.201&  0.299&  0.305\\
20494 &      3&   140&     12&      0&      0\phantom{00,}&     12&  2.653&  2.690&  0.119&  0.132&  0.470&  0.480\\
23255 &      3&   140&     10&      0&      0\phantom{00,}&     10&  2.655&  2.688&  0.095&  0.113&  0.460&  0.469\\
\\
\hline
\end{tabular}
\end{table}

Given the cautionary statements we have given about the ``small
families'', what is the value of the ``tiny'' ones? To understand
this, it is useful to check the zones/regions where these have been
found: 3 tiny families belong to zone 5, 1 to zone 6, 12 to zone 4
high inclination, 9 to zone 3 high inclination. Indeed, groupings with
such small numbers can be statistically significant only in the
regions where the number density is very low.

These families satisfy the requirements to be considered
statistically reliable according to the standard HCM procedure
adopted in the above zones. It should be noted that, due to the low
total number of objects present in these regions, the adopted
minimum number $N_{min}$ of required members to form a family turns
out to be fairly low, and its choice can be more important with
respect to more densely populated regions. In the case of high-I
asteroids, \cite{bojan_highi} included in their analysis a large
number of unnumbered objects which we are not considering in the
present paper. The nominal application of the HCM procedure leads to
accept the groups listed in Table~\ref{tab:tinyfam} as families, but
it is clear that their reliability will have to be tested in the
future when larger numbers will be available in these
zones.

Thus, each one of these groups is only a proposed family, in need of
confirmation. There is an important difference with most of the
small families listed in Table~\ref{tab:smallfam}: there are two
reasons why the number densities are much lower in these regions,
one being the lower number density of actual asteroids, for the same
diameters; the other being the presence of strong observational
biases which favor discovering only objects of comparatively large
size. In the case of the high inclination asteroids the
observational bias is due to most asteroid surveys looking more
often near the ecliptic, because there more asteroids can be found
with the same observational effort. For the more distant asteroids
the apparent magnitude for the same diameter is fainter because of
both larger distance and lower average albedo.

If a family is small because of observational bias, it grows very
slowly in membership unless the observational bias is removed, which
means more telescope time is allocated for asteroid search/recovery at
high ecliptic latitude and more powerful instruments are devoted to
find more distant/dark asteroids. Unfortunately, there is no way to be
sure that these resources will be available, thus some ``tiny''
families may remain tiny for quite some time. In conclusion, the list
of Table~\ref{tab:tinyfam} is like a bet to be adjudicated in a
comparatively distant future.
we can already confirm that many of these tiny families
are slowly increasing in numbers. As already mentioned, while this
paper was being completed, the proper elements catalog has already
been updated, the automatic step 4 was completed, resulting in a new
classification with a $4\%$ increase in family membership.  In this
upgrade $14$ out of $25$ tiny families have increased their
membership, although in most cases by just $1\div 2$.

\section{Use of absolute magnitude data}
\label{sec:absol_mag}

For most asteroids a direct measurement of the size is not available,
whereas all the asteroids with an accurate orbit have some estimated
value of absolute magnitude. To convert absolute magnitude data into
diameter $D$, we need the albedo, thus $D$ can be accurately estimated
only for the objects for which physical observations, such as WISE
data or polarimetric data, are available. However, it is known that
families are generally found to be quite homogeneous in terms of
albedo and spectral reflectance properties
\cite{CellinoAstIII}. Therefore, by assuming an average albedo value
to be assigned to all the members of a given family, we can derive the
size of each object from its value of absolute magnitude. This
requires that a family has one or more members with a known albedo,
and we have reasons to exclude that they are interlopers.
%(e.g., not Hertha)
%Families resulting from cratering events are especially suitable for
%this, since the physical parameters of the large parent body are
%generally known.

The main applications of the statistical information on diameter $D$
of family members are three: estimation of the total volume of
the family, of the age of a collisional family, and of the size
distribution.

\subsection{The volume of the families}
\label{sec:volume}

In case of a total fragmentation, the total volume of a collisional
family can be used to give a lower bound for the size of the parent
body. For cratering, the volume computed without considering the
parent body can be used to constrain from below the size of the
corresponding crater. In case of dubious origin, the total volume can
be used to discard some possible sources if they are too small.

As an example let us choose the very large family of (4) Vesta. The
albedo of Vesta has been measured as $0.423$ \cite{IRAS}, but more
recently an albedo of $0.37$ has been reported by \cite{shevted06},
while a value around 0.30 is found by the most recent polarimetric
investigations \cite{ELSXI}\footnote{By using WISE albedos, it can be
  shown that the most frequent albedos for members of family 4 are in
  the range spanning these measurements, see
  Figure~\ref{fig:vesta_back_albedo_hist}}.
Before computing the volume of fragments we need to remove the biggest
interlopers, because they could seriously contaminate the result:
asteroids (556) and (1145) are found to be interlopers because they
are too big for their position with respect to the parent body, as
discussed in the next subsection.
If we assume albedo $0.423$ for all family 4 members, we can compute
the volume of all the family members at $32\,500$ km$^3$. The volume
would be $54\,500$ with albedo $0.3$, thus the volume of the known
family can be estimated to be in this range. On Vesta there are two
very large craters, Rheasilvia and Veneneia, with volumes of $>1$
million km$^3$. Thus it is possible to find some source crater.

Another example of cratering on a large asteroid is the family of (10)
Hygiea.  After removing Hygiea and interlopers with $D>40$ km which
should be too large for being ejected from a crater, and assuming a
common albedo equal to the IRAS measure for (10), namely $0.072$, we
get a
%that is asteroids (100), (108), (758) (too large for cratering, they
%do not fit in the V-shape discussed in Section~\ref{sec:ages}, and
%have WISE albedo at least double that of Hygiea), and assuming a
%common albedo equal to the WISE measure for (10), which is $0.058\pm
%0.005$,
total volume of the family as $550\,000$ km$^3$. This implies on the
surface of (10) Hygiea there should be a crater with a volume
at least as large as for Rheasilvia.
%\footnote{Given that the family contains a good number of members with
%  $D\sim 20$ km, the crater should be deeper than Rheasilvia.}.
Still the known family corresponds to only $1.3\%$ of the volume of (10)
Hygiea.

\subsection{Family Ages}
\label{sec:ages}

The computation of family ages is a high priority goal. As a matter of
principle it can be achieved by using V-shape plots such as
Figure~\ref{fig:20_vshapea}, for the families old enough to have
Yarkovsky effect dominating the spread of proper $a$. The basic
procedure is as follows: as in the previous section by assuming a
common geometric albedo $p_v$, from the absolute magnitudes $H$ we can
compute\footnote{$D=1\,329\times 10^{-H/5}\, \times 1/\sqrt{p_v}$} the
diameters $D$. The Yarkovsky secular effect in proper $a$ is
$da/dt=c\;\cos(\phi)/D$, with $\phi$ the obliquity (angle of the spin
axis with the normal to the orbit plane), and $c$ a calibration
depending upon density, thermal conductivity and spin rate. As a
matter of fact $c$ is weakly dependent upon $D$, but this cannot be
handled by a general formula since the dependence of $c$ from thermal
conductivity is highly nonlinear \cite[Figure 1]{vok2000}.  Thus, as
shown in \cite[Appendix A]{chesley_bennu}, the power law expressing
the dependence of $c$ upon $D$ changes from case to case. For cases in
which we have poor information on the thermal properties (almost all
cases) we are forced to use just the $1/D$ dependency.

\begin{figure}[h!]
  \figfig{12cm}{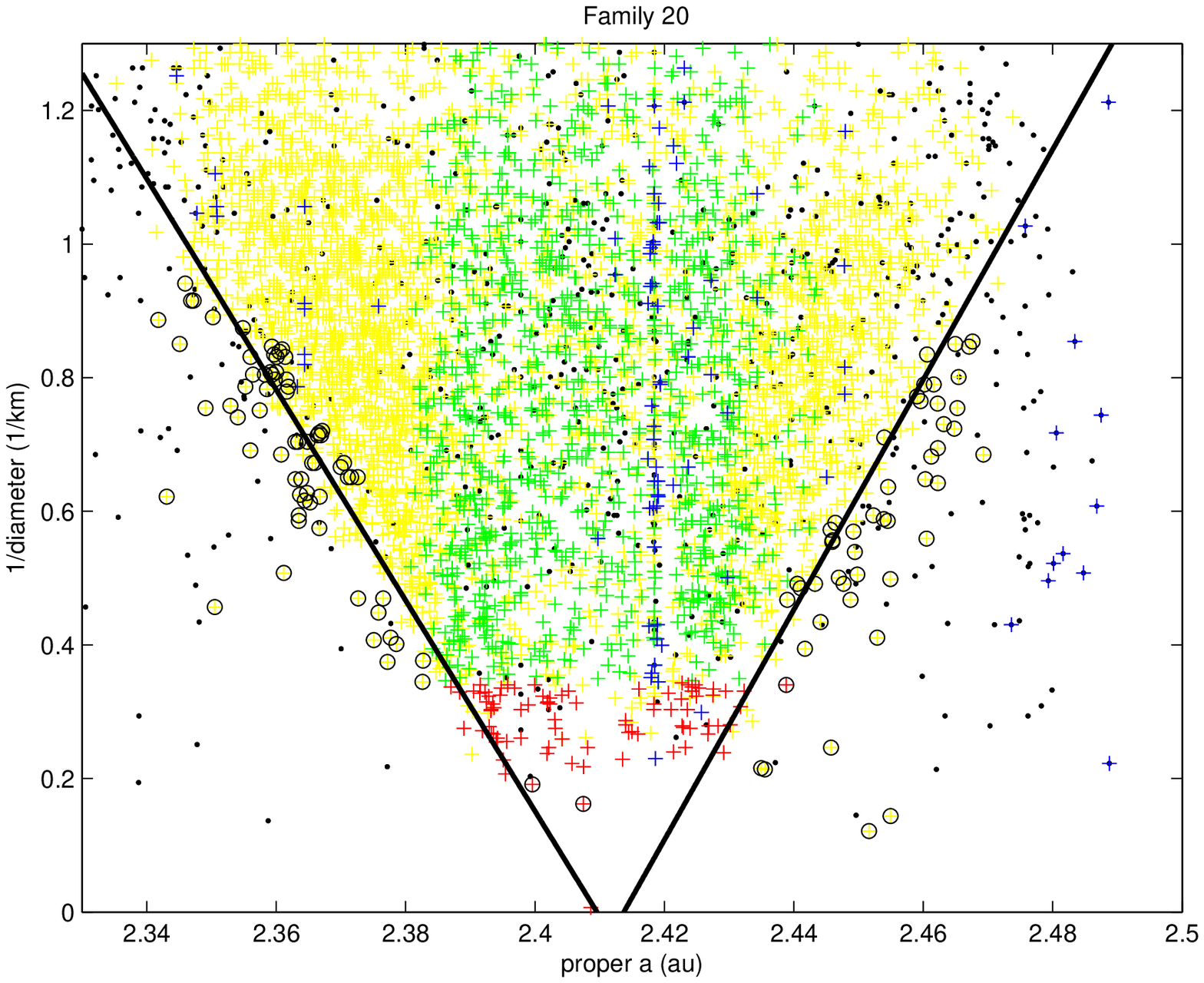}{V-shape of the family 20 in the proper $a$
    vs. $1/D$ plane. The black lines are the best fit on the two
    sides; the black circles indicate the outliers.}
\end{figure}

Then in a plot showing proper $a$ vs. $1/D$ for asteroids formed by
the same collisional event we get straight lines for the same
$\phi$. We can try to fit to the data two straight lines representing
the prograde spin and retrograde spin states ($\phi=0^\circ$ and
$\phi=180^\circ$). The slopes of these lines contain information on the
family age. Note that this procedure can give accurate results only if
the family members cover a sufficient interval of $D$, which now is
true for a large set of dynamical families thanks to the inclusion of
many smaller objects (represented by green and yellow points in all
the figures).

As an example in Figure~\ref{fig:20_vshapea} we show two such lines
for the Massalia family on both the low proper $a$ and the high proper
$a$ side, that is representing the above mentioned retrograde and
direct spin rotation state, respectively. This is what we call
\emph{V-shape}, which has to be analyzed to obtain an age estimate.

A method of age computation based on the V-shape has already been used
to compute the age of the Hungaria family \cite[Figure
20]{hungaria}. In principle, a similar method could be applied to all
the large (and several medium) families. However, a procedure capable
of handling a number of cases with different properties needs to be
more robust, taking into account the following problems.

\begin{figure}[h!]
  \figfig{12cm}{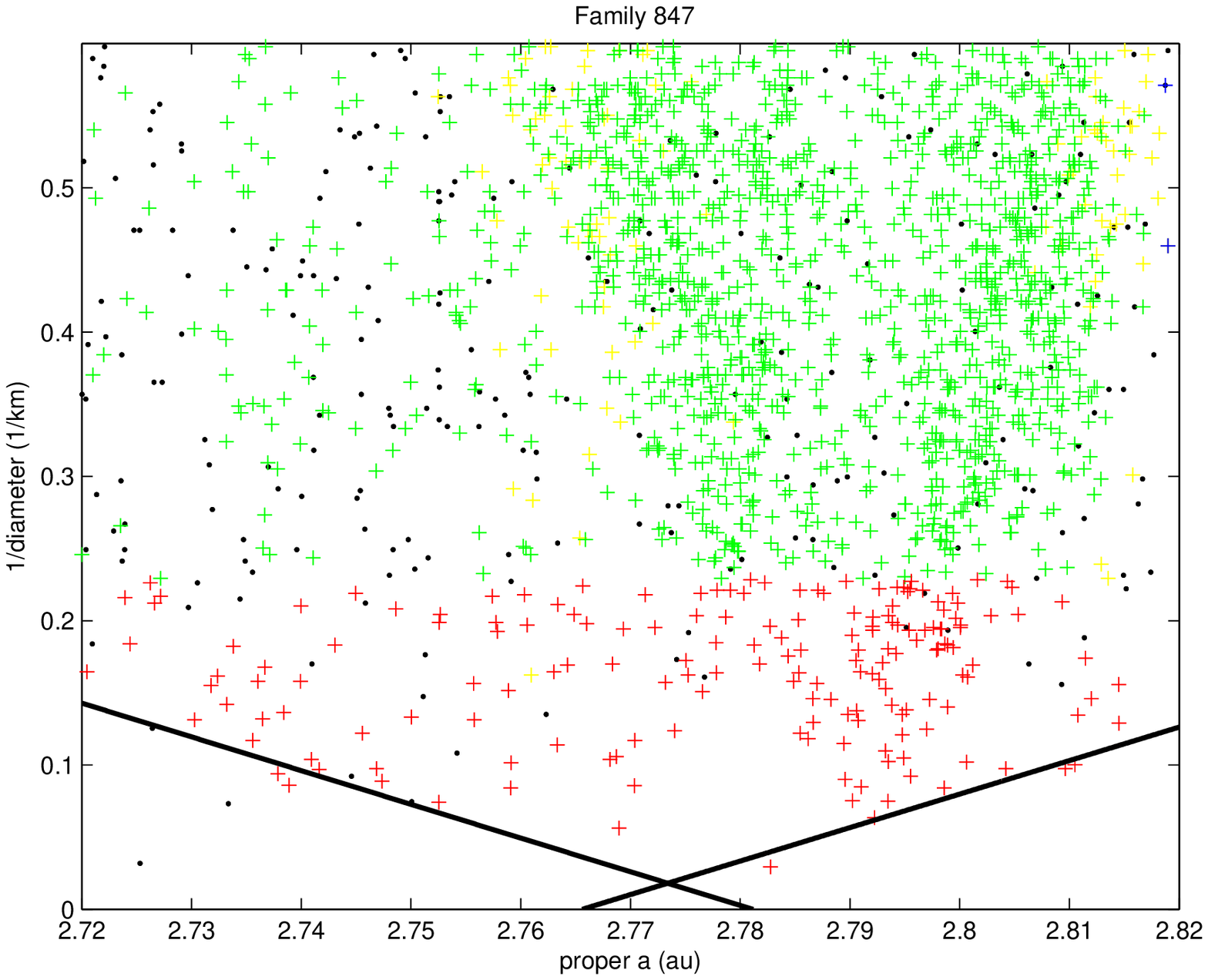}{V-shape of the family 847 in the proper
    $a$ vs. $1/D$ plane. The most striking feature is the presence of
    much denser substructure, which exhibits its own V-shape, shown in
    Figure~\ref{fig:3395_vshapea}.}
\end{figure}

\begin{itemize}

\item The method assumes all the family members have the same age,
  that is, it assumes the coincidence of the dynamical family with the
  collisional family. If this is not the case, the procedure is much
  more complicated: see Figure~\ref{fig:847_vshapea}, which shows two
  superimposed V-shapes (the outer one is marked with a fit for the
  boundary) for the Agnia family, indicating at least two collisional
  events with different ages. Thus a careful examination of the family
  shape, not just in the $a,1/D$ plot but also in all projections, is
  required to first decide on the minimum number of collisional events
  generating each dynamical family. If substructures are found, with
  shape such that interpretation as the outcome of a separate
  collisional event is possible, their ages may in some cases be
  computed.

\item To compute the age we use the inverse slope $\Delta a(D)/(1/D)$,
  with $D$ in km, of one of the two lines forming the V-shape, which
  is the same as the value of $\Delta a$ for an hypothetical asteroid
  with $D=1$ km along the same line. This is divided by the value of
  the secular drift $da/dt$ for the same hypothetical asteroid, giving
  the estimated age. However, the number of main belt asteroids for
  which we have a measured value of secular $da/dt$ is zero. There are
  $>20$ Near Earth Asteroids for which the Yarkovsky drift has been
  reliably measured (with $S/N>3$) from the orbit determination
  \cite[Table 2]{yarko_all}.  It is indeed possible to estimate the
  calibration $c$, thus the expected value of $da/dt$ for an asteroid
  with a given $D, a, e, \phi$, by scaling the result for another
  asteroid, in practice a Near Earth one, with different $D, a, e,
  \phi$. However, to derive a suitable error model for this scaling is
  very complicated; see e.g. \cite{farnocchia_apophis} for a full
  fledged Monte Carlo computation.

\item The data points $(1/D, a)$ in the V-shape are not to be taken as
  exact measurements. The proper $a$ coordinate is quite accurate,
  with the chaotic diffusion due to asteroid-asteroid interaction
  below $0.001$ au \cite{laskar2}, and anyway below the Yarkovsky
  secular contribution for $D<19$ km; the error in the proper elements
  computation (with the synthetic method) gives an even smaller
  contribution. To the contrary, the value of $D$ is quite inaccurate.
  Thus a point in the $1/D, a$ plane has to be considered as
  measurement with a significant error, especially in $1/D$, and the
  V-shape needs to be determined by a least squares fit, allowing also
  for outlier rejection.

\item Most families are bounded, on the low-$a$ side and/or on the
  high-$a$ side, by resonances strong enough to eject the family
  members reaching the resonant value of $a$ by Yarkovsky, into
  unstable orbits (at least most of them). Thus the V-shape is cut by
  vertical lines at the resonant values of proper $a$. The computation
  of $\Delta a$ must therefore be done at values of $1/D$ below
  the intersection of one of the slanted sides of the V and the
  vertical line at the resonant value of $a$. For several families
  this significantly restricts the range of $1/D$ for which the V-shape
  can be measured.

\item The dynamical families always contain interlopers, which should
  be few in number, but not necessarily representing a small fraction
  of the mass (the size distribution of the background asteroids is less
  steep). The removal of large interlopers is necessary not to spoil
  the computation of the slopes, and also of centers of mass.

\end{itemize}
As a consequence of the above arguments, we have decide to develop a
new method, which is more objective than the previous one we have
used, because the slope of the two sides of the V-shape is computed in
a fully automated way as a least squares fit, with equally automatic
outlier rejection. The following points explain the main features of
this new method.

\begin{enumerate}

\item For each family we set up the minimum and maximum values of the
  proper semimajor axis and of the diameter for the fit. We may use
  different values for the inner and the outer side of the V, taking
  into account the possibility that they measure different ages. Note
  this is the only ``manual'' intervention.

\item We divide the $1/D$-axis into bins, which are created in such a
  way to contain about the same number of objects.  Hence:

\begin{itemize}

\item the bins, which correspond to small values of $1/D$ are bigger
  than the ones which correspond to large values of $1/D$;
%  (Example figure???)

\item the inner side and the outer side of the family may have
  different bins.

\end{itemize}

%\item For each bin we choose one family member with coordinates
%  $(\bar{a},1/\bar{D})$ where $\bar{a}$ is the minimum value
%  (for low-$a$ or inner side) or the maximum value
%  (for high-$a$ or outer side) of all the values of the
%  proper semimajor axis in the bin.

\item We implement a linear regression for both sides. The method is
  iterative. For each iteration we calculate the residuals, the
  outliers and the kurtosis of the distribution of residuals. A point
  is an outlier if its residual is greater than $3 \sigma$. Then we
  remove the outliers and repeat the linear regression. Note that the
  outliers for the fit can be interlopers in the family, but also
  family members with low accuracy diameters.

\item We say that the method converges if the kurtosis of the
  residuals is $3 \pm 0.3$ or if there exists an iteration without
  additional outliers.

\item The two straight lines on the sides of the V-shape are computed
  independently, thus they do not need to cross in a point on the
  horizontal axis. We compute the \emph{V-base} as the difference in the
  $a$ coordinate of the intersection with the $a$ axis of the outer
  side and of the inner side. This quantity has an interpretation we
  shall discuss in Section~\ref{sec:collmodel}.

\end{enumerate}

\begin{table}[h!]
\footnotesize
 \centering
 \caption{Results of the fit for the low $a$ (IN) and high $a$ (OUT)
   sides for each considered family: number of iterations, minimum
   diameter $D$ (in km) used in the fit, number of bins, number of
   outliers, value of the kurtosis and standard deviation of the
   residuals in $1/D$, and the value of the inverse of the slope (in
   au).}
  \label{tab:tableslopes}
\medskip
  \begin{tabular}{lrrrrcccc}
  \hline
family &  &\# iter.  & min. D fit  &  \# bins  & \# outliers &  kurtosis & RMS(resid) & 1/slope\\
\hline
\\
20  &   IN&   7&  1.0&  20&  69&  2.98&  0.0371&            -0.063\\
20  &  OUT&   8&  1.1&  12&  48&  3.12&  0.0118&  \phantom{-}0.058\\
\\
4   &   IN&   2&   2.0&  23&   9&  3.27&  0.0364&            -0.267\\
4   &  OUT&   2&   3.6&   6&   2&  1.49&  0.0060&  \phantom{-}0.537\\
\\
15  &   IN&   2&   5.0&  11&   1&  3.22&  0.0021&            -0.659\\
15  &  OUT&   2&   6.7&  16&   1&  2.84&  0.0027&  \phantom{-}0.502\\
\\
158 &   IN&   2&  10.0&   9&   2&  3.03&  0.0013&            -0.442\\ 
158 &  OUT&   2&   5.9&  10&   0&  1.99&  0.0024&  \phantom{-}0.606\\
\\
847 &   IN&   2&   6.7&   5&   0&  1.25&  0.0004&            -0.428\\
847 &  OUT&   2&   9.1&   7&   0&  1.65&  0.0003&  \phantom{-}0.431\\
\\
3395&   IN&  10&   1.8&   7&  43&  2.92&  0.0041&            -0.045\\
3395&  OUT&   9&   2.2&   8&  49&  3.20&  0.0053&  \phantom{-}0.045\\
\\
\hline
\end{tabular}
\end{table}

\begin{table}[h!]
\footnotesize
 \centering
  \caption{V-base and center of the V-base.}  
  \label{tab:tableVbase}
\medskip
  \begin{tabular}{lcc}
  \hline
family & V-base & center of V-base\\
\hline
\\
20  &  \phantom{-}0.004&  2.4117\\
4   &            -0.025&        \\
15  &  \phantom{-}0.009&  2.6389\\
158 &  \phantom{-}0.021&  2.8774\\
847 &            -0.016&  2.7734\\
3395&            -0.008&  2.7900\\
\\
\hline
\end{tabular}
\end{table}

Let us emphasize that our main goal in this paper is to introduce
methods which are objective and take as thoroughly as possible into
account all the problems described above. Also we pay a special
attention to the computation of quantities like $\Delta a$ and the
slopes, used to estimate the family age, but the determination of the
ages themselves, involving the complicated calibration of the Yarkovsky
effect, is performed only as an example, to demonstrate what we
believe should be a rigorous procedure.

\subsubsection{Massalia}

One of the best examples of dynamical family for which the computation
of a single age for a crater is possible is the one of (20) Massalia,
see Figure~\ref{fig:massalia_ae}. Massalia has an albedo $0.21$
measured by IRAS\footnote{Massalia does not have a WISE albedo, but it
  is possible to use WISE data to confirm that $65\%$ of the members
  of family 20 have albedo between $0.17$ and $0.32$.}.

The two slopes of the inner and outer side of V-shape
(Figure~\ref{fig:20_vshapea}) have of course opposite sign, with the
absolute value different by $9\%$, see
Table~\ref{tab:tableslopes}. Taking into account that there is some
dependence of the calibration on $a$, this indicates an accurate
determination of the slope. This is due to the fact that the fit can
be pushed down to comparatively small diameters, around $1$ km, because
the family is not cut by a resonance on the low $a$ side, and is
affected by the $3/1$ resonance with Jupiter on the high $a$ side, but
only for $D<1$ km.  The V-base is small and positive
(Table~\ref{tab:tableVbase}).

The internal structure of family 20 is further discussed in
Section~\ref{sec:massalia}.

%\begin{figure}[h]
%\figfig{14cm}{euphrosyne_a1D}{caption}
%\end{figure}
%family (31) Euphrosyne: Using H, assuming albedo=0.045 (WISE) we get
%Figure~\ref{fig:euphrosyne_a1D}. No Yarkovsky calibration, but
%assuming $da/dt=-10^{-10}$ at 10 km diameter would give 650 My.

%useful case of medium family 808 using albedo=0.165 (+-0.021 WISE)
%from a-1/D we find very good V-shape, also thanks to the halo but
%high-a outlier 1327, confirmed by WISE alb 0.05+-0.01. Other two
%potential outliers on low-a side are 25701 Alexkeeler (WISE albedo
%0.256 +- 0.197) and 78288 2002PC50 (no albedo) cannot be ruled out
%(also no SDSS colors). Age between 160-190 My (excluding 25701 and
%78288)

%1128 good v-shape, gives age 120 My for $da/dt=2.5e-10$ au/y

\subsubsection{Vesta}

\begin{figure}[h!]
  \figfig{10cm}{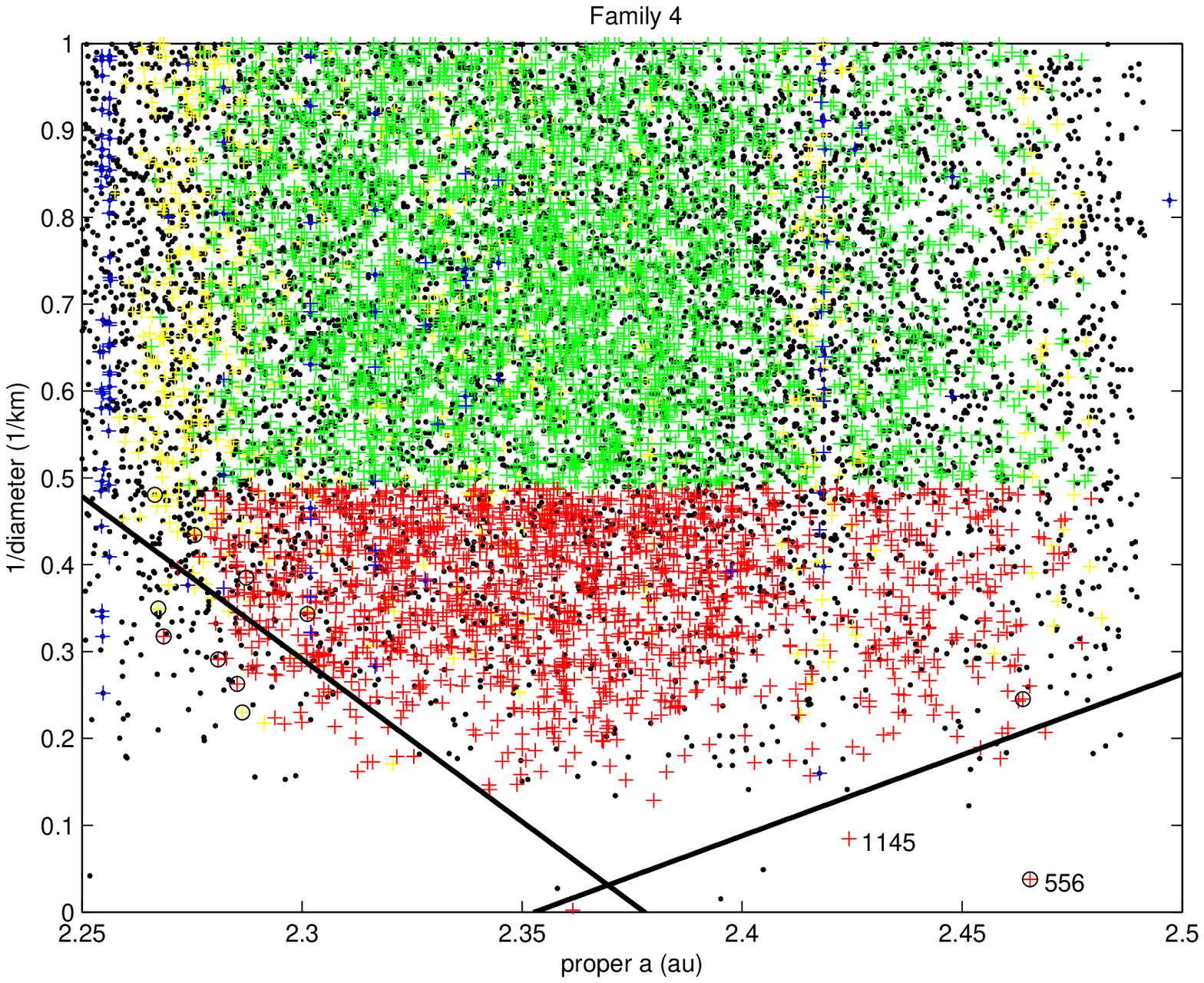}{V-shape of the family 4. The lines
    identified by the fit have different slopes on the two sides; for
    the explanation see Section~\ref{sec:vesta}.}
\end{figure}

For the V-shape and slope fit (Figure~\ref{fig:4_vshapea}) we have
used as common albedo $0.423$. 
The Vesta family has a complex structure, which is discussed in
Section~\ref{sec:vesta}: thus the presence of two different slopes on
the two sides (Table~\ref{tab:tableslopes}) indicates that we are
measuring the age of two different collisional events, the one forming
the high $a$ boundary of the family being older. In theory two
additional slopes exist, for the outer boundary of the inner subfamily
and for the inner boundary of the outer family, but they cannot be
measured because of the significant overlap of the two substructures.
Thus the negative V-base appearing in the figure has no meaning.

The family is cut sharply by the $3/1$ resonance with Jupiter on the
high $a$ side and somewhat less affected by the $7/2$ on the low $a$
side. As a result the outer side slope fit is somewhat less robust,
because the range of sizes is not as large. The fact that the slope is
lower (the age is older) on the high $a$ side is a reliable
conclusion, but the ratio is not estimated accurately.
The calibration constant is not well known, but should be similar
for the two subfamilies with the same composition (with only a small
difference due to the relative difference in $a$), thus the ratio of
the ages is $\sim 2/1$.

For the computation of the barycenter (Table~\ref{tab:tablebar_crat})
it is important to remove the interlopers (556) and (1145), which
clearly stick out from the V-shape on the outer side, although (1145)
is not rejected automatically by the fit\footnote{By using the
  smaller WISE albedos, these two are even
  larger than shown in Figure~\ref{fig:4_vshapea}.}.

\subsubsection{Eunomia}

\begin{figure}[h!]
  \figfig{10cm}{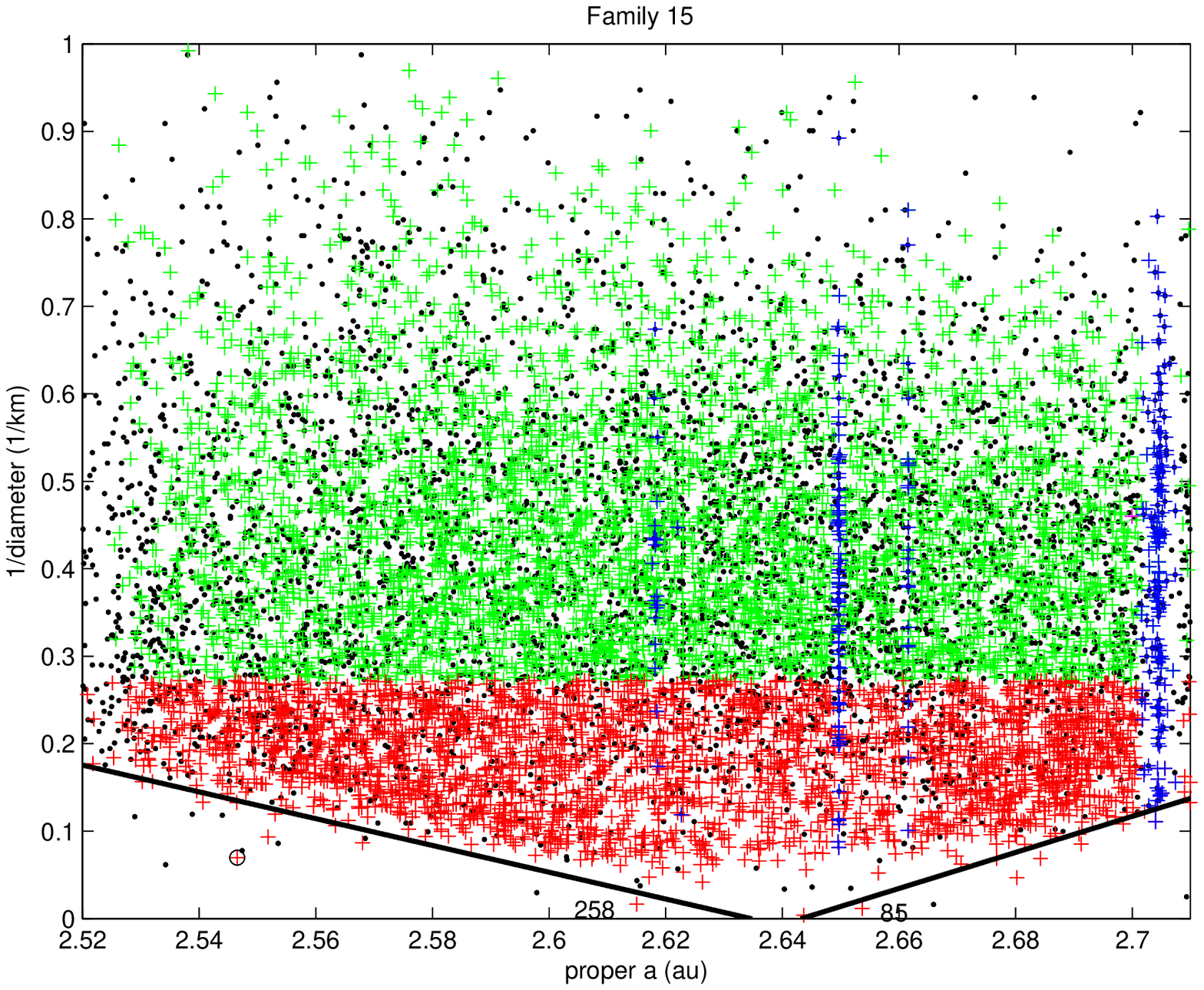}{V-shape of the family 15 exhibiting two
    rather different slopes; for the explanation see
    Section~\ref{sec:eunomia}.}
\end{figure}

For the V-shape plot of Figure~\ref{fig:15_vshapea} we have used as
common albedo the IRAS value for Eunomia which is $0.209$.  The inner
and outer slopes of the V-shape for the Eunomia family are different
by $31\%$. The slope on the outer side is affected by the $8/3$
resonance with Jupiter, forcing us to cut the fit already at $D=6.7$
km, thus the value may be somewhat less accurate. On the contrary the
inner slope appears well defined by using $D>5$ km, although the $3/1$
resonance with Jupiter is eating up the family at lower diameters. 
The V-base is small and positive (Table~\ref{tab:tableVbase}).
The possibility of an internal structure, affecting the interpretation
of the slopes and ages, is discussed in Section~\ref{sec:eunomia}.

For the computation of the barycenter (Table~\ref{tab:tablebar_crat})
it is important to remove the interlopers (85) and (258) which stick
out from the V-shape, on the right and on the left, respectively, with
the largest diameters after (15), see
Figure~\ref{fig:15_vshapea}\footnote{Moreover, the albedo of (85) Io
  is well known (both from IRAS and from WISE) to be incompatible with
  (15) Eunomia as parent body.}.

\begin{table}[h!]
\footnotesize
 \centering
 \caption{Cratering families: family, proper $a$, $e$ and $\sin{I}$ of
   the barycenter, position of the barycenter with respect to the
   parent body, escape velocity from the parent body. The barycenter
   is computed by removing the parent body, the interlopers and the
   outliers.}
  \label{tab:tablebar_crat}
\medskip
  \begin{tabular}{lccccccc}
  \hline
family & a$_b$ & e$_b$ & $\sin{I}_b$ & a$_b-$a$_0$ & e$_b-$e$_0$ & $\sin{I}_b$ & $v_e$ (m$/$s)\\
       &       &      &             &             &     & $-\sin{I}_0$      &\\ 
\hline
\\
20                &  2.4061&  0.1622&  0.0252&            -0.0025&  \phantom{-}0.0004&  \phantom{-}0.0004&  102\\
4                 &  2.3637&  0.1000&  0.1153&  \phantom{-}0.0022&  \phantom{-}0.0012&  \phantom{-}0.0040&  363\\
4 (N$\ne$1145)    &  2.3621&  0.0993&  0.1153&  \phantom{-}0.0006&  \phantom{-}0.0005&  \phantom{-}0.0040&     \\
4 low $e$         &  2.3435&  0.0936&  0.1169&            -0.0180&            -0.0052&  \phantom{-}0.0056&     \\
4 high $e$        &  2.3951&  0.1094&  0.1124&  \phantom{-}0.0336&  \phantom{-}0.0106&  \phantom{-}0.0011&     \\
15                &  2.6346&  0.1528&  0.2276&            -0.0091&  \phantom{-}0.0042&  \phantom{-}0.0010&  176\\
15 (N$\ne$85, 258)&  2.6286&  0.1495&  0.2282&            -0.0168&  \phantom{-}0.0010&  \phantom{-}0.0016&     \\
15 low $a$        &  2.6090&  0.1494&  0.2294&            -0.0347&  \phantom{-}0.0008&  \phantom{-}0.0028&     \\
15 high $a$       &  2.6808&  0.1501&  0.2246&  \phantom{-}0.0371&  \phantom{-}0.0015&            -0.0021&     \\
\\
\hline
\end{tabular}
\end{table}

\subsubsection{Koronis}

\begin{figure}[h!]
\figfig{10cm}{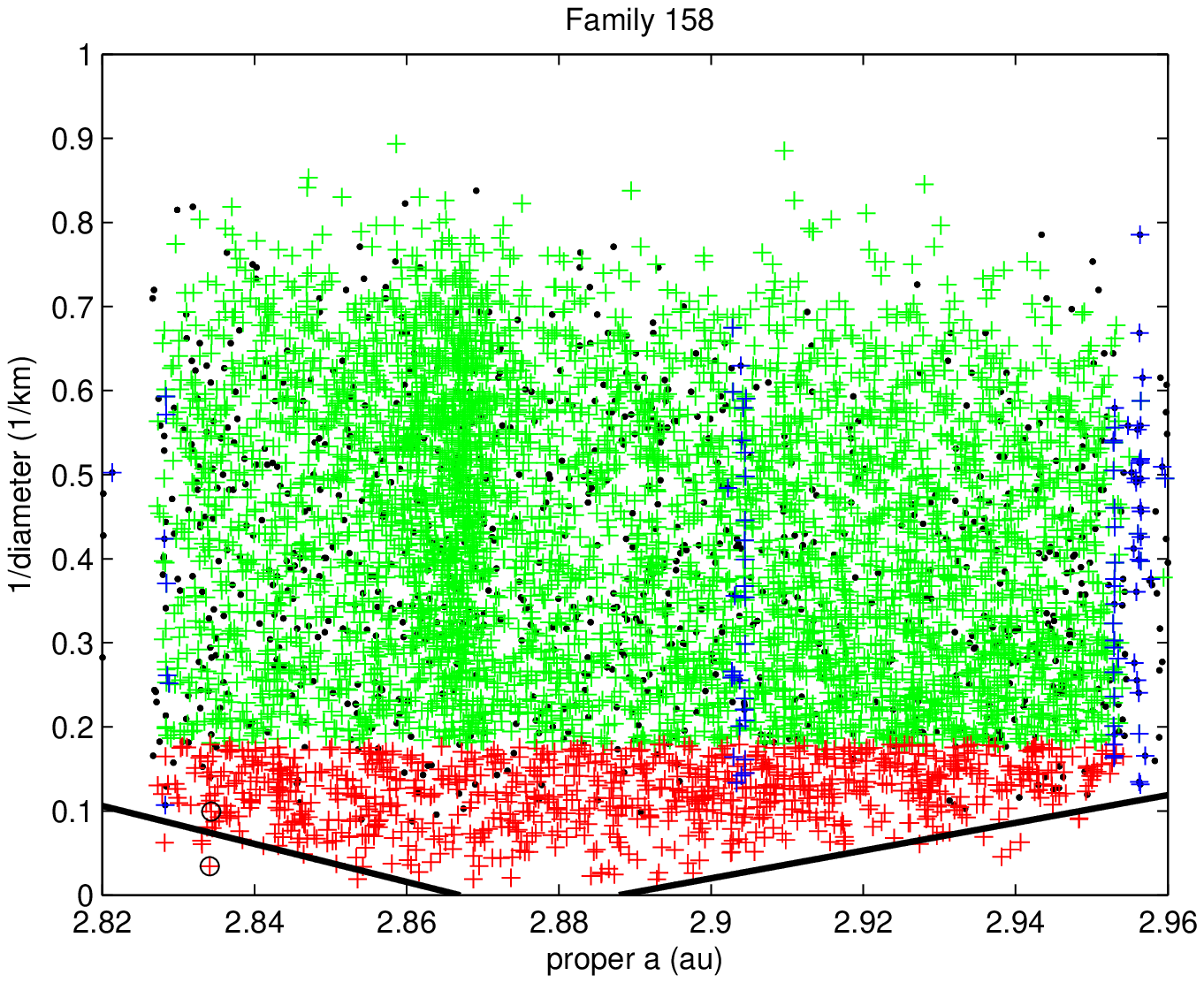}{V-shape of the family 158. The Karin
    subfamily is clearly visible at about $a=2.865$ au.}
\end{figure}

The Koronis family has a V-shape sharply cut by the $5/2$ resonance
with Jupiter on the low $a$ side, by the $7/3$ on the high $a$
side (Figure~\ref{fig:158_vshapea}). This results in a short range of
diameters usable to compute the slope, especially on the low $a$ side,
where we have been forced to cut the fit at $D=10$ km. This could be
the consequence of an already well known phenomenon, by which leakage
by Yarkovsky effect from family 158 into the $5/2$ resonance occurs
even for comparatively large objects \cite{vysh1,vysh2}.

This implies a less accurate slope estimate on the inner side. This
could explain the discrepancy by $37\%$ of the two slopes, since we
have no evidence for substructures which could affect the
V-shape\footnote{There is a well known substructure, the Karin
  subfamily, which is perfectly visible in
  Figure~\ref{fig:158_vshapea}, but does not affect the two sides of
  the V-shape.}. Anyway we recommend to use the outer slope for the
age estimation.

\subsubsection{Agnia}

The Agnia family, as shown by Figure~\ref{fig:847_vshapea}, has a
prominent subfamily forming a V-shape inside the wider V-shape of the
entire family. We call this structure the subfamily of (3395) Jitka.

\begin{figure}[h!]
  \figfig{10cm}{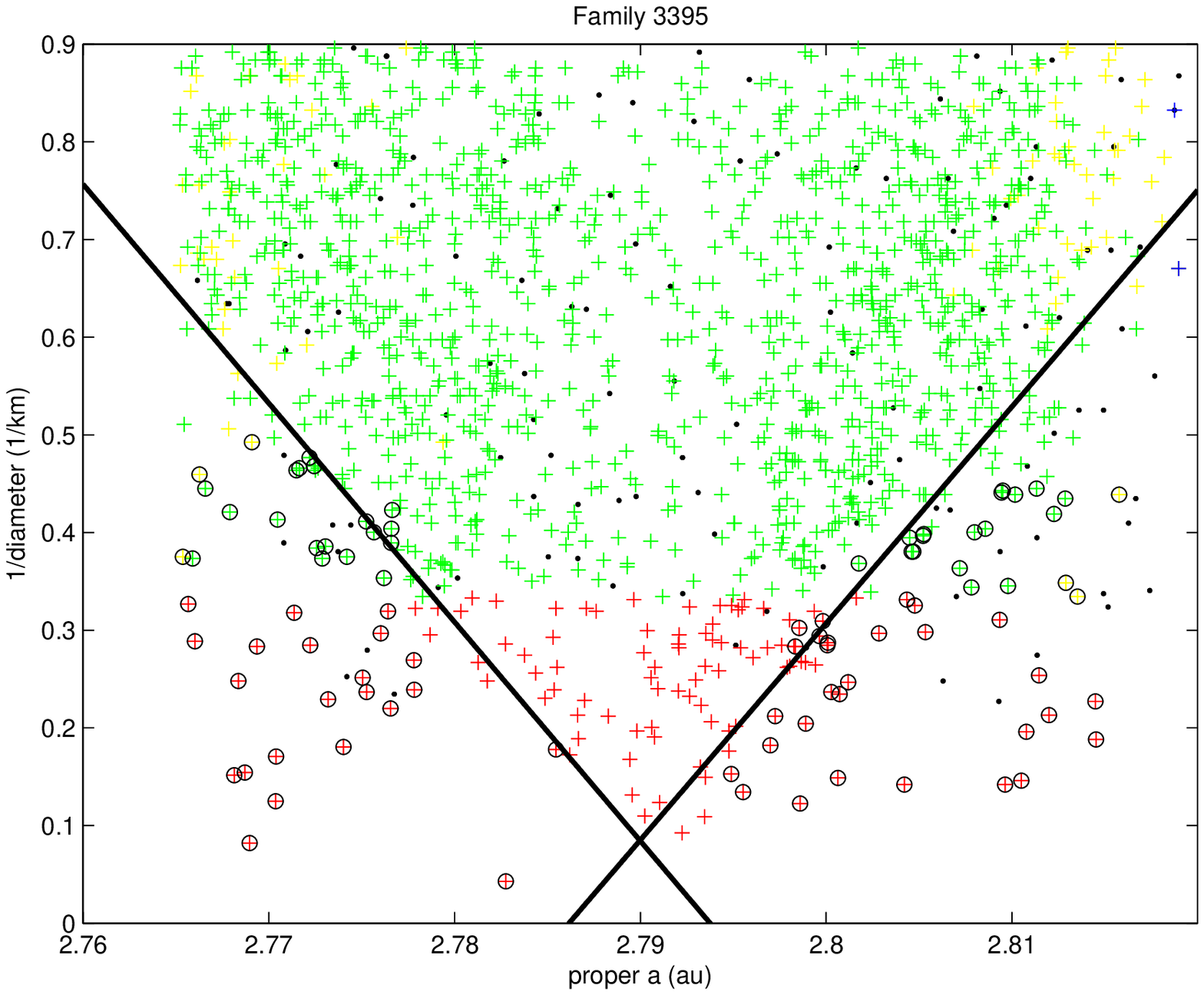}{V-shape of the subfamily of (3395)
    Jitka. Most of the outliers, marked by a black circle, are members
    of the larger Agnia family, but not of the subfamily.}
\end{figure}

For the entire family, almost identical values of the slopes on the
two sides (Table~\ref{tab:tableslopes}) appear to correspond to the
much older age of a wider and less dense family.  The two slopes of
the Jitka subfamily are also identical, but with inverse slopes lower
by a factor $>9$. The V-base is negative in both cases
(Table~\ref{tab:tableVbase}).

The Jitka subfamily shows in the V-shape plot
(Figure~\ref{fig:3395_vshapea}) a depletion of the central portion,
which should correspond to obliquities $\gamma$ not far from
$90^\circ$. This can be interpreted as a signature of the YORP effect,
in that most members with $\gamma\sim 90^\circ$ would have had their
spin axes evolved by YORP towards one of the two stable states,
$\gamma=0^\circ, 180^\circ$. 

If the two collisional families belong to parent bodies with similar
composition, then the ratio of the inverse slopes correspond to the
ratio of the ages, independently from the calibration. Thus Jitka
could be a catastrophic fragmentation of a fragment from another
fragmentation $9$ times older.

However, there are some problems if we use the WISE albedo data for
family 847 members; there are $114$ albedos with $S/N>3$, which
introduces some risk of small number statistics. Anyway, they indicate
that the two subgroups, the 3395 subfamily and the rest of the 847
dynamical family, have a similar distribution of albedos, including
dark interlopers. However, the albedo of (847) Agnia $0.147\pm 0.012$
is lower than most family members, while (3395) Jitka has $0.313\pm
0.045$ which is more compatible with the family. Thus it is not clear
whether (847) Agnia is the largest remnant or an interloper, and
whether the parent body of the Jitka subfamily did belong to the
first generation family.

\begin{table}[h!]
\footnotesize
 \centering
 \caption{Fragmentation families: family, proper $a$, $e$ and
   $\sin{I}$ of the barycenter. The barycenter is computed by removing
   the outliers.}
  \label{tab:tablebar_frag}
\medskip
  \begin{tabular}{lccc}
  \hline
family & a$_b$ & e$_b$ &  $\sin{I}_b$ \\
\hline
\\
158           &  2.8807&  0.0488&  0.0371\\
847           &  2.7799&  0.0715&  0.0664\\
847 (w/o 3395)&  2.7462&  0.0725&  0.0654\\
3395          &  2.7911&  0.0728&  0.0669\\
\\
\hline
\end{tabular}
\end{table}

\subsubsection{Yarkovsky effect calibration and family age estimation}

Recalling that there is not a single measurement of the Yarkovsky
effect for the main belt asteroids, thus also for the families, we can
perform the necessary calibration only by using the available
measurements for the Near Earth Asteroids.

Thus, here the age estimation is obtained by scaling the results for
the asteroid for which there is the best Yarkovsky effect
determination \cite{yarko_all}, namely the low-albedo asteroid
(101955) Bennu, with scaling taking into account the different values
of $D,a,e, \rho$ and $A$, where $\rho$ is the density and $A$ is the
Bond albedo. The $da/dt$ value for (101955) Bennu has a $S/N=197.7$,
thus a relative uncertainty $<1\%$. The scaling formula we have used
is:
\[
\frac{da}{dt} = \left.\frac{da}{dt}\right|_{Bennu}
\frac{\sqrt{a}_{(Bennu)}(1-e^2_{Bennu})}{\sqrt{a}(1-e^2)}
\frac{D_{Bennu}}{D}\frac{\rho_{Bennu}}{\rho}
\frac{\cos(\phi)}{\cos(\phi_{Bennu})}\frac{1-A}{1-A_{Bennu}}
\]
where $D=1$ km used in this scaling formula is not the diameter
of an actual asteroid, but is due to the use of the inverse
slope and $\cos(\phi)=\pm 1$, as explained in the description
of the method above.

It may appear that the use of the Yarkovsky effect measurements for
asteroids more similar in composition to the considered families than
Bennu would be more appropriate. So, for example, the asteroid (2062)
Aten has the best determined $da/dt$ value of all S-type asteroids. It
has a $S/N=6.3$, thus a relative uncertainty $0.16$, which has to be
taken into account in the calibration error. As for the scaling
formula above, it introduces additional uncertainty, especially since
there is no scaling term accounting for the different thermal
properties.

Thus, using an S-class asteroid for scaling may not result in a better
calibration, because the S-type asteroids are not all the same, e.g.,
densities and thermal properties may be different. In the case of
(2062) Aten there is an error term due to the lack of knowledge on the
obliquity $\phi$, which can contribute an additional relative error up
to $0.2$. The other two S-type asteroids with measured Yarkovsky are
(1685) Toro and (1620) Geographos, but with $S/N= 3.7$ and $3.0$,
respectively. 

In the same way, for the family of (4) Vesta one would expect that the
use of the Yarkovsky measurement for asteroid (3908) Nyx, presumably
of V-type \cite{binzel}, should represent a natural choice for
calibration. In fact, the same authors warn that this asteroid belongs
to a small group of objects with ``sufficiently unusual or relatively
low $S/N$ spectra'', thus the taxonomic class may be different from
nominal. This suspicion is further strengthened by the value of
geometric albedo of only $0.16 \pm 0.06$ reported by \cite{benner},
which is significantly lower than the typical value ($\sim 0.35$) for
a Vestoid. (3908) Nyx is apparently of extremely low density
(Farnocchia and Chesley, private communication), thus it has too many
properties inconsistent with Vestoids.

This is why we have decided to use (101955) Bennu as benchmark to
be scaled for the Yarkovsky calibration of all families, because it
is the known case with both the best estimate of Yarkovsky and  best
known properties, including obliquity, density, and size.

%\subsubsection{Family age estimation}

\begin{table}[h!]
\footnotesize
 \centering
 \caption{Family age estimation: family, $da/dt$ for $D=1$ km obtained
   using (101955) Bennu for the calibration, for the two sides of the
   V-shape, and corresponding family age estimation.}
  \label{tab:tableage}
\medskip
  \begin{tabular}{lrcc}
  \hline
family & & da/dt  & $\Delta$t \\
       & & ($10^{-10}$ au/y)& (Gy)\\
\hline
\\
20  &   IN&           -3.64& 0.173\\
20  &  OUT& \phantom{-}3.55& 0.163\\
\\
4   &   IN&           -2.65& 1.010\\
4   &  OUT& \phantom{-}2.57& 2.090\\
\\
15  &   IN&           -3.49& 1.890\\
15  &  OUT& \phantom{-}3.39& 1.480\\
\\
158 &   IN&           -3.13& 1.410\\
158 &  OUT& \phantom{-}3.08& 1.970\\
\\
847 &   IN&           -3.37& 1.270\\
847 &  OUT& \phantom{-}3.33& 1.300\\
\\
3395&   IN&           -3.35& 0.134\\
3395&  OUT& \phantom{-}3.33& 0.135\\
\\
\hline
\end{tabular}
\end{table}

The results of our age computation for the considered families are given in
Table~\ref{tab:tableage}, for the two sides of the V-shape.

As one can appreciate from these data, the estimations of the age of
Massalia family from the two slopes differ by only a small
amount, and they are also in good agreement with results obtained with a
quite different method by \cite{vokyorp}.

The Vesta family case is particularly interesting as the lower age
appears to be compatible with the estimated age of one of the two
largest craters on Vesta, Rheasilvia ($\sim 1$ Gy)
\cite{dawn_marchi}. An age for the other big crater, Veneneia, has not
been estimated, although it must be older because this crater is below
the other.  Our estimated $\sim 2/1$ ratio for the collisional
families ages is an interesting result, although it should not be
considered as a proof that the sources  are the
two known largest craters.

The difference of the values of the inner and outer slopes of the
V-shape for the Eunomia could be interpreted as the age of two different
events, see Section~\ref{sec:eunomia}. There is no previous estimate
of the age of Eunomia we are aware of, a ``young age'' being
suggested on the basis of simulations of size distribution evolution
by \cite{michel}. 

The estimation of the age of Koronis family as inferred from the
longer outer side of the V-shape is consistent with the age ($\leq 2$
Gy) reported previously by \cite{marzari}, based on the observed size
distribution of larger members, and by \cite{chapman}, based on the
crater count on the surface of the Koronis family member (243)
Ida. \cite{bottke2001} give $2.5\pm 1$ Gy, which is also consistent.

The age estimate for the Agnia family of $<140$ My, provided
by \cite{agnia_vok}, is in a very good agreement with our result for
the Jitka subfamily; the older age for the entire Agnia family has not
been found previously because the low $a$ component identified by us
was not included in the family.

The two youngest according to our estimates, family 20 and subfamily
3395, have in common the presence of a lower density central region of
the V-shape, more pronounced for Jitka, barely visible for
Massalia. This suggests the following: the time scale for the YORP
effect to reorient the rotation axis towards either full spin up or
full spin down is much smaller than the time scale for randomization
of a significant fraction of the spin states, which would fill the
central gap.

We need to stress that the main uncertainty in the age computation is not
due to the estimate of the slope (apart for the ``bad'' case of the
inner slope of Koronis). The main error term is due to the
calibration; we do not yet have enough information to derive a formal
standard deviation estimate, but the relative uncertainty of the age
should be of the order of $0.2\div 0.3$.

% 17/114 alb<0.1
% alb(847)=0.145+-0.012 alb(3395)=0.313+-0.045 mica tantop compatibili

\subsubsection{Collisional Models and the interpretation of V-shapes}
\label{sec:collmodel}

The method we have proposed for the computation of family ages has the
advantage of using an objective measurement of the family V-shape,
rather than using a line placed ``by eye'' on a plot. However, because
two parameters are fit for each boundary line, that is the slope and
the intersection with a-axis, whenever both sides are accessible the
output is a set of four parameters: the inverse slopes on both sides
(Table~\ref{tab:tableslopes}), the V-base and the center of V-base
(Table~\ref{tab:tableVbase}). To force the lines to pass from a single
vertex on the horizontal axis would remove one fit parameter, to
assign also the proper $a$ of this vertex (e.g., at the value of some
barycenter) would remove two parameters: by doing this we would bias
the results and contradict the claimed objectivity of the procedure,
which has to be defined by the family membership only.

On the other hand, our procedure does not use the V-base and its
center to estimate the family age. Only the slopes of the leading edge
of family members (for either high or low $a$) are used for the
ages. This leads to two questions: which information is contained in
the two parameters we are not using, and is it appropriate to obtain,
e.g., a negative V-base, or is this an indication of a poor fit of at
least one of the two lines?

The interpretation of the V--shape plots is not straightforward,
because they are the outcome of a game involving three major players,
each one producing its own effect. These players are: (1) the
collisional history of a family, including the possible presence of
overlapping multi-generation events; (2) the Yarkovsky effect, which
in turn is influenced by the YORP effect; (3) the original field of
fragment ejection velocities at the epoch of family formation. In
addition, also the possible presence of strong nearby resonances plays
an important role.  Note also that in the present list of families
several ones have been created by a cratering and not by a
catastrophic disruption.

As for the effect (3), the existence of a correlation between the size
and the dispersion in semi-major axis of family members has been known
for several years. In the past, pre--Yarkovsky era (and with most of
the recognized families resulting from catastrophic events), this
correlation was assumed to be a direct consequence of the distribution
of original ejection velocities, with smaller fragments being ejected
at higher speeds. The ejection velocities derived from observed proper
element differences, however, turned out to be too high to be
consistent with the experiments, since they implied
collisional energies sufficient to thoroughly pulverize the parent
bodies.

Later, the knowledge of the Yarkovsky effect and the availability of
more detailed hydrodynamic simulations of catastrophic fragmentation
family--forming events (see, for instance, ~\cite{michel}) suggested a
different scenario: most family members would be reaccumulated
conglomerates, issued from merging of many fragments ejected at
moderate velocities. In this scenario, the original ejection
velocities give a moderate contribution to the observed dispersion of
proper elements. Then the V--shape plots discussed in the previous
subsections would be essentially a consequence of such
Yarkovsky--driven evolution (see~\cite{bottke2002} for a general
reference). The extension of the above scenario to families formed by
craterization events is not obvious, nor --at the present time--
supported by numerical simulations, which are not yet capable to reach
the required resolution~\cite{jutzi}.  However, the interpretation of
the V--shape as a consequence of Yarkovsky effect should hold also for
them.

Unfortunately, a fully satisfactory interpretation of the observed
V--shape plots can hardly be achieved in such a purely
Yarkovsky--dominated scenario: the original ejection velocities of
fragments cannot be totally disregarded.  For the Eos family
\cite{brozmorby}, \cite{eos_vok}, assume, for bodies of a size of $5$
km, average asymptotic relative velocities $v_\infty$ of about $90$
m/s.  This is even more true for the families formed by cratering
events on very large asteroids, since ejection velocities $v_0$ must
be $>v_e$ (escape velocity) as to overcome the gravitational well of
the parent body, and the $v_\infty$ of the family members are both
large and widely dispersed (see Section~\ref{sec:craters}).

Due to the original dispersion of the family members, we cannot expect
that the two sides of any given V--plot exactly intersect on the
horizontal axis, as one might expect for a "pure" Yarkovsky model.
The original extension of the family depends on the ejection
velocities $v_\infty$ of the bodies, while the Yarkovsky effect on
every body of a given size depends on the orientation of the spin
vector. If velocities and spin vectors are not correlated, the two
terms should combine as independent distributions.  If the Yarkovsky
term is assumed to be the dominant signal, the original velocities
provide a noise term; the noise/signal value is certainly significant
for large objects, thus the two lines of the "V" should not intersect
at $1/D=0$, but in the halfplane $D<0$. The "V--base" has therefore to
be positive.  Yet, this is not the case in $3$ out of $6$ examples
presented in this paper. How to possibly explain this?

A more physical explanation may be tentatively suggested, based
on an argument which has been previously discussed in the literature
\cite{laspina}, \cite{paolicchi2008} but yet not fully explored.
According to the results of some laboratory fragmentation experiments
\cite{fuji}, \cite{holsapple} the fragments ejected from a
catastrophic disruption rotate, and the sense of spin is related to
the ejection geometry: the fragments rotate away from the side of
higher ejection velocity.  Such behavior is clearly represented
in \cite[Fig. 1]{fuji}. This experimental evidence was used in
developing the so-called semi-empirical model
\cite{paolicchi1989,paolicchi1996}, assuming that fragment rotations
are produced by the anisotropy in the velocity ejection field.
\begin{figure}[h!]
  \figfig{9cm}{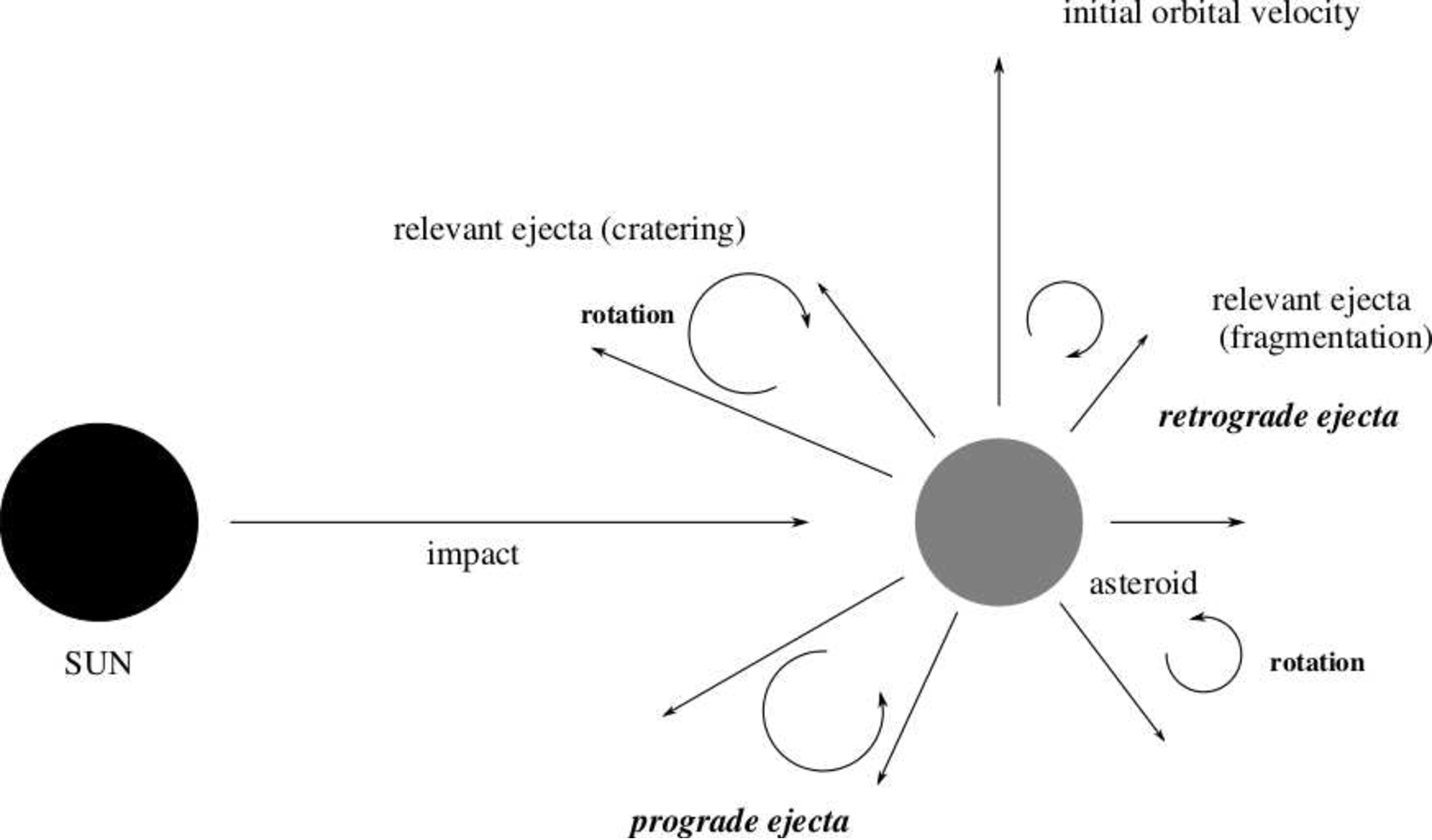}{The possible correlation spin--ejection
    velocity for a radial impact from the interior of the Solar
    System. This is a projection on the orbital plane. For a cratering
    impact the ejecta are in the same hemisphere as the impact
    point, while for a catastrophic disruption the crater zone is
    pulverized, most sizable fragments are ejected from the antipodal
    hemisphere. In both cases the ejection velocity decreases with the
    angular distance from the impact point.  If the rotation is
    connected to the velocity shear the fragment with a positive along
    track velocity (top of the figure) have a retrograde (clockwise)
    rotation, and viceversa. This is true both for the front side
    ejecta (cratering) and for the rear side fragments
    (disruption). In this case the correlation between the initial
    $\Delta a$ and $\cos\gamma$ is negative, and the Yarkovsky effect
    tends initially to shrink the family in $a$.}
\end{figure}

In this scenario, the rotation of fragments created in a catastrophic
process can be strongly correlated with the ejection velocity. For
what concerns cratering events, as far as we know, there is not in
this respect any experimental evidence mentioned in the
literature. However, also in cratering events the ejection velocity
field is strongly anisotropic (see, for instance, the popular Z--model
by \cite{maxvell}), and a similar correlation between ejection velocity
and spin rate can be expected for the fragments. It is not obvious how
significantly the reaccumulation of ejecta (a process which certainly
is very important after catastrophic events) can affect this
correlation. There are very few simulations taking into account the
rotation of fragments recorded in the literature \cite{richardson},
\cite{michel}, they are all about fragmentations, and their results do
not solve the present question. However, if the fragments which
stick together were ejected from nearby regions of the parent body, an
original correlation might be preserved.

If this is the case, different impact geometries will result in
different evolutions of the semi-major axis spread of the family. To
model the geometry of the impact, let us call \emph{crater radiant}
the normal $\hat n$ to the smoothed terrain before the crater is
excavated (at the impact point). What matters are the angles between
$\hat n$ and the directions $\hat v$ of the orbital velocity of the
parent body, and $\hat s$ towards the Sun (both at the epoch of the
impact).  

If $\hat n\cdot \hat s>0$ (impact on the inner side) with $\hat n\cdot
\hat v\simeq 0$ (impact radiant close to normal to the velocity) there
are preferentially retrograde fragments on the side where ejection
velocity adds up with orbital velocity, thus giving rise to larger a
of fragments, preferentially prograde on the opposite, lower a
side. This implies that the spread in proper $a$ of the family
initially decreases (ejection velocity and Yarkovsky term act in the
opposite sense), then increases again, and the V-base is negative.
If $\hat n\cdot \hat s<0$ (impact on the outer side) with $\hat n\cdot
\hat v\simeq 0$ there are preferentially prograde fragments at larger
$a$, preferentially retrograde at lower $a$. This results in a large
spread, even after a short time, of the family in proper $a$ (ejection
velocity and Yarkovsky term add), and the V-base is positive.
Finally, in case of negligible $\hat n\cdot \hat s$, the
original ejection velocities and Yarkovsky drift add up as a
noise terms, the latter dominating in the long run; the V-base is
positive but small. Note that, as shown by Figure~\ref{fig:Fig_fam},
this argument applies equally to cratering and to fragmentation
cases.

Thus, in principle, the properties of the V--base and of the family
barycenter (Tables~\ref{tab:tableVbase}, \ref{tab:tablebar_crat}, and
\ref{tab:tablebar_frag}) contain information on the impact geometry
and on the original distribution of $v_\infty$.  However, the
interpretation of these data is not easy. A quantitative model of the
ejection of fragments, describing the distribution of $v_\infty$, the
direction of $v_\infty$, $\cos\gamma$, and $D$, taking into account
all the correlations, is simply not available. We have just shown that
some of these correlations (between direction and $\cos\gamma$) are
not negligible at all, but all the variables can be correlated. Even
less we have information on shapes, which are known to be critical for
the YORP effect.

This does introduce error terms in our age estimates.  The main
problem is the dependence of the Yarkovsky drift in $a$, averaged over
very long times, from $D$.  According to the basic YORP theories
(see~\cite{bottke2006} for a general reference) the bodies should
preferentially align their rotation axes close to the normal to the
orbital plane (both prograde and retrograde), with a timescale
strongly dependent on the size. This result is also supported by the
recent statistical work on the spin vector
catalog~\cite{paolicchi2012}. Consequently, there should be a
substantial fraction of the small bodies moving towards the borders of
the V--plot, especially after times long with respect to the time
scale for the YORP-driven evolution to the spin up/spin down stable
states. Using a different database~\cite{vokyorp} have found for most
families, a number density distribution in accordance to this idea.
However, the maxima are not at the edges, but somewhere in between the
symmetry axis of the V-shape and the edges: e.g., see
Figure~\ref{fig:3395_vshapea}. It is not easy to draw general
conclusions from this kind of data, because in most familes the
portions near the extreme values of proper $a$ are affected by
resonances and/or by the merging of step 3 families as haloes.

There are many models proposed in the literature to account for a form
of randomization of the spin state, resulting in something like a
Brownian motion along the $a$ axis over very long time scales; e.g.,
\cite{statler} and \cite{cotto} show that the YORP effect can be suddenly
altered. Thus after a long enough time span, most family members may
be random-walking between the two sides of the V-shape, and the
central area is never emptied. However, what we are measuring is not the
evolution in $a$ of the majority of family members, but the evolution
of the members fastest in changing $a$. Our method, indeed any method
using only the low and high $a$ boundaries of the family, should be
insensitive to this effect for large enough families. In a random
effect a portion of the family with a spin state remaining stable at
$\cos\gamma\simeq \pm 1$ will be maintained for a very long time, and
this portion is the one used in the V-shape fit.

Our method is mathematically rigorous in extracting from the family
data two components of the evolution of proper $a$ after the family
formation, a term which is constant in time (from the original
distribution of velocities) and independent from $D$, and a term which
is proportional to $1/D$ and to the time elapsed. If the situation is
much more complicated, with a larger number of terms with different
dependence on both $D$ and $t$, we doubt that the current dataset is
capable of providing information on all of them, independently from
the method used. Moreover, some terms may not be discriminated at all,
such as an $1/D$ dependency not due to a pure Yarkovsky term $\Delta
t/D$. 

\subsection{Size distributions}

\begin{figure}[h!]
  \figfig{11cm}{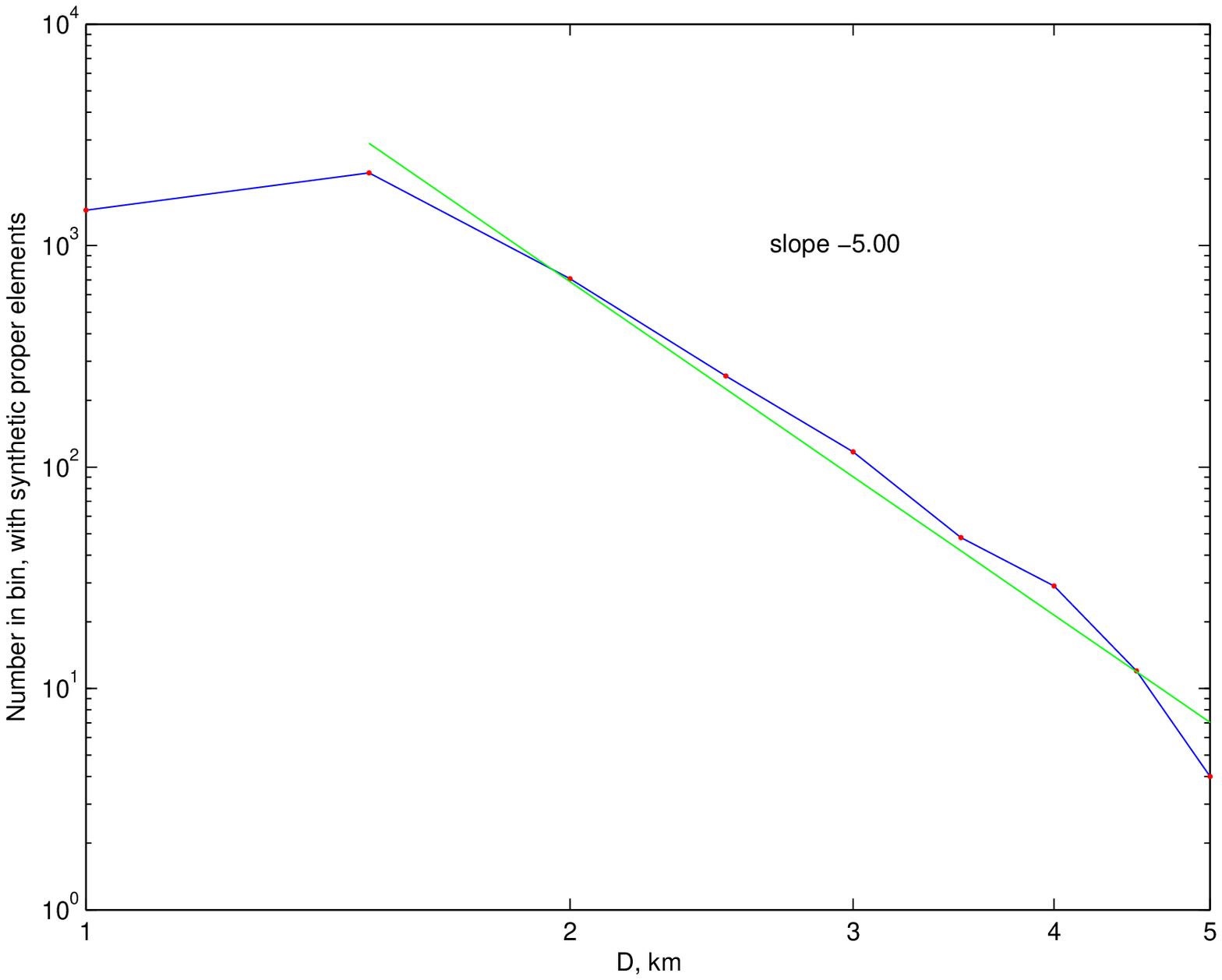}{Size distribution of family 20 using
    the range $1.5<D<5$ km.}
\end{figure}

Another use of the diameters deduced from absolute magnitudes assuming
uniform albedo is the possibility of computing a size
distribution. This is a very delicate computation, depending strongly
upon the range of diameters used. Numbers of members at too small
diameters are affected by incompleteness, while too large diameters
are affected by small number statistics, especially for cratering
events. The utmost caution should be used in these
estimates for families less numerous and/or with a more complex
structure.

In Figure~\ref{fig:massalia_sizefit} we show the result of a size
distribution power law fit for family 20, by using the range from
$D=1.5$ to $5$ km, thus excluding (20) and the two outliers identified
above as well as two others above $5$ km. The resulting best fit
differential power law is proportional to $1/D^{5}$, that is the
cumulative distribution is proportional to $1/D^{4}$; this value suggests
that the fragments are not yet in collisional equilibrium, thus
supporting a comparatively young age for the family. The results are
somewhat dependent upon the range of diameters considered, as it is
clear by the much lower value for diameters $D< 1$ km with respect to
the fit line (in green): this is a clear signature of observational
incompleteness. 
  
\begin{figure}[h!]
\figfig{11cm}{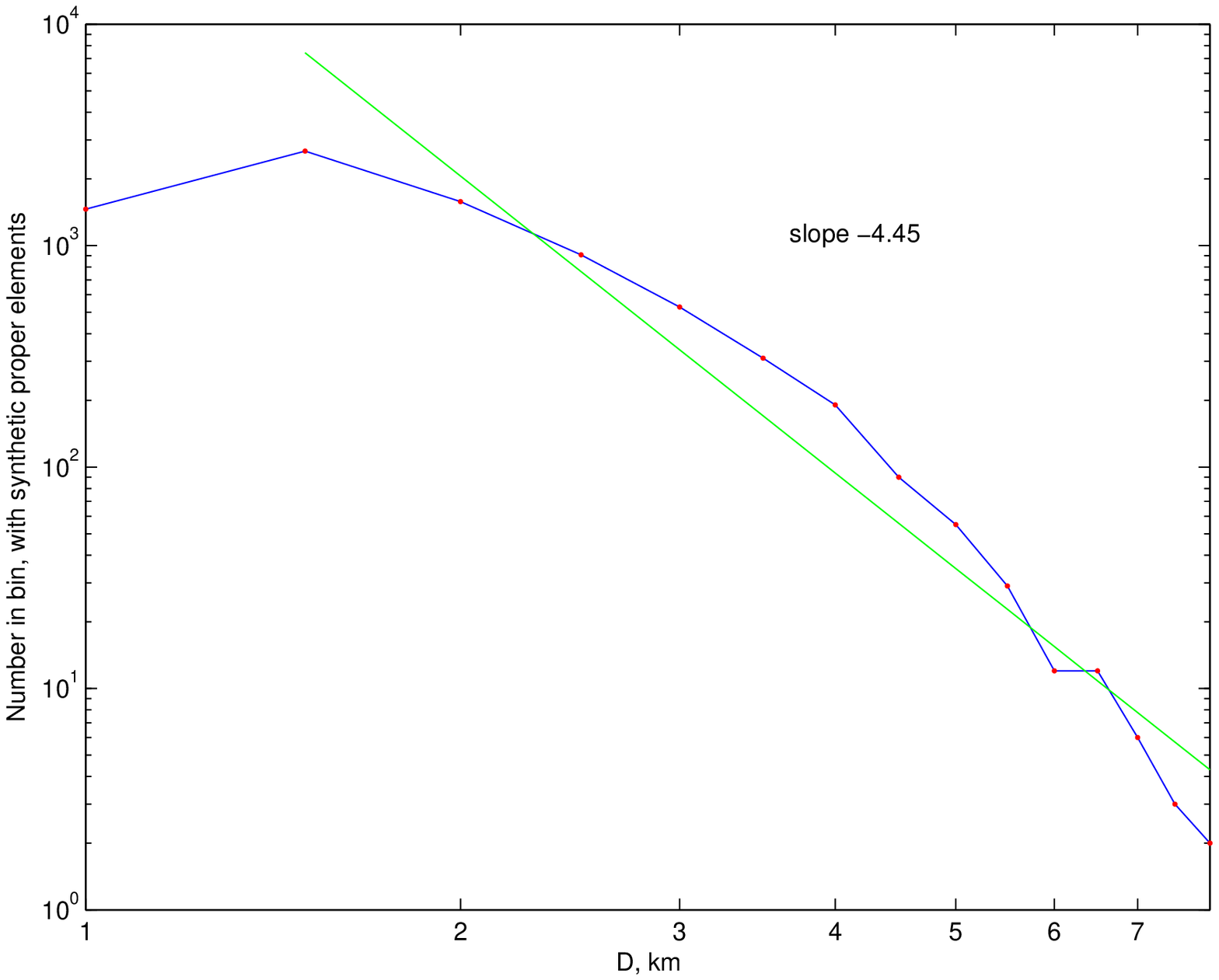}{Size distribution of family 4 using
    the range $1.5<D<8$ km.}
\end{figure}

In Figure~\ref{fig:vesta_sizefit} we show the size distribution power
law fit for family 4; to improve the estimates of the diameters, we
have used as a common albedo $0.35$ because it is more representative
for the small members of the family, see
Figure~\ref{fig:vesta_back_albedo_hist}.  We have used the range from
$D=1.5$ to $8$ km, thus excluding (4) and the interlopers (556) and
(1145) as well as another member marginally larger than $D=8$. The
differential power law is $1/D^{4.5}$, that is the cumulative is
$1/D^{3.5}$; this value also suggests that collisional equilibrium has
not been reached, the family appears somewhat less steep, which should
mean it is older, in agreement with the estimates from the previous
subsection.  Because of the larger span in $D$ the figure shows,
besides the incompleteness of family members with diameter $D<2$ km,
an additional phenomenon, namely a tendency to decrease the number of
members with respect to the power law at comparatively large diameters
$D>5$. However, a complete interpretation of the size distribution
should not be attempted without taking into account the results of
Section~\ref{sec:vesta}, proposing a complex structure for the family.

The fact that the size distribution of families formed in
cratering events usually exhibits such kind of concavity, is in
agreement with available fragmentation models \cite{Tangaetal99,
Durdaetal2007}. The concavity in the size distribution tends to
disappear when the parent body to largest member size ratio
decreases, and this can be qualitatively explained in terms of
available volume for the production of larger fragments
\cite{Tangaetal99}.

%These two features of the size distribution can be understood as
%physical constraints which do not allow the size distribution to be
%a power law over a very wide range of $D$. The constraint on the
%large $D$ side is due to the fact that a cratering event is unlikely
%to generate fragments with diameter larger than the depth of the
%crater. This is not likely to be a sharp transition, but a smooth
%one, by which the slope becomes gradually steeper as the critical
%value $D=$ depth is approached.

Another constraint on the low $D$ side comes from the fact
that a slope larger than $4$ for the differential size distribution
corresponds to an infinite total volume of all the fragments of
diameter in the interval $D_{max}>D>0$. This implies that, even if
there was no observational bias, the slope must decrease below some
critical value of $D$. The problem is, we do not know what this
critical size is; thus we cannot discriminate between observational
bias and possible detection of a real change in slope.

\section{Refinement with physical data}
\label{sec:use_physical}

The data from physical observations of asteroids, especially if they
are available in large and consistent catalogs like WISE and SDSS,
are very useful to solve some of the problems left open by purely
dynamical classifications such as the one discussed in the previous
sections. This happens when there are enough consistent and quality
controlled data, and when the albedo and/or colors can
discriminate, either between subsets inside a family, or between a
family and the local background, or between nearby families. 
(See also examples in Section~\ref{sec:craters}.)

\subsection{The Hertha--Polana--Burdett complex family}

The most illustrative example of discrimination inside a dynamical
family is the case of the family with lowest numbered asteroid (135)
Hertha; when defined by purely dynamical arguments, it is the largest
family with $11\,428$ members. Its shape is very regular in the $a,sin
I$ proper element projection, but has a peculiar $>$-shape in the
$a,e$ projection (Figure~\ref{fig:hertha_ae_wise}), which has been
strongly enhanced by the addition of the smaller asteroids of the halo
(in yellow).

\begin{figure}[h!]
  \figfig{12cm}{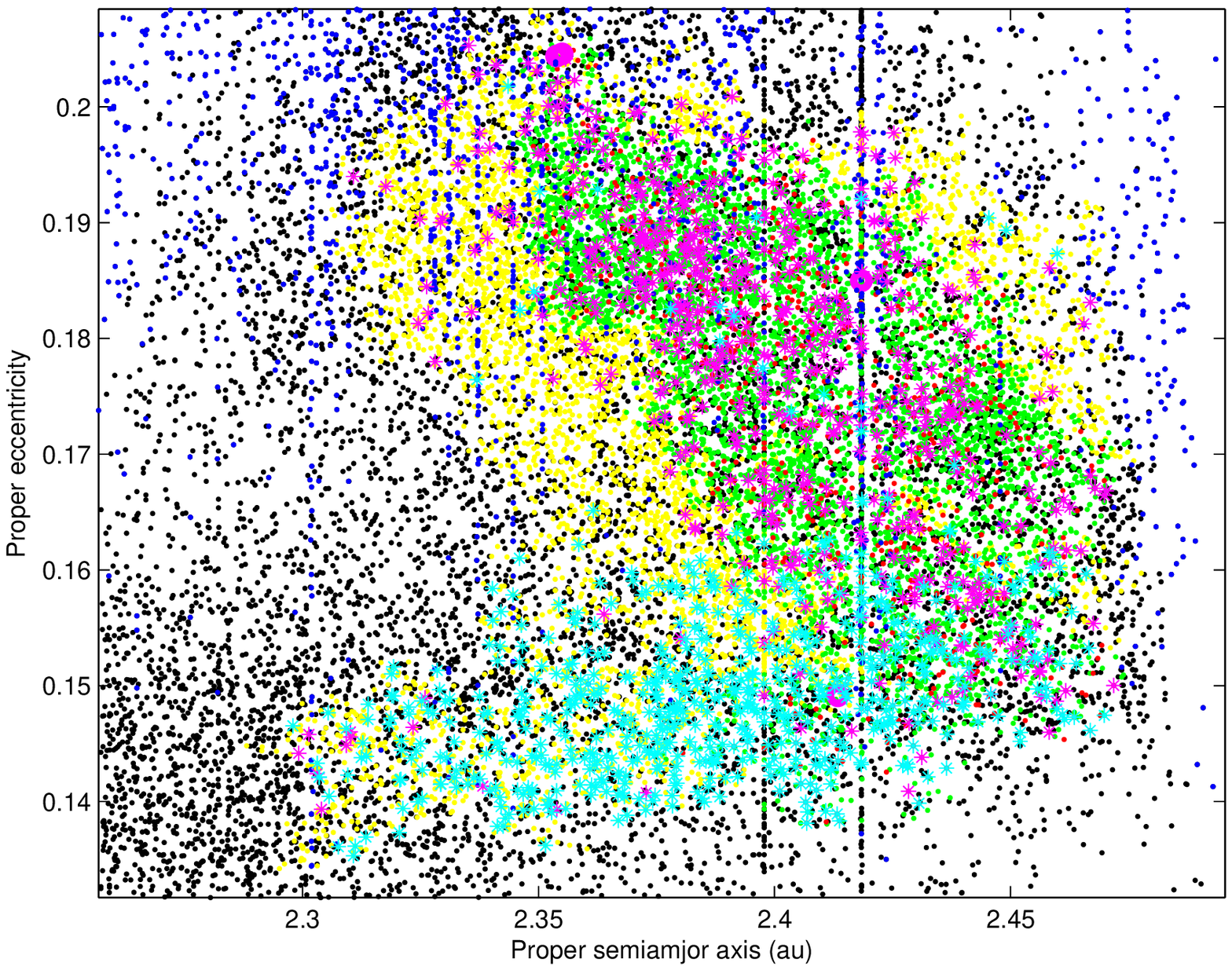}{The Hertha dynamical family in proper
    $a, e$ plane. Bright objects (magenta stars) and dark
    objects (cyan stars) forming a characteristic $>$-shape indicate 
    two partly overlapping collisional families. }
\end{figure}

Already by using absolute magnitude information some suspicion
arises from the V-shape plot, from which it appears possible to
derive a consistent ``slope'' neither from the inner nor for the
outer edge. Problems are already well known to arise in this family
from the very top, that is (142) Polana which is dark (WISE albedo
$0.045$) and diameter  $D\simeq 60$ km, and (135) Hertha which is of
intermediate albedo $0.152$ and $D\simeq 80$ km, also known to be an
M type asteroid, but exhibiting the 3 $\mu$m spectral feature of
hydrated silicates \cite{Rivkinetal2000}.

\begin{figure}[h!]
  \figfig{12cm}{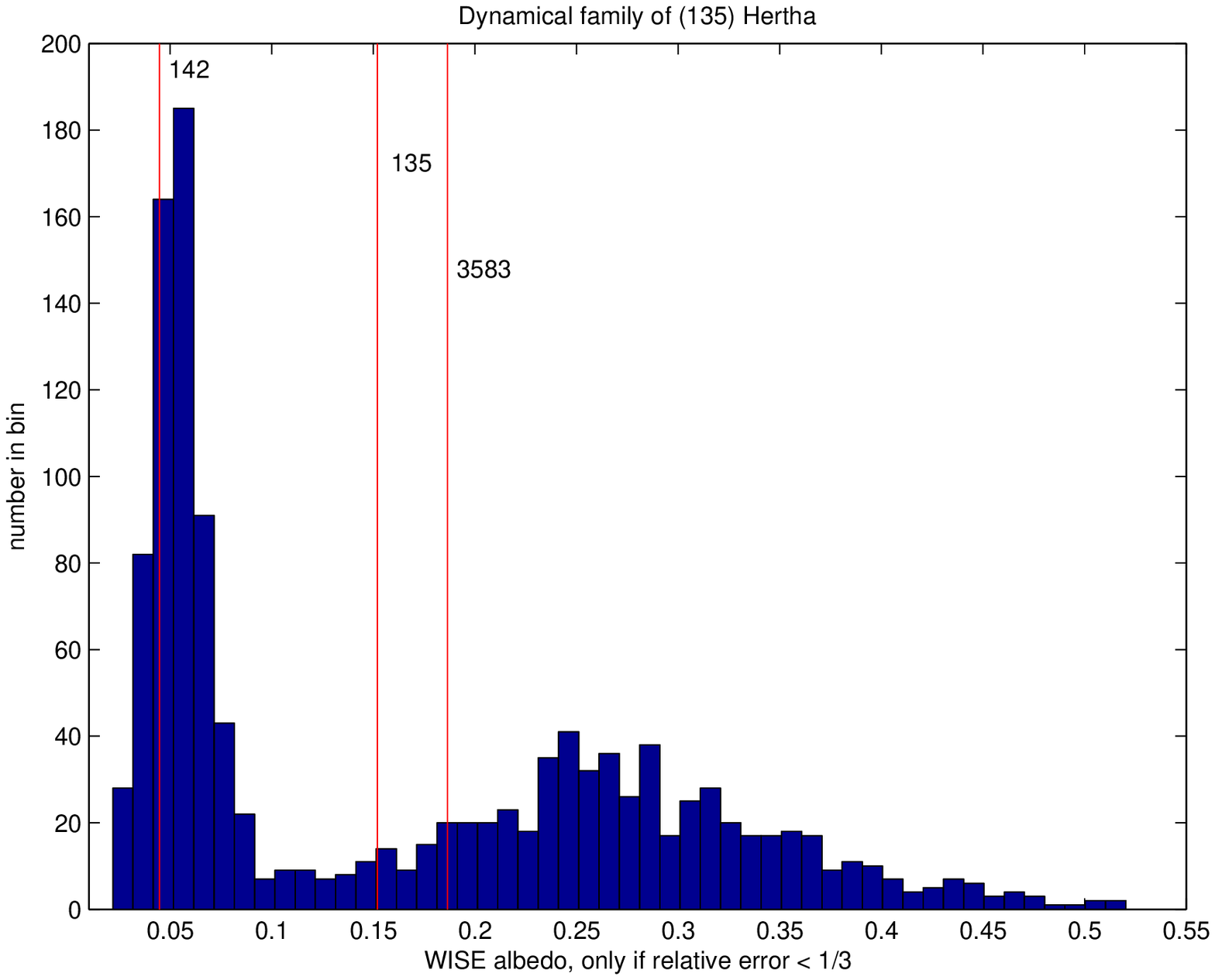}{The distribution of WISE albedos for
    the 135 dynamical family with the locations of the three namesakes
    indicated by red lines. The distribution is clearly bimodal
    supporting the scenario with two collisional families.}
\end{figure}

By using systematically the WISE albedo, limited to the asteroids for
which the albedo uncertainty is less than $1/3$ of the nominal value
($1\,247$ such data points in the 135 dynamical family), we find the
sharply bimodal distribution of Figure~\ref{fig:hertha_hist_alb}.
(142) Polana is by far the largest of the ``dark'' population (for the
purpose of this discussion defined as albedo $<0.09$, $611$ asteroids)
as well as the lowest numbered. The ``bright'' population (albedo
$>0.16$, $568$ asteroids) does not have a dominant large member, the
largest being (3583) Burdett (albedo $0.186\pm 0.02$, $D\simeq 7.6$
km)\footnote{The asteroid (878) Mildred was previously cited as
  namesake of a family in the same region: Mildred is very likely to
  be ``bright'', although the WISE data are not conclusive (albedo
  $=0.40\pm 0.22$), but is very small ($D\simeq 2.5$ km). The fact
  that (878) was imprudently numbered in 1926 after the discovery and
  then lost is a curious historical accident explaining a low numbered
  asteroid which is anomalously small. Thus we are going to use
  Burdett as the namesake of the ``bright'' component.}

In Figure ~\ref{fig:hertha_ae_wise} we have plotted with magenta stars
the ``bright'', with cyan stars the ``dark'', and it is clear that
they are distributed in such a way that the $>$-shape in the proper
$a,e$ plane can be explained by the presence of two separate
collisional families, the Polana family and the Burdett family, with a
significant overlap in the high $a$, low $e$ portion of the Hertha
dynamical family. Because the WISE dataset is smaller than the proper
elements dataset, we cannot split the list of members of the 135
dynamical family into Polana and Burdett, because such a list would
contain an overwhelming majority of ``don't know''. Erosion of the
original clouds of fragments by the $3/1$ resonance with Jupiter must
have been considerable, thus we can see only a portion of each of the
two clouds of fragments. Based on the total volume of the objects for
which there are good albedo data, the parent body of Polana must
have had $D> 76$ km, the one of Burdett $D> 30$ km.

Note that we could get the same conclusion by using the $a^*$
parameter of the SDSS survey: among the $1\,019$ asteroids in the
135 dynamical family with SDSS colors and $a^*$ uncertainty less
than $1/3$ of the nominal value, $184$ have $-0.3<a^*<-0.05$ and
$835$ have $+0.05< a^*< 0.3$, thus there is also a bimodal
distribution, which corresponds to the same two regions marked in
magenta and cyan in Figure~\ref{fig:hertha_ae_wise}, with negative
$a^*$ corresponding to low albedo and positive $a^*$ corresponding
to high albedo, as expected \cite[Figure 3]{parker2008}. The lower
fraction of ``dark'' contained in the SDSS catalog, with respect to
the WISE catalog, is an observation selection effect: dark objects
are less observable in visible light but well observable in the
infrared.

%\textbf{[Lists of Polana and Burdett members have been
%  prepared. Should we display these with a separate caption
%  ``collisional family''? What of (135) Hertha? Anyway we must wait
%  for the publication of the paper.]}

Because of its very different composition (135) Hertha can be presumed
to belong to neither the one nor the other collisional family,
although strictly speaking this conclusion cannot be proven from the
data we are using, listed in Section 2, but requires some additional
information (e.g., taxonomic classification of Hertha) and suitable
modeling (e.g., excluding that a metallic asteroid can be the core of
a parent body with ordinary chondritic mantle). All these conclusions
are a confirmation, based on a statistically very significant
information, of the results obtained by \cite{cellino2001} on the
basis of a much more limited dataset (spectra of just 20
asteroids). Other authors, such as \cite{masiero2013}, have first
split the asteroids by albedo then formed families by proper elements,
and they get the same conclusion on two overlapping families, but the
total number of family members is lower by a factor $\sim 3$.

\subsection{The Eos family boundaries}

Unfortunately, in many other cases in which the dynamical
classification raises doubts on the family structure the physical
observations databases are not sufficient to solve the problem. An
example is the family of (221) Eos, which on the basis of the proper
elements only cannot be neatly separated from the two smaller
families of (507) Laodica and (31811) 1999 NA41 (there are a few
members in common), as discussed in Section~\ref{sec:result_dyn}.
The Eos family mostly contains intermediate albedo asteroids,
including (221) Eos (WISE albedo $=0.165\pm 0.034$)
belonging to the unusual K taxonomic class, but the
surrounding region predominantly contains low albedo objects,
including (31811) (albedo $0.063\pm 0.014$) and the majority of the
members of the 507 family for which WISE data are available; (507)
Laodica itself has an albedo $=0.133\pm 0.009$ which is compatible
with the Eos family. From this we are able to give the negative
conclusion: attaching families 507 and 31811 to 221 would not be
supported by physical observations, but leaving them separate is not
much better.

\subsection{Watsonia and the Barbarians}

The family of (729) Watsonia had been already identified in the past
by \cite{bojan_highi}, who adopted a proper element data base
including also a significant number of still unnumbered,
high-inclination asteroids, not considered in our present analysis.
This family is interesting because it includes objects called
``Barbarians'', see \cite{Barbara}, which are known to exhibit unusual
polarization properties. Two of us (AC and BN) have recently obtained
VLT polarimetric observations \cite{barbarians} showing that
seven out of nine observed members of the Watsonia family exhibit the
Barbarian behavior. This result strongly confirms a common origin of
the members of the Watsonia family. On the other hand, the presence of
another large (around 100 km in size) Barbarian, (387) Aquitania,
which has proper semi-major axis and inclination within the limits of
the Watsonia family, but shows a difference in proper eccentricity of
about $0.1$, exceedingly large to include it in the family, indicates
that the situation can be fairly complex and opens interpretation
problems, including a variety of possible scenarios which are beyond
the scopes of the present paper.

%If it was possible to confirm, with high statistical confidence,
%that all members  of the family 729 share this unusual observable
%property, then it would become possible to follow the asteroids as
%they are spread even outside the recognized family by Yarkovsky
%effect.

\section{Cratering families}
\label{sec:craters}

As a result of the availability of accurate proper elements for
smaller asteroids, our classification contains a large
fraction of families formed by cratering events.

Modeling of the formation of cratering families needs to take into
account the escape velocity $v_e$ from the parent body, which results
in the parent body not being at the center of the family as seen in
proper elements space. This is due to the fact that fragments which do
not fall back on the parent body need to have an initial relative
velocity $v_0>v_e$, and because of the formula giving the final
relative velocity $v_\infty=\sqrt{v_0^2-v_e^2}$ the values of
$v_\infty$ have a wide distribution even for a distribution of $v_0$
peaking just above $v_e$. The mean value of $v_\infty$ is expected to
be smaller than $v_e$, at most of the same order. Thus immediately
after the cratering event, the family appears in the proper elements
space as a region similar to an ellipsoid, which is centered at a
distance $d$ of the order of $v_e$ from the parent body. Of course
this effect is most significant for the very largest parent bodies.

Moreover, it is important not to forget that cratering events
typically occur multiple times over the age of the solar system,
since the target keeps the same impact cross section. The outcomes
can appear either as separate dynamical families or as structures
inside a single one.

We use as criterion for identification of a cratering family that the
fragments should add up to $\leq 10\%$ of the parent body volume; we
have tested only the large and medium families, and used the common
albedo hypothesis to compare volumes. In this way 12 cratering
families have been identified with the asteroids (2), (3), (4), (5),
(10), (15), (20), (31), (87), (96), (110), (179), and (283) as parent
bodies. Other large asteroids do not appear to belong to families.  We
will discuss some interesting examples.

% missing among H<7: 1,6,7,8,9,11, 13, 14, 16, 18, 22, 27, 29, 39, 52, 57, 65,68, 89,
%  92,130, 196, 324, 349, 354, 451, 471, 511, 532, 704 (the 5th largest after 1,2,4,10)
% altri grossi 24 frag; 63 cratering but attached to Vesta

\subsection{The Massalia family}
\label{sec:massalia}

Although the V-shape plot (Figure~\ref{fig:20_vshapea}) does not
suggest any internal structure for the family 20, the inspection of
the shape of the family in the space of all three proper elements
suggests otherwise.

The distribution of semimajor axes is roughly symmetrical with respect
to (20) Massalia, while these of eccentricity and inclination are, on
the contrary, rather asymmetrical. The eccentricity distribution is skewed
towards higher eccentricities (third moment positive), this is apparent
from Figure~\ref{fig:massalia_ae} as a decrease of number density for
$e<0.157$; the inclination one is skewed towards lower inclinations
(third moment negative).

Thus the barycenter of the ejected objects appears quite close to the
parent body (20), see Table~\ref{tab:tablebar_crat}: if the
differences in $e, \sin{I}$ are scaled by the orbital velocity they
correspond to about $7$ m/s, which is much smaller than the escape
velocity. Even if the distributions are skewed in number density,
fragments appear to have been launched in all directions, and this is
not possible for a single cratering event.

These arguments lead us to suspect a multiple collision origin of the
dynamical family. At $e<0.157$ there seems to be a  portion of a
family with less members which does not overlap the other, more dense
collisional family. The more dense family has been ejected in a
direction such that $e$ increases and $\sin{I}$ decreases, the other
in a direction with roughly opposite effect. 

However, the presence of the low $e$ subfamily does not affect the age
computation, which only applies to the high $e$ subfamily, due to the
fact that the extreme values of $a$ are reached in the high $e$
region. Thus there are two concordant values for the slopes on the two
sides, and a single value of the age we can compute, which refers only
to the larger, high $e$ subfamily.

\subsection{The Vesta family substructures}
\label{sec:vesta}

\begin{figure}[h!]
  \figfig{12cm}{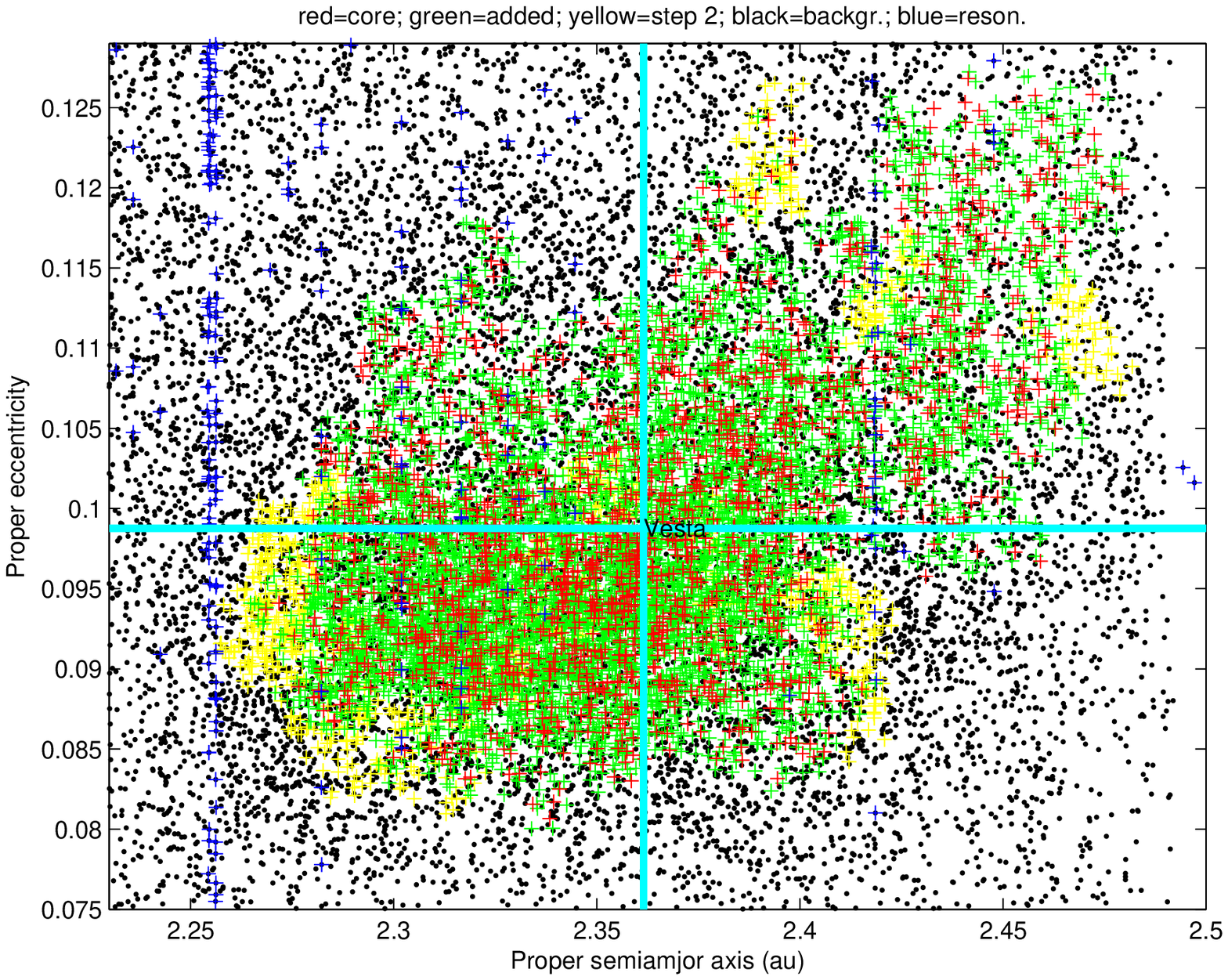}{The family 4 shown in the proper $a,e$
    plane. The halo families merged in step 5 of our classification
    procedure (yellow dots) extend the family closer to the $3/1$
    resonance with Jupiter on the right and to the $7/2$ resonance on
    the left. The position of (4) Vesta is indicated by the cyan
    cross, showing that the parent body is at the center of neither of
    the two concentrations of members, at lower $e$ and at higher
    $e$. This because of the strongly anisotropic distribution of
    velocities $v_\infty$ for a cratering event.}
\end{figure}

The Vesta family has a curious shape in the proper $a,e$ plane, see
Figure~\ref{fig:4_ae}, which is even more curious if we
consider the position of (4) Vesta in that plane.

In proper $a$, the family 4 is bound by the $3/1$ resonance with
Jupiter on the outside and by the $7/2$ inside. Closer inspection
reveals the role of another resonance at $a\simeq 2.417$ au, which is
the $1/2$ with Mars. Indeed, the low $e$ portion of the family has the
outside boundary at the $1/2$ resonance with Mars. By stressing the
position of Vesta, as we have done with the cyan cross in
Figure~\ref{fig:4_ae}, we can appreciate the existence of a
group of roughly oval shape with proper $e$ lower than, or only
slightly above, the one of Vesta (which is $0.099$). This can be
confirmed by a histogram of the proper $a$ for family 4 members,
showing for $2.3<a<2.395$ a denser core containing about $2/3$ of the
family members.  We can define a subgroup as the family 4 members with
$a<2.417$ and $e<0.102$, conditions satisfied by $5\,324$ members. We
shall call this group with the non-committing name of ``low $e$
subfamily''.

By assuming for sake of simplicity that albedo and density are the
same, we can compute the center of mass, which is located at
$a=2.3435$ and $e=0.0936$. To get to such values the relative velocity
components after escape from Vesta should have been $-76$ and $-98$
m/s, respectively\footnote{The negative sign indicates a direction
  opposite to the orbital velocity for $a$, and a direction, depending
  upon the true anomaly at the collision, resulting in decrease
  of $e$.}. Since the escape velocity from Vesta surface is $\sim 363$
m/s, this is compatible with the formation of the low $e$ subfamily
from a single cratering event, followed by a Yarkovsky evolution
significant for all, since no member has $D>8$ km.

What is then the interpretation of the rest of the family 4? We shall
call ``high $e$ subfamily'' all the members not belonging to the low
$e$ portion defined above, excluding also asteroids (556) and (1145)
which have been found to be interlopers. This leaves $2\,538$ members,
again with size $D<8$ km.  It is also possible to compute a center of
mass: the necessary relative velocities after escape are larger by a
factor $\sim 2$, still comparable to the escape velocity from Vesta,
although this estimate is contaminated by the possible inclusion of
low $e$, low $a$ members into the low $e$ subfamily. Anyway, the shape
of this subfamily is not as simple as the other one, thus there could
have been multiple cratering events to generate it.

This decomposition provides an interpretation of the results from
Section~\ref{sec:ages}, in which there was a large discrepancy, by a
factor $\sim 2$, between the age as computed from the low $a$ side and
from the high $a$ side of the V-shape in $a, 1/D$. Indeed, if the low
$e$ subfamily ends for $a<2.417$, while the high $e$ subfamily ends at
$a\sim 2.482$, then the right side of the V-shape belongs to the high
$e$ subfamily. From Figure~\ref{fig:4_ae} we see that the
low $a$ side of the family appears to be dominated by the low $e$
subfamily.

As a consequence of this model, the two discordant ages computed in
Section~\ref{sec:ages} belong to two different cratering events.  This
interpretation is consistent with the expectation that cratering
events, large enough to generate an observable family, occur multiple
times on the same target.

As for the uncertainties of these ages, they are dominated by the poor
a priori knowledge of the Yarkovsky calibration constant $c$ for the
Vesta family. Still the conclusion that the two ages should be
different by a factor $\sim 2$ appears robust. From the DAWN images,
the age of the crater Rheasilvia on Vesta has been estimated at about
$1$ Gy \cite{dawn_marchi}, while the underling crater Veneneia must be
older, its age being weakly constrained. Thus both the younger age and
the ratio of the ages we have estimated in Section~\ref{sec:ages} are
compatible with the hypothesis that the low $e$ subfamily corresponds
to Rheasilvia, the high $e$ subfamily (or at least most of it)
corresponds to Veneneia. We are not claiming we have a proof of this
identification.

Unfortunately, for now there are no data to disentangle the portions
of the two collisional families which overlap in the proper elements
space.  Thus we can compute only with low accuracy the barycenter of the two
separate collisional families, and to model the initial distributions
of velocities would be too difficult. However, there are some
indications \cite{bus_vesta} that discrimination of the two
subfamilies by physical observations may be possible.

In conclusion, the current family of Vesta has to be the outcome of
at least two, and possibly more, cratering events on Vesta,
not including even older events which should not have left
visible remnants in the family region as we see it today.

\subsection{Vesta Interlopers and lost Vestoids}
\label{sec:lost}

\begin{figure}[h!]
  \figfig{12cm}{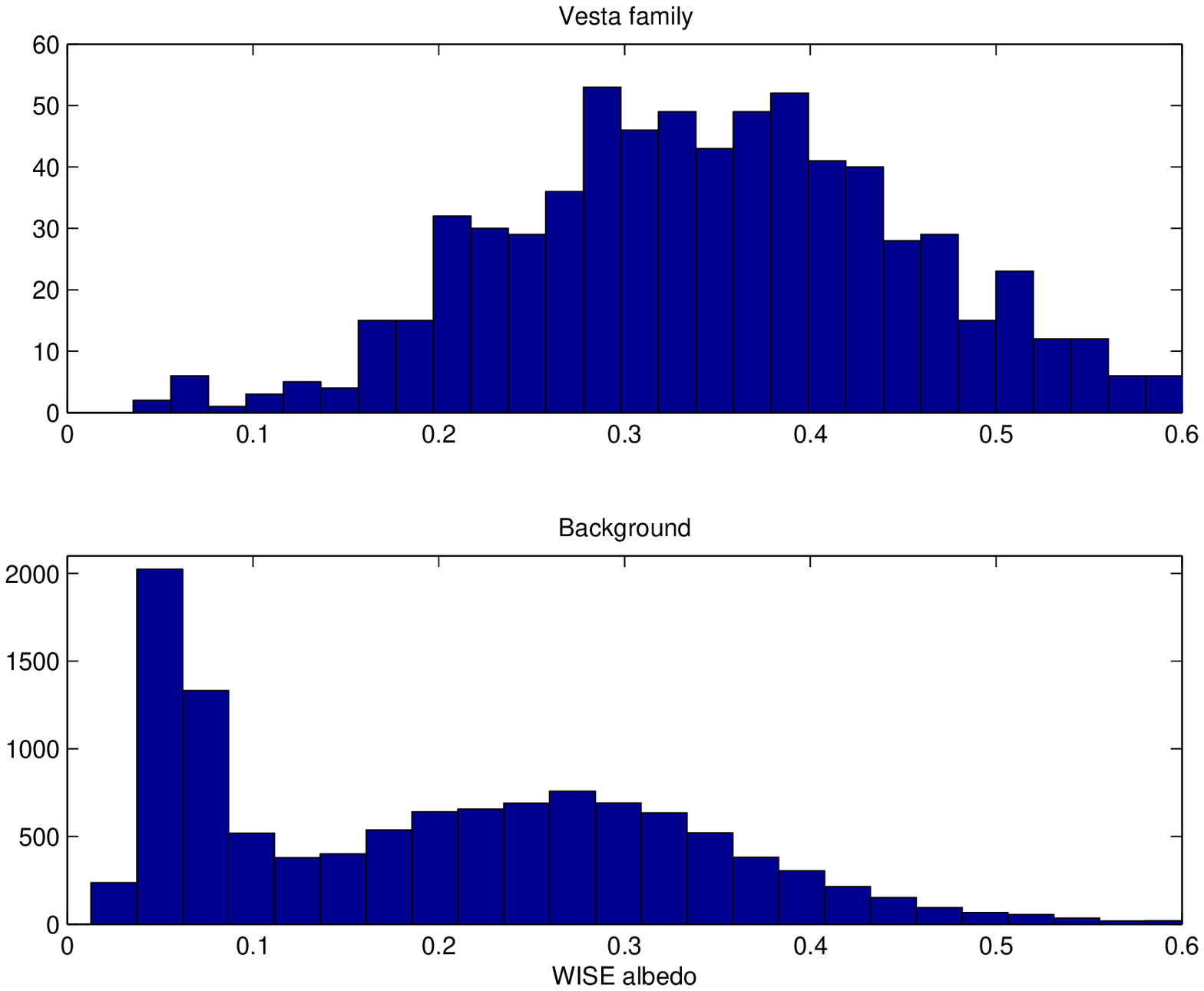}{Histogram of albedo measured
    by WISE with $S/N>3$: above for the asteroids belonging to family
    4; below for the background asteroids with $2.2 <$ proper $a<2.5$
    au. The uneven distribution of ``dark'' asteroids (albedo $<0.1$)
    is apparent. The asteroids with $0.27 <$ albedo $<0.45$,
    corresponding to the bulk of the family 4, are present, but as a
    smaller fraction, in the background population.}
\end{figure}

Another possible procedure of family analysis is to find interlopers,
that is asteroids classified as members of a dynamical family, not
belonging to the same collisional family, because of discordant
physical properties; see as an example of this procedure Figure 25 and
Table 3 of \cite{hungaria}.

In the dynamical family of (4) Vesta there are $695$ asteroids with
reasonable (as afore) WISE albedo data.  We find the following 10
asteroids with albedo $<0.1$: (556), (11056), (12691), (13411),
(13109), (17703), (92804), (96672), (247818), (253684); the first is
too large ($D\simeq 41$ km) and was already excluded, the next 3 are
larger than $7.5$ km, that is marginally too large for typical
Vestoids; we had also excluded in Section~\ref{sec:volume}
(1145), which has an intermediate albedo but is also too large
($D\simeq 23$ km). We think these $11$ are reliably identified as
interlopers, of which $10$ belong to the C-complex. By scaling to the
total number $7\,865$ of dynamic family members, we would expect a
total number of interlopers belonging to the C-complex $\simeq 120$.

The problem is how to identify the interlopers belonging to the
S-complex, which would be expected to be more numerous. For this task
the WISE albedo data are not enough, as shown by
Figure~\ref{fig:vesta_back_albedo_hist}.  The albedos of most family
members are in the range between $0.16$ and $0.5$, which overlaps the
expected for the S-complex, but there is no ostensible bimodality in
this range.

The background asteroids, with $2.2<a<2.5$, for which significant WISE
albedos are available, clearly have a dark component, $34\%$ of them
with albedo $<0.1$, but the majority have albedos compatible with the
S-complex, a large fraction also compatible with V-type.  

The estimated value of the albedo is derived from an assumed absolute
magnitude, which typically has an error of $0.3$ magnitudes (or
worse). This propagates to a relative error of $0.3$ in the albedo.
Thus the values of albedo for S and V type are mixed up as a result of
the measurement errors, both in the infrared and in the visible
photometry.
The only class of objects which are clearly identified from the albedo
data are the dark C-complex ones, because the main errors in the
albedo are relative ones, thus an albedo estimated at $<0.1$ cannot
correspond to an S-type, even less to a V-type. 

In conclusion, by using only the available albedo data there is no way
to count the interlopers in the Vesta family belonging to the
S-complex; it is also not possible to identify ``lost Vestoids'',
originated from Vesta but not included in the dynamical family.

\begin{figure}[h!]
  \figfig{12cm}{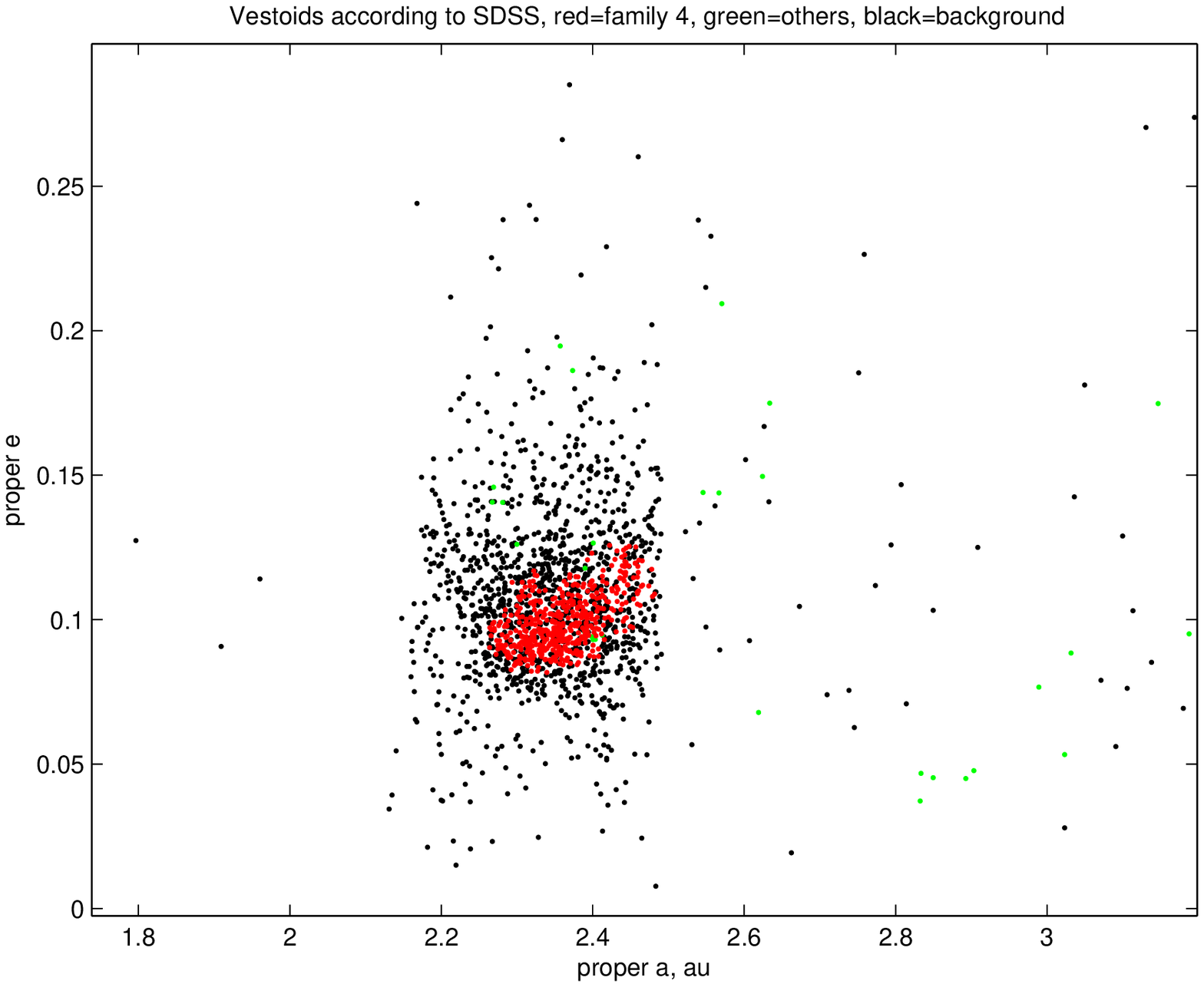}{Asteroids complying with the
    \cite{parker2008} criterion for V-type. Red points: members of
    family 4, green: members of other families, black: background
    asteroids. The background asteroids apparently matching the color
    properties of Vestoids are, among objects with significant SDSS
    data, at least twice more numerous than the family 4 members with
    the same colors.}
\end{figure}

The question arises whether it would be possible to use the SDSS data
to solve these two problems. According to \cite{parker2008} the V-type
objects should correspond to the region with $a^*>0$ and $i-z<-0.15$
in the plane of these two photometric parameters.  However, as it is
clear from \cite[Figure 3]{parker2008}, these lines are not sharp
boundaries, but just probabilistic ones. Thus this criterion is
suitable to reliably identify neither family 4 interlopers, nor lost Vestoids. 

On the other hand, the Parker et al. criterion can be used to estimate
the V-type population in a statistical sense.  To select the asteroids
which have a large probability of being V-type we require
$a^*-2\,STD(a^*)>0$ and $i-z +2\,STD(i-z)<-0.15$; we find $1\,758$
asteroids, of which $55$ with $a>2.5$ au; they are plotted on the
proper $a,e$ plane in Figure~\ref{fig:vesta_ae_SDSS}. The number of
asteroids of V-type beyond the $3/1$ resonance with Jupiter should be
very small, anyway $55$ is an upper bound on the number of false
positive for the V-type criterion in that region.

Of the V-type with $a<2.5$ au, $504$ are members of the dynamical
family 4 and $1\,199$ are not. In conclusion, even taking into account
the possible number of false positive, there are at least twice as
many V-types in the inner belt outside of the dynamical family rather
than inside. 

Conversely, if we define ``non-V type'' by the criterion either
$a^*+2\,STD(a^*)<0$ or $i-z -2\,STD(i-z)>-0.15$ we find in the inner
belt $a<2.5$ as many as $8\,558$ non-V, out of which only $42$ belong
to the dynamical family 4, which means the number of S-type
interlopers is too small to be accurately estimated, given the
possibility of ``false negative'' in the V-type test.

%We can estimate a volume of all these non-family
%vestoids, again assuming they have the same albedo as (4), at
%$11\,000$ km$^3$. This is of course a lower bound, but it is right at
%the order of magnitude level because the objects without good SDSS data
%tend to be smaller.

This gives an answer to another open question: where are the ``lost
Vestoids'', remnants of cratering events on Vesta which occurred
billions of years ago? The answer is everywhere, as shown by
Figure~\ref{fig:vesta_ae_SDSS}, although much more in the inner belt
than in the outer belt, because the $3/1$ barrier deflects most of the
Vestoids into planet crossing orbits, from which most end up in the
Sun, in impacts on the terrestrial planets, etc. Still there is no
portion of the asteroid main belt which cannot be reached, under the
effect of billions of years of Yarkovsky effect and chaotic diffusion
combined. We should not even try to find families composed with them,
because they are too widely dispersed. All but the last two (possibly
three) family-forming cratering events have completely disappeared
from the Vesta family.

\subsection{The Eunomia Family}
\label{sec:eunomia}

The number frequency distributions of the family members' proper
elements indicate that some multiple collisions interpretation is
plausible: the distribution of semimajor axes exhibits a gap around
$a=2.66$~au, close to where Eunomia itself is located\footnote{A
  narrow resonance occurs at about the location of the gap, but it
  does not appear strong enough to explain it.}. The distribution of
family members on all sides of the parent body for all three proper
elements, and the barycenter of the family (not including Eunomia)
very close to (15), are discordant with the supposed anisotropic
distribution of velocities of a single cratering event.

All these pieces of evidence indicate that a single collisional event
is not enough to explain the shape of the dynamical family 15.  Then
the discrepancy in the slopes on the two sides could be interpreted as
the presence of two collisional families with different ages.  Since
the subfamily with proper $a>2.66$ dominates the outer edge of the
V-shape, while the inner edge is made only from the rest of the
family, we could adopt the younger age as that of the high $a$
subfamily, the older as the age of the low $a$ subfamily. However, the
lower range of diameters, starting only from $D>6.7$ km on the outer
edge, and the ratio of ages too close to $1$ result in a difference of
ages which is poorly constrained.

Still the most likely interpretation is that the Eunomia dynamical
family was generated by two cratering events, with roughly opposite
crater radiants, such that one of the two collisional families has
barycenter at $a>a(15)$, the other at $a<a(15)$, see
Table~\ref{tab:tablebar_crat}.
The WISE albedo distributions of the two subfamilies are practically
the same, which helps in excluding more complex interpretations in
which one of the two subfamilies has a different parent body.
In conclusion, the interpretation we are proposing is similar to the
one of the Vesta family. 

%(704) Interamnia is the fifth largest asteroid after (1), (2), (4),
%(10). It has no family at all! sin(I)=0.322, very low density
%region. There is the $11/5$ resonance with Jupiter, reinforced by 4
%3-body resonances, at $a$ higher by 0.015 au. On the low a side there
%could be a very small group.]

\subsection{The missing Ceres family}
\label{sec:ceres}

(1) Ceres in our dynamical classification does not belong to any
family, still there could be a family originated from Ceres.
The escape velocity from Ceres is $v_e\sim 510$ m/s, while the QRL
velocity used to form families in zone 3 was $90$ m/s. An ejection
velocity $v_0$ just above $v_e$ would results in a velocity at
infinity larger than $90$ m/s: $v_0=518$ m/s is enough.

Thus every family moderately distant from Ceres, such that the
relative velocity needed to change the proper elements is $<v_e$, is a
candidate for a family from Ceres. Family 93 is one such
candidate\footnote{Other authors have proposed family classifications
  in which (1) is a member of a family largely overlapping with our
  family 93.}.  By computing the distance $d$ between the proper
element set of (1) and all the family 93 members, we find the minimum
possible $d=153$ m/s for the distance (in terms of the standard
metric) between (1) and (28911). Although the relationship between $d$
and $v_\infty$ is not a simple one (depending upon the true anomaly at
the impact), anyway $v_\infty$ would be of the same order as $d$, thus
corresponding approximately to $v_0=532$ m/s.

\begin{figure}[h!]
  \figfig{11cm}{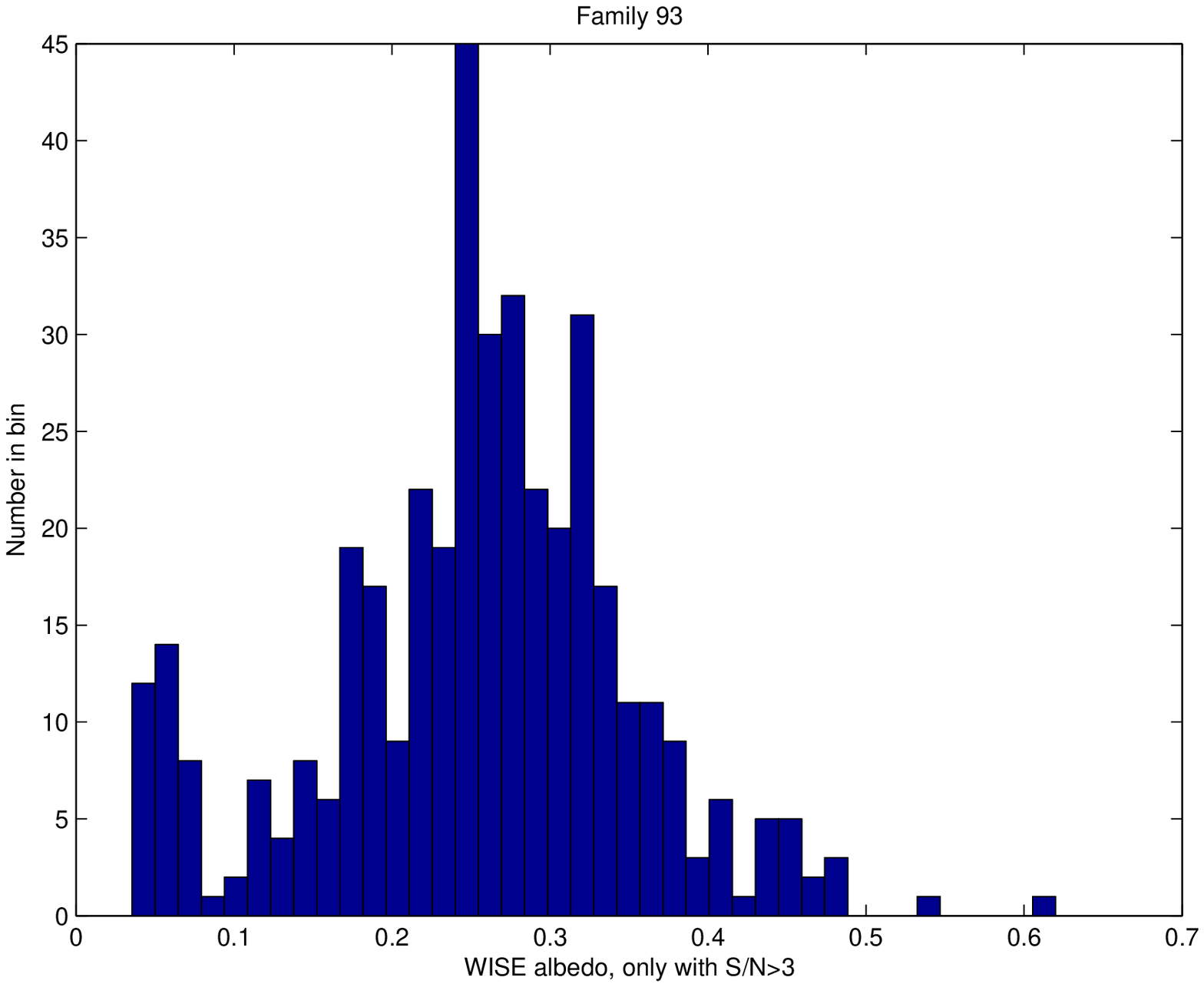}{Histogram of the albedos measured by
    WISE with $S/N>3$ among the members of the family 93. There is an
    obvious ``dark'' subgroup with albedo $<0.1$ and a large spread of
    higher estimated albedos.  Most members have intermediate albedos
    typical of the S-complex.}
\end{figure}

This is a hypothesis, for which we seek confirmation by using absolute
magnitudes and other physical observations, and here comes the
problem.

The albedo of (1) is $0.090\pm 0.0033$ according to \cite{Lietal2006}.
The surface of Ceres has albedo inhomogenities, but according to the
HST data reported by \cite{Lietal2006} the differences do not exceed
$8\%$ of the value.

The WISE albedos of the family 93 (again accepting only the 403 data
with $S/N>3$) are much brighter than that of Ceres, apart from a small
minority: only 37, that is $9\%$, have albedo $<0.1$.  (93) has albedo
$0.073$ from IRAS, but we see no way to eject a $D \sim 150$ km
asteroid in one piece from a crater; also (255) belongs to the dark
minority, and is too large for a crater ejecta.  No other family
member, for which there are good WISE data, has diameter $D \geq 20$
km. Actually, from Figure~\ref{fig:minerva_albhist} the albedo of
Ceres is a minimum in the histogram. By using the SDSS data we also
get a large majority of family 93 members in the S-complex region.

\begin{figure}[h!]
  \figfig{11cm}{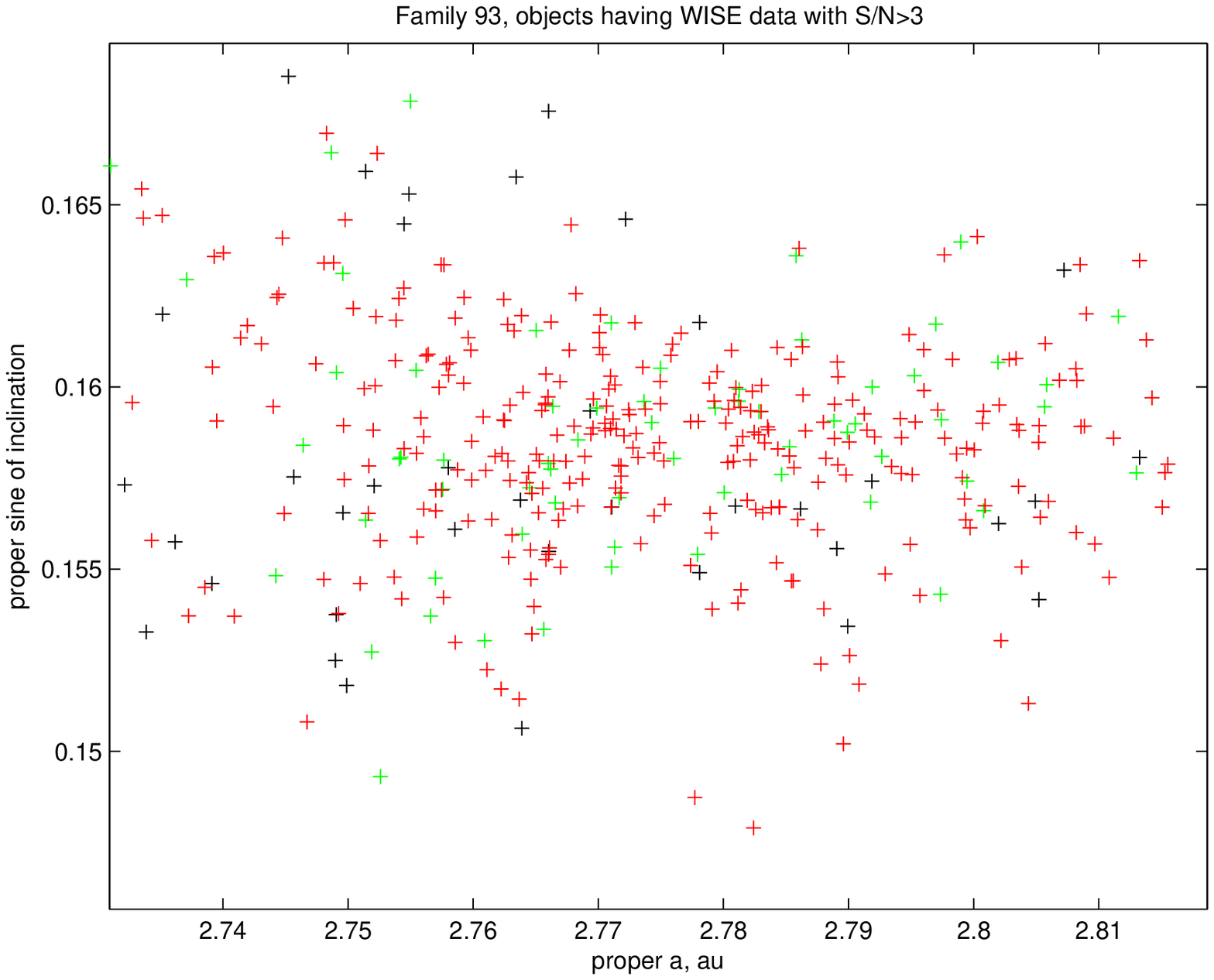}{The members of the dynamical family 93 for
    which significant WISE data are available, plotted in the proper
    $a, \sin{I}$ plane. Red= albedo $>0.2$, green=albedo between $0.1$
    and $0.2$, black=albedo $<0.1$.}
\end{figure}

We cannot use V-shape diagrams composed with the same method used in
Section~\ref{sec:ages}, because the assumption of uniform albedo is
completely wrong, as shown by Figure~\ref{fig:minerva_albhist}. As an
alternative, to study possible internal structures we use only the 403
objects with good WISE data, and use a color code to distinguish low
albedo, high albedo and intermediate. The Figure~\ref{fig:93_wise_aI}
in the proper $a,\sin{I}$ plane and the one in the proper $a,e$ plane
show no concentration of objects with low, not even with intermediate,
albedo. Thus there appears to be no family originated from Ceres, but
only a family of bright and (possibly) intermediate asteroids, having to
do neither with (1), nor with (93), nor with (255).

Thus the family 93 is the only one suitable, for its position in
proper elements space, to be a cratering family from Ceres, after
removing the two large outliers. However, physical observations
(albedo and colors) contradict this origin for an overwhelming
majority of family members.  Should we accept the idea that the
bright/intermediate component of family 93 is the result of
a catastrophic fragmentation of some S-complex asteroid, and the fact of
being very near Ceres is just a coincidence?

Moreover, why Ceres would not have any family? This could not be
explained by assuming that Ceres has been only bombarded by much
smaller projectiles than those impacting on (2), (4), (5) and (10):
Ceres has a cross section comparable to the sum of all these four
others. 

"How often have I said to you that when you have eliminated the
impossible, whatever remains, however improbable, must be the
truth?"\footnote{Sherlock Holmes to dr. Watson, in: C. Doyle,
  \emph{The Sign of the Four}, 1890, page 111.}.

Following the advise above, we cannot discard the ``coincidence'' that
a family of bright asteroids is near Ceres, but then all the dark
family members have to be interlopers. 
%\textbf{[$9\%$ of interlopers is not too much? Can someone find a way
%  to compute an expected number of interlopers, to confirm this?]}
This means we can assign $366$ ``bright and intermediate'' (with
albedo $>0.1$ measured by WISE with $S/N>3$) members of family 93 to
the Gerantina collisional family, from the lowest numbered member,
(1433) Gerantina, which has albedo $0.191\pm 0.017$, from which
$D\simeq 15$. The volume of these $366$ is estimated at $49\, 000$
km$^3$, equivalent to a sphere with $D\simeq 45$ km. This can be
interpreted as a catastrophic fragmentation of some S-complex asteroid
with a diameter $D>50$ km, given that a good fraction of the fragments
has disappeared in the $5/2$ resonance.

As for the missing family from Ceres, before attempting an explanation
let's find out how significant are our family classification data,
that is whether our classification could just have missed a Ceres
family. 

This leads to a question, to which we can give only a low accuracy
answer: which is the largest family from Ceres which could have
escaped our classification? This question has to be answered in two
steps: first, how large could be a family resulting from cratering on
Ceres which could be superimposed to the family 93?
$37$ dark interlopers out of $403$ with good WISE data could be
roughly extrapolated to $168$ out of $1\,833$ members of family
93\footnote{This is an upper bound, since some dark interlopers from the
background are expected to be there: in the range of proper semimajor
axis $2.66< a <2.86$ au we find that $59\%$ of background asteroids
have albedo $<0.1$.}.

Second, how large could be a Ceres family separate from 93 and not
detected by our classification procedure? In the low proper $I$ zone
3, the three smallest families in our classification have $93$, $92$
and $75$ members. Although we do not have a formal derivation of this
limit, we think that a family with about $100$ members could not be
missed. 
%\textbf{[Comment on the possibility of getting another step 3
%  family if we were to increase the QRL from 40 to 50-60 m/s]}

By combining these two answers, families from Ceres cratering with up
to about $100$ members could have been missed. The comparison with the
family of (10) Hygiea, with $2\,402$ members, and the two subfamilies
of Vesta, the smallest with $>2,538$ members, suggests that the loss
in efficiency in the generation of family members in the specific case
of Ceres is at least by a factor $20$, possibly much more.

Thus as an explanation for the missing Ceres family, the only
possibility would be that there is some physical mechanism preventing
most fragments which would be observable, currently those with $D>2$
km, either from being created by a craterization, or from being
ejected with $v_0>510$ m/s, or from surviving as asteroids.  We are
aware of only two possible models.

One model can be found in \cite[section 6.3]{Lietal2006}: ``The lack
of a dynamical family of small asteroids associated with Ceres, unlike
Vesta's Vestoids, is consistent with an icy crust that would not
produce such a family.'' We have some doubts on the definition of an
``icy'' crust with albedo $<0.1$, thus we generalize the model from
\cite{Lietal2006} by assuming a comparatively thin crust effectively
shielding the mantle volatiles, with whatever composition is necessary
to achieve the measured albedo. When Ceres is impacted by a large
asteroid, $20<D<50$ km, the crust is pierced and a much deeper crater
is excavated in an icy mantle. Thus the efficiency in the generation
of family asteroids (with albedo similar to the crust) is decreased by
a factor close to the ratio between average crater depth and crust
thickness. The ejected material from the mantle forms a family of main
belt comets, if they have enough water content they quickly dissipate
by sublimation and splitting.  This would be one of the most
spectacular events of solar system history; unfortunately, it is such
a rare event that we have no significant chance of seeing
it\footnote{The exceptional presence of some $10\,000$ comets could
  perhaps be detected in an extrasolar planetary system, but since we
  know only of the order of $1\,000$ extrasolar systems, also this is
  a low probability event.}. Thus a crust thickness of few km would be
enough to explain a loss of efficiency by a factor $>20$ and the
possible Ceres family could be too small to be found.

A second possible model is that there is a critical value for the
ejection velocity $v_0$ beyond which asteroids with $D>2$ km cannot be
launched without going into pieces. If this critical velocity $v_c$ is
$>363$ m/s for V-type asteroids, but is $<510$ m/s for the composition
of Ceres ejecta (presumably with much lower material strength), then
Vesta can have a family but Ceres can not. Of course it is also
possible that the number of large ejecta with $v_0>510$ is not zero
but very small. Thus even a very large impact on Ceres does generate
few observable objects, it does not matter whether they are asteroids
or comets, leading to a family too small to be detected.  However, if
the crater is deeper than the crust, Ceres itself behaves like a main
belt comet for some time, until the crater is ``plugged'' by
dirt thick enough to stop sublimation. This would be spectacular too.

The fact is, little is known of the composition and geological
structure of Ceres. This situation is going to change abruptly in
2015, with the visit by the DAWN spacecraft. Then what these two
models would predict for the DAWN data? 

With both models, larger impacts would leave only a scar, resulting
from the plugging of the mantle portion of the crater (because Ceres
does not have large active spots).  What such a big scar looks like is
difficult to be predicted, they could be shallow if the mantle
restores the equilibrium shape, but still be observable as
albedo/color variations.  
If there is a thin crust, then craters should have low maximum depth
and a moderate maximum diameter. At larger diameters only scars would
be seen.
If the family generation is limited by the maximum ejection velocity,
the crust could be thicker: there would be anyway craters and scars,
but the craters can be larger and the scars would be left only by the
very large impact basins.

\section{Binaries and couples}
\label{binaries}

\begin{table}[h]
%\footnotesize
 \centering
  \caption{Binary (or multiple) systems belonging to some family identified in the
  present paper.
  Confirmed binary systems are listed in bold.}
  \label{tab:binaries}
\medskip
  \begin{tabular}{rlr}
  \hline
\multicolumn{2}{c}{Binary asteroid} & Family\\
\hline
\\
\textbf{(87)} & \textbf{Sylvia} & \textbf{87} \\
\textbf{(90)} & \textbf{Antiope} & \textbf{24} \\
\textbf{(93)} & \textbf{Minerva} & \textbf{93} \\
\textbf{(243)} & \textbf{Ida} & \textbf{158} \\
\textbf{(283)} & \textbf{Emma} & \textbf{283} \\
\textbf{(379)} & \textbf{Huenna} & \textbf{24} \\
\textbf{(1338)} & \textbf{Duponta} & \textbf{1338} \\
(3703) & Volkonskaya & 4 \\
(3782) & Celle & 4 \\
\textbf{(5477)} & \textbf{Holmes} & \textbf{434} \\
(5481) & Kiuchi & 4 \\
(10208) & Germanicus & 883 \\
(11264) & Claudiomaccone & 5 \\
(15268) & Wendelinefroger & 135 \\
\textbf{(17246)} & \textbf{2000 GL74} & \textbf{158} \\
\textbf{(22899)} & \textbf{1999 TO14} & \textbf{158} \\
\textbf{(76818)} & \textbf{2000 RG79} & \textbf{434} \\
\\
\hline
\end{tabular}
\end{table}

Having defined a list of asteroid families, an interesting
investigation is to look at the list of currently known binary or
multiple asteroid systems, to see whether these systems tend to be
frequent among family members. This because a possible origin of
binary systems is related to collisional events. This is true
especially for objects above some size limit, for which rotational
fission mechanisms related to the YORP effect become less likely.

By limiting our attention to main belt asteroids, the list of binary
systems includes currently a total of $88$ asteroids\footnote{A list
  of identified binary systems is maintained by R. Johnston
  at the URL http://johnstonsarchive.net/astro/asteroidmoons.html},
including $34$ systems which are not definitively confirmed. Among
them, $17$ (including $6$ binaries still needing confirmation) are
family members according to our results; see Table
\ref{tab:binaries}.

The data set of definitively confirmed is still fairly limited, but
it is interesting to note that $17$ out of $88$, or $11$ out $54$
objects if we consider only certain binary systems, turn out to be
family members. The relative abundance is of the order of $20$\%, so
we cannot conclude that binary asteroids are particularly frequent
among families. Looking more in detail at the data shown in Table
\ref{tab:binaries}, we find that binaries tend to be more abundant
in the Koronis, Hungaria and Themis families, while the situation is
not so clear in the case of Vesta, due to the still uncertain binary
nature of some objects. We also note that binary asteroids tend also
to be found among the biggest
members of some families, including Sylvia, Minerva and Duponta.

\subsection{Couples}
\label{sec:couples}

One step of our family classification procedure is the computation of
the distance in proper elements space between each couple of
asteroids; the computation is needed only if the distance is less than
some control $d_{min}$. If the value of $d_{min}$ is chosen to be much
smaller than the QRL values used in the family classification, a new
phenomenon appears, namely the \emph{very close couples}, with
differences in proper elements corresponding to few m/s.

A hypothesis for the interpretation of asteroid couples, very close
in proper elements, has been proposed long ago, see \cite{trojan}[p.
166-167]. The idea, which was proposed by the late P.\ Farinella, is
the following: the pairs could be obtained after \emph{an
intermediate stage as binary}, terminated by a low velocity escape
through the so-called fuzzy boundary, generated by the heteroclinic
tangle at the collinear Lagrangian points.

\begin{table}[h!]
\footnotesize
 \centering
  \caption{Very close couples among the numbered asteroids: the $d$
    distance corresponds to $<0.5$ m/s. }
\medskip
  \begin{tabular}{rrrrrrrr}
  \hline
name & H &name&H& $d$ & $\delta a_p/a_p$ & $\delta e_p$ & $\delta \sin I_p$\\
\hline
%  name      H     name         H      sigma       da/a        de        dsinI   
92652    & 15.11 & 194083  &  16.62 & 0.1557213 &  0.0000059 & -0.0000011 &  0.0000011\\
27265    & 14.72 & 306069  &  16.75 & 0.2272011 &  0.0000076 &  0.0000021 & -0.0000029\\
88259    & 14.86 & 337181  &  16.99 & 0.2622311 & -0.0000091 &  0.0000019 & -0.0000011\\
180906   & 17.41 & 217266  &  17.44 & 0.2649047 &  0.0000069 & -0.0000049 & -0.0000013\\
60677    & 15.68 & 142131  &  16.05 & 0.3064294 &  0.0000090 &  0.0000021 &  0.0000052\\
165389   & 16.31 & 282206  &  16.85 & 0.3286134 &  0.0000019 &  0.0000080 &  0.0000022\\
188754   & 16.29 & 188782  &  16.90 & 0.3384864 & -0.0000059 & -0.0000019 &  0.0000087\\
21436    & 15.05 & 334916  &  18.14 & 0.3483815 & -0.0000041 & -0.0000081 &  0.0000016\\
145516   & 15.39 & 146704  &  15.57 & 0.4131796 & -0.0000142 & -0.0000062 &  0.0000046\\
76111    & 14.55 & 354652  &  16.55 & 0.4137084 &  0.0000174 &  0.0000033 & -0.0000011\\
67620    & 15.35 & 335688  &  16.91 & 0.4225815 &  0.0000017 &  0.0000027 &  0.0000105\\
52009    & 15.15 & 326187  &  17.22 & 0.4459464 &  0.0000079 &  0.0000115 & -0.0000020\\
39991    & 14.15 & 340225  &  17.90 & 0.4495984 &  0.0000003 & -0.0000101 & -0.0000061\\
64165    & 15.14 & 79035   &  14.54 & 0.4501940 & -0.0000018 &  0.0000010 &  0.0000127\\
57202    & 15.34 & 276353  &  17.45 & 0.4543033 & -0.0000027 &  0.0000103 &  0.0000052\\
180255   & 16.85 & 209570  &  17.08 & 0.4686305 &  0.0000185 & -0.0000026 &  0.0000003\\
39991    & 14.15 & 349730  &  17.35 & 0.4686824 &  0.0000182 & -0.0000020 & -0.0000043\\
56285    & 14.98 & 273138  &  16.72 & 0.4958970 &  0.0000199 & -0.0000061 &  0.0000025\\

\hline
\end{tabular}\label{tab:couples}
\end{table}

The procedure to actually prove that a given couple is indeed the
product of the split of a binary is complex, typically involving a
sequence of filtering steps, followed by numerical integrations (with
a differential Yarkovsky effect, given the differences in size) to find
an epoch for the split and confirm that the relative velocity was
indeed very small. Many authors have worked on this, including
ourselves \cite{hungaria} and \cite{Delloroetal12}, who analyzed the
efficiency of non-disruptive collisional events in leading to binary
``evaporation''. Our goal for this paper is not to confirm a large
number of split couples, but just to offer the data for confirmation
by other authors, in the form of a very large list of couples with
very similar proper elements.

Currently we are offering a dataset of $14\,627$ couples with distance
$<10$ m/s, available from
AstDyS\footnote{http://hamilton.dm.unipi.it/~astdys2/propsynth/numb.near}.
A small sample of these couples is given in Table~\ref{tab:couples},
for $d<0.5$ m/s.

To assess the probability of finding real couples in this large
sample, it is enough to draw an histogram of the distance $d$. It
shows the superposition of two components, one growing quadratically
with $d$ and one growing linearly. Since the incremental growth of
volume is quadratic in $d$, this should correspond to the fraction of
couples which are random and to the ones which result from a very
different phenomenon, such as split followed by drift due to
differential Yarkovsky. It turns out that from the histogram it is
possible to compute that, out of $14\, 627$ couples, about half
belong to the ``random'' sub-population, half to the linear growth
component.

\section{Conclusions and Future Work}
\label{sec:conclusion}

By performing an asteroid family classification with a
very enlarged dataset the results are not just ``more families'', but
there are interesting qualitative changes. These are due to the large
number statistics, but also to the larger fraction of smaller objects
contained in recently numbered asteroids and to the accuracy
allowing to see many structures inside the families\footnote{Note that
  these three are reasons to discard the usage of physical
  observation data as primary parameters for classification.
  Consistent catalogs of physical observation are available for less
  asteroids, they are also observationally biased against smaller
  asteroids, which are either absent or present with very low accuracy
  data.}.
 
Another remarkable change is that we intend to keep this
classification up to date, by the (partially automated) procedures we
have introduced.

In this section we would like to summarize some of these changes, and
also to identify the research efforts which can give the most
important contribution to this field. Note that we do not necessarily
intend to do all this research ourselves, given our complete open data
policy this is not necessary.

\subsection{How to use HCM}

The size increase of the dataset of proper elements has had a negative
effect on the perception about the HCM method in the scientific
community, because of the chaining effect which tends to join even
obviously distinct families. Thus some authors have either reduced the
dataset by asking that some physical observations are available, or
used QRL values variable for each family, or even reverted to visual
methods.

We believe that this paper shows that there is no need to abandon the
HCM method, provided a more complex multistep procedure is adopted. In
short, our procedure amounts to using a truncation QRL for the larger
members belonging to the core family different from the one used for
smaller members. This is justified from the statistical point of view,
because the smaller asteroids have larger number density. 

We are convinced that our method is effective in adding many smaller
asteroids to the core families, without expanding the families with
larger members. As a result we have a large number of families with
very well defined V-shapes, thus with a good possibility of age
estimation. We have also succeeded in identifying many families formed
only with small asteroids, or at most with very few large ones, as
expected for cratering.

The portion of the procedure on which we intend to work more is the
step 5, merging, which is still quite subjective (indeed, even visual
inspection plays a significant role).  This procedure is cumbersome
and not automated at all, and may become even more difficult in the
future with the expected large increase of the dataset size, thus further
improvements are needed.

\subsection{Stability of the classification}

We have established a procedure to maintain our family classification
up to date, by adding the newly discovered asteroids, as soon as their
orbital elements are stable enough because determined with many
observations. First the proper elements catalogs are updated, then we
attribute some of the new entries as new members to the already known
families. This last step 6 (see Section~\ref{sec:method}) is performed
in a fully automated way, just by running a single computer code (with
CPU times of the order of 1 hour), which is actually the same code
used for steps 2 and 4.

We already have done this for the proper elements catalog update of
April 2013, and we continue with periodic updates. The results are
available, as soon as computed, on the AstDyS information system.

The classification is meant to be methodologically stable but
frequently updated in the dataset.  In this way, the users of our
classification can download the current version. They can find a full
description of the methods in this paper, even if specific results
(e.g., number and list of members of some family) should not be taken
from the paper but from the updated web site.

%\textbf{For your eyes only: abrupt increase of dataset due to
%  computation of proper elements for more known asteroids, e.g.,
%  multiopposition: we actually plan to do this in the near future, but
%  not to say it in the paper.}

Some changes in the classification such as mergers are not automatic,
thus we are committed to apply them periodically, until when we will
be able to automatize this too.

\subsection{Magnitudes}

As shown in Table~\ref{tab:infocount}, the absolute magnitudes
computed with the incidental photometry could contribute a significant
fraction of the information used in family classification and
analysis. However, the accuracy is poor, in most cases not
estimated; better data are available for smaller samples, not
enough for statistical purposes, such as the analysis of
Section~\ref{sec:absol_mag}.
This does not affect the classification, but has a very negative
impact on the attempt to compute ages, size distributions, and
volumes. It also affects the accuracy of the albedos, and has serious
implications on the population models.

The question is what should be done to improve the situation. One
possibility would be a statistically rigorous debiasing and weighting
of the photometry collected with the astrometry. The problem is, the
errors in the photometry, in particular the ones due to difference in
filters and star/asteroid colors, are too complex for a simple
statistical model, given that we have no access to a complete
information on the photometric reduction process.

Thus we think that the only reliable solution is to have an optimally
designed photometric survey, with state of the art data processing,
including the new models of the phase effect \cite{muinonen}. This
requires a large number (of the order of $100$) photometric
measurements per asteroid per opposition, with a wide filter, and with
enough S/N for most numbered main belt asteroids. These appear tough
requirements, and a dedicated survey does not appear a realistic
proposal.  However, it turns out that these requirements are the same
needed to collect enough astrometry for a NEO wide survey, aiming at
discovering asteroids/comets on the occasion of close approaches to
the Earth. Thus a ``magnitude survey'' could be a byproduct of a NEO
discovery survey, provided the use of many different filters is
avoided.

\subsection{Yarkovsky effect and ages}

One of our main results is that for most
families, large enough for statistically significant analysis of the
shape, the V-shape is clearly visible. 

We have developed our method to compute ages, which we believe is
better than those used previously (including by ourselves,
\cite{hungaria}) because it is more objective and takes into account
the error in the estimate of the diameter by the common albedo
hypothesis, which is substantial. 

We believe our new method tackles in an appropriate way all the
difficulties of the age estimation discussed in
Section~\ref{sec:ages}, but for one: the calibration. The difficulty
in estimating the Yarkovsky calibration, due to the need to
extrapolate from NEAs with measured $da/dt$ to main belt asteroids, is
in most cases the main limitation to the accuracy of the age
estimation. 

Thus the research effort which could most contribute to the
improvement of age estimation (for a large set of families) would be
either the direct measurement of Yarkovsky effect for some family
members (with known obliquity) or the measurement of the most
important unknown quantities affecting the scaling from NEA, such as
thermal conductivity and/or density. These appear ambitious goals, but
they may become feasible by using advanced observation techniques, in
particular from space, e.g., GAIA astrometry, Kuiper infrared
observations, and radar from the ground.

\subsection{Use of physical observations}

In this paper we have made the choice of using the dynamical data
first, to build the family classification, then use all the available
physical data to check and refine the dynamical families.

This is best illustrated by the example of the Hertha/Polana/Burdett
complex dynamical family, in which the identification of the two
collisional families Polana and Burdett can be obtained only with the
physical data. This leaves us with more understanding, but with an
incomplete classification because the majority of the members of the
Hertha dynamical family there are no physical data. If we had a much
larger database of accurate albedos and/or colors, we would be very
glad to use the separation by physical properties as part of the
primary family classification. The same argument could apply to the
separation of the two collisional families of Vesta.

In other words, the use of dynamical parameters first is not an
ideological choice, but is dictated by the availability and accuracy
of the data. We would very much welcome larger catalogs of physical
data, including much smaller asteroids and with improved S/N for those
already included. However, we are aware that this would require larger
aperture telescopes. 

\subsection{Cratering vs. Fragmentation}
\label{sec:crat_frag}

Our procedure, being very efficient in the inclusion of
small asteroids in the haloes of core families, has allowed to
identify new cratering families and very large increases
of membership for the already known ones.

As a result of the observational selection effect operating against
the cratering families, because they contain predominantly small
asteroids (e.g., $D<5$ km), in the past the cratering events have been
less studied than the catastrophic fragmentations. On the contrary,
elementary logic suggests that there should be more cratering than
fragmentation families, because the target of a cratering remains
available for successive impacts, with the same cross section. The
number of recognized cratering families is reduced because small
asteroids from craterizations have a shorter lifetime, thus the
observable cratering families have a limited age. As the observational
bias against small asteroids is progressively mitigated, we expect
that the results on cratering will become more and more important.

This argument also implies that multiple cratering collisional
families should be the rule rather than the exception. They
necessarily intersect because of the common origin but do not overlap
completely because of the different crater radiants. Although we have
studied only some of the cratering families (listed
in Section~\ref{sec:craters}), all the examples we
have analyzed, namely the families of (4), (20), (15), show
a complex internal structure.

For the catastrophic fragmentation families, the two examples we have
analyzed, (158) and (847), appear to contain significant substructures
(named after Karin and Jitka). The general argument could be used that
fragments are smaller, thus should have collisional lifetimes shorter
than the parent body of the original fragmentation family. Thus we
should expect that as fragmentation families become larger in
membership and include smaller asteroids, substructures could emerge
in most of the families.

A full fledged collisional model for the ejection velocities and
rotations of fragments from both cratering and fragmentation needs to
be developed, possibly along the lines of the qualitative model
of Section~\ref{sec:collmodel}.  This will contribute both to the
age determination and to the understanding of the family formation
process.
As for future research, there is the need to search for substructures and to
classify as cratering/fragmentation, this for all the big families of
Table~\ref{tab:bigfam} and many of the medium families of
Table~\ref{tab:mediumfam}.

\subsection{Comparison with space mission data}

When on-site data from spacecraft, such as DAWN, become available for
some big asteroid, a family classification should match the evidence,
in particular from craters on the surface. For Vesta the main problem
is the relationship between the dynamical family 4 and the two main
craters Rheasilvia and Veneneia. The solution we are suggesting is
that the two subfamilies found from
internal structure of family 4 correspond to the two main craters. We
have found no contradiction with this hypothesis in the data we can
extract from the family, including ages. However, to prove this
identification of the source craters requires more work:
more accurate age estimates for both subfamilies and
craters, and more sophisticated models of how the ejecta from a large
crater are partitioned between ejecta
reaccumulated, fragments in independent orbits but
too small, and detected family members.

Because the problem of the missing Ceres family is difficult, due to
apparently discordant data, we have tried to build a consistent model,
with an interpretation of the dynamical family 93 (without the
namesake) and two possible physical explanations of the
inefficiency of Ceres in generating families. We are not claiming
these are the the only possible explanations, but they appear
plausible.  The data from DAWN in 2015 should sharply
decrease the possibilities and should lead to a well
constrained solution.

\section*{Acknowledgments}

The authors have been supported for this research by: 
%the Italian Space Agency, under the contract ASI/INAF I/015/07/0, (A.M., A.C); 
the Ministry of Science and Technological Development of Serbia, under
the project 176011 (Z.K. and B.N.). We thank S. Marchi, M. Micheli and S. Bus
for discussions which have contributed to some of the ideas of this paper.

\end{document}